\documentclass[twocolumn,showpacs,showkeys,preprintnumbers,amsmath,amssymb]{revtex4}
 \usepackage{dcolumn}
 \usepackage{bm}
 \usepackage{graphicx}
 \usepackage{subfigure}
 \usepackage{color}
 \usepackage{multirow}

\begin{document}
 \title{Energy and entropy compensation, phase transition and kinetics of four dimensional charged Gauss-Bonnet Anti-de Sitter black holes on the underlying free energy landscape}

\author{Ran Li$^{1,2}$}
\thanks{liran@htu.edu.cn}

\author{Jin Wang$^{2,3}$}
\thanks{Corresponding Author: jin.wang.1@stonybrook.edu}

\affiliation{
 $^1$ School of Physics, Henan Normal University, Xinxiang 453007, China\\
 $^2$ Department of Chemistry, State University of New York at Stony Brook, Stony Brook, New York 11794, USA\\
 $^3$ Department of Physics and Astronomy, State University of New York at Stony Brook, Stony Brook, New York 11794, USA}

\begin{abstract}
By treating black hole as a state in canonical ensemble or in grand canonical ensemble, we study the phase transition and the kinetics of the four dimensional charged Anti-de Sitter (AdS) black hole in Gauss-Bonnet (GB) gravity based on the free energy landscape. Below the critical temperature, the free energy landscape topography has the shape of double basins with each representing one stable/unstable black hole phase (the small or the large black hole). The thermodynamic small/large black hole phase transition is determined by the equal depths of the basins. We also demonstrate the underlying kinetics of the phase transition by studying the time evolution of the probability distribution of the state in the ensemble as well as the mean first passage time (MFPT) and the kinetic fluctuations of the state switching process caused by the thermal fluctuations. The results show that the final distribution is determined by the Boltzmann law and the MFPT and its fluctuation are closely related to the free energy landscape topography through the barrier heights between the small and the large black hole states and the ensemble temperature. Furthermore, we provide a complete description of the kinetics of phase transition by investigating the temperature dependence of the MFPT and the kinetic fluctuations with different physical parameters. It is shown that the free energy is the result of the delicate balance and competition between the two relatively large numbers, the energy and entropy multiplied by temperature compared to $kT$. Low energy (mass) and low entropy can give rise to a stable thermodynamic state in terms of free energy minimum (energy/mass preferred) while the high energy (mass) and high entropy (entropy preferred) can also give rise to a stable state in terms of free energy minimum. The comparable free energy barrier with respect to $kT$ makes it possible for the switching from small size black hole state to the large size black hole state under thermal fluctuations and vice versa. When the GB coupling coupling constant increases, or the electric charge (potential) increases, or the pressure (absolute value of cosmological constant) decreases, it is easier for the small black hole state to escape to the large black hole state. Meanwhile, the inverse process becomes harder, i.e. the small (large) black hole state becomes less (more) stable.
\end{abstract}

\maketitle

\section{Introduction}

Black holes are the fascinating objects in Einstein's general relativity. However, the breakdown of general relativity is expected to occur near the singularity inside the black hole, which promotes the studies of the modified theory of gravitation. GB gravity is an important alternative theory of general relativity. It modifies the Einstein-Hilbert action to include the GB term. However, the GB gravity is usually studied in high dimensions because the GB term in four dimensions is a topological invariant and has no contribution to the gravitational dynamics. Recently, a non-trivial four dimensional GB gravity was proposed \cite{Glavan:2019inb} by rescaling the coupling constant as $\alpha\rightarrow \frac{\alpha}{d-4}$ and then considering the limit $d\rightarrow 4$, which is shown to bypass the Lovelock's theorem \cite{Lovelock} and avoid Ostrogradsky instability \cite{Woodard}. Furthermore, for a positive GB coupling constant, the spherically symmetric black hole solution is free from the singularity problem. However, it is shown that the above regularization scheme cannot lead to a well-defined theory \cite{Gurses:2020ofy,Hennigar:2020lsl,Bonifacio:2020vbk,Arrechea:2020evj,Mahapatra:2020rds}. A consistent description of GB gravity in four dimensions was proposed in \cite{Aoki:2020lig}.

Due to these features, the four dimensional GB gravity has recently attracted great attentions, including the studies of thermodynamics and phase transitions of the four dimensional GB black holes. It should be noted that a number of black hole solutions of the original proposal \cite{Glavan:2019inb} also satisfy the field equations of the theory suggested in \cite{Aoki:2020lig}. In \cite{Fernandes:2020rpa}, the vacuum solution found by Glavan and Lin in \cite{Glavan:2019inb} was generalized to be charged in AdS space. Then, the thermodynamics and phase transitions in extended phase space, where the cosmological constant is taken as the thermodynamic pressure, was studied in \cite{Hegde:2020xlv,Singh:2020xju,HosseiniMansoori:2020yfj,Wei:2020poh,
EslamPanah:2020hoj,Singh:2020mty,Li:2020vpo,Wang:2020pmb,Shaymatov:2020yte}. Especially, it is shown in \cite{Wei:2020poh} that there exists the small-large black hole phase transition both in the canonical ensemble and in the grand canonical ensemble, which has an analogy of the van der Walls liquid-gas phase transition.

This work is intended to study the phase transition and the underlying kinetics of the four dimensional charged GB black holes based on the free energy landscape formulism. For various systems in physics, chemistry, and biology, free energy landscape is a very important concept and tool that can be used to study the phase transition phenomenon \cite{FSW,FW,NG,JW}. Free energy landscape formulism was initiated to study the Hawking-Page phase transitions in Einstein gravity and massive gravity in \cite{Li:2020khm}. Then, it was also applied to investigate the small-large Reissner-Nordstr\"{o}m-AdS (RNAdS) black hole phase transition \cite{Li:2020nsy}. These studies provide us significant insights into the understanding of the underlying thermodynamics and kinetics of the black hole phase transition. This motivates us to study the phase transition of GB black holes and the associated kinetics by using the approach of the free energy landscape and investigate the influence of the higher derivative terms on the black hole phase transition.

In order to apply the free energy landscape to study the kinetics of black hole phase transition, we have made a series of assumptions \cite{Li:2020khm,Li:2020nsy}, which are summarized as follows. Firstly, we regard the horizon radius as the order parameter to describe the microscopic degree of freedom or the black hole molecules \cite{WeiLiuPRL,WeiLiuMann}. Then, we construct the canonical ensemble at the specific temperature which is composed of a series of black hole spacetimes with the horizon radius ranging from zero to infinity. The generalized Gibbs free energy for every spacetime state in the ensemble was proposed as the function of the order parameter and the ensemble temperature. In terms of the Gibbs free energy topography, we can investigate the thermodynamic stability of the emergent phases and the associated phase diagram. Furthermore, the stochastic dynamics of black hole phase transition under the thermal fluctuations can be studied by using the associated probabilistic Fokker-Planck equation on the free energy landscape \cite{NSM,WangPRE,WangJCP,BW}. The underlying kinetics of phase transition is then revealed by the first passage problem of the thermal free energy barrier crossing process.

In the present work, our aim is to apply the free energy landscape and the kinetics formulism to study the van der Waals type phase transition of the novel four dimensional charged GB AdS black holes. It should be noted that, for the RNAdS black hole in the Einstein gravity, there is no small-large black hole phase transition in the grand canonical ensemble \cite{Mann}. Therefore, our previous study was performed in the canonical ensemble \cite{Li:2020nsy}. However, for the four dimensional charged GB AdS black hole, there is a small-large black hole phase transitions in both the canonical ensemble and the grand canonical ensemble \cite{Wei:2020poh}. Therefore, we will generalize the free energy landscape constructed in the canonical ensemble to the grand canonical ensemble by defining the corresponding free energy function and study the thermodynamics and the underlying kinetics of the black holes in the canonical ensemble as well as in the grand canonical ensemble. Furthermore, we will also provide a complete description of the kinetics of GB AdS black hole phase transition by studying the temperature dependence of the mean first passage time and the kinetic fluctuations with different physical parameters (GB coupling constant, electric charge or electric potential, and thermodynamic pressure). It is shown that the free energy is the result of the delicate balance and competition between the two relatively large numbers, the energy and entropy multiplied by temperature compared to $kT$. Low energy (mass) and low entropy can give rise to a stable thermodynamic state in terms of free energy minimum (energy/mass preferred) while the high energy (mass) and high entropy (entropy preferred) can also give rise to a stable state in terms of free energy minimum. The comparable free energy barrier with respect to $kT$ makes it possible for the switching from small size black hole state to the large size black hole state under thermal fluctuations and vice versa.

This paper is organized as follows. In section II, we give a brief review of the four dimensional GB gravity theory and the thermodynamics of the four dimensional charged GB AdS black hole. In section III, the thermodynamics and the kinetics of the small-large GB AdS black hole phase transition are studied based on the free energy landscape in the canonical ensemble. In section IV, we perform the analysis in the grand canonical ensemble. The conclusion and discussion are presented in the last section.

\section{Thermodynamics of Four Dimensional Gauss-Bonnet Black Hole}

Consider the action of the $d$-dimensional Einstein-Maxwell theory with a GB term and a
cosmological constant
\begin{eqnarray}\label{action}
S=\frac{1}{16\pi} \int d^d x \sqrt{-g} \left[ R-2\Lambda
+\frac{\alpha}{d-4}\mathcal{L}_{GB}-F^{\mu\nu}F_{\mu\nu}\right],
\end{eqnarray}
where $R$ is Ricci scalar, $\Lambda=-\frac{(d-1)(d-2)}{2L^2}$ is the $d$-dimensional cosmological constant, and $\alpha$ is the GB coupling constant. The GB Lagrangian $\mathcal{L}_{GB}$ is given by
\begin{eqnarray}
\mathcal{L}_{GB}=R^2-4 R^{\mu\nu}R_{\mu\nu}+R^{\mu\nu\lambda\rho}R_{\mu\nu\lambda\rho}\;.
\end{eqnarray}
$F_{\mu\nu}=\partial_\mu A_{\nu}-\partial_\nu A_{\mu}$ is the Maxwell field strength and $A_{\mu}$ is the $U(1)$ electromagnetic gauge potential.

The spherically symmetric solution for the action (\ref{action}) is
\begin{eqnarray}
ds^2=-f(r)dt^2+f(r)^{-1}dr^2+r^2 d\Omega_{d-2}^2\;,
\end{eqnarray}
where $d\Omega_{d-2}$ is the line element of the $(d-2)$-dimensional sphere. In the limit $d\rightarrow 4$, the lapse function $f(r)$ is given by \cite{Fernandes:2020rpa}
\begin{eqnarray}
f(r)=1+\frac{r^2}{2\alpha}\left[1-\sqrt{1+4\alpha\left(\frac{2M}{r^{3}}
-\frac{1}{L^2}-\frac{Q^2}{r^{4}}\right)}\right]\;,
\end{eqnarray}
where $M$ and $Q$ are the mass and the charge of the black hole. This solution is supported by an $U(1)$ electromagnetic gauge potential given by
\begin{eqnarray}
A_t=\frac{Q}{r}\;.
\end{eqnarray}

The black hole event horizon $r_h$ is determined by the largest real root of the equation $f(r)=0$. The mass of the black hole can then expressed in terms of the black hole horizon $r_h$ as
\begin{eqnarray}
M=\frac{r_h}{2}\left(1+\frac{r_h^2}{L^2}+\frac{\alpha}{r_h^2}+\frac{Q^2}{r_h^2}\right)\;.
\end{eqnarray}
The Hawking temperature of the black hole which can be calculated from the surface gravity is given by \cite{Fernandes:2020rpa}
\begin{eqnarray}\label{HawkingTem}
T_H=\frac{3r_h^4/L^2+r_h^2-Q^2-\alpha}{4\pi r_h(r_h^2+2\alpha)}\;.
\end{eqnarray}

We discuss the thermodynamics of the black hole in an extended phase space, where the
cosmological constant $\Lambda$ is treated as the thermodynamic pressure $P$ by using the relation
\cite{reviewPVcriticality}
\begin{eqnarray}
P=-\frac{\Lambda}{8\pi}=\frac{3}{8\pi L^2}\;.
\end{eqnarray}
When treating the GB coupling constant $\alpha$ as a new thermodynamic variable, the first law of black hole thermodynamics can be written as \cite{Wei:2020poh}
\begin{eqnarray}
dM=T_HdS+\Phi dQ+VdP+\mathcal{A} d\alpha\;,
\end{eqnarray}
where $\Phi=A_t(r_h)=Q/r_h$ is the electrostatic potential on the horizon, and $V$ and $A$ are
the conjugate variables of the thermodynamic pressure $P$ and GB coupling parameter $\alpha$. From the first law, the black hole entropy can be calculated as \cite{Fernandes:2020rpa}
\begin{eqnarray}
S=\pi r_h^2+4\pi\alpha \log(r_h)\;,
\end{eqnarray}
where the integral constant is taken as $1$ without loss of generality.

\section{Phase transition and Kinetics in canonical ensemble}

\subsection{Phase transition based on the free energy landscape}

Now, Let us analyze the thermodynamic stability and the phase transition of the four dimensional GB AdS black hole based on the free energy landscape in the canonical ensemble. In this subsection, we will show the emergence of the small black hole and the large black hole states as well as the phase transition by using the free energy landscape. We consider the canonical ensemble at the specific temperature $T$, which is composed of a series of black hole spacetime states with an arbitrary horizon radius. We define the generalized off-shell Gibbs free energy for the transient black hole state as \cite{Li:2020nsy}
\begin{eqnarray}\label{Gibbs}
G&=&M-TS\nonumber\\
&=&\frac{r_h}{2}\left(1+\frac{8\pi P}{3}r_h^2+\frac{\alpha}{r_h^2}+\frac{Q^2}{r_h^2}\right)
\nonumber\\
&&-T\left(\pi r_h^2+4\pi\alpha \log(r_h)\right)\;.
\end{eqnarray}
Note that the horizon radius $r_h$ can take all values from zero to infinity, because $r_h$ is considered to be the order parameter describes the microscopic degree of freedoms. It should also be noted that the mass of an AdS black hole is identified as the enthalpy of the spacetime \cite{Kastor:2009wy,Kastor:2010gq,Kastor:2011qp}. In this way, we can formulate the free energy landscape for the charged GB AdS black holes in the canonical ensemble. It should also be noted that all the information about the emergence of the thermodynamically stable black hole states as well as the phase transition between these states can be obtained from the generalized off-shell Gibbs free energy.

The thermodynamically stable black hole states corresponds to the local extremals of the Gibbs free energy function, which are determined by the equation
\begin{eqnarray}\label{pGibbs}
\frac{\partial G}{\partial r_h}=\frac{1}{2}+4\pi P r_h^2-\frac{\alpha+Q^2}{2r_h^2}-2\pi T r_h
-\frac{4\pi \alpha T}{r_h}=0\;.
\end{eqnarray}
When solving for the ensemble temperature $T$, the above equation gives the expression of the Hawking temperature $T_H$ as shown in Eq.(\ref{HawkingTem}). This implies that the on-shell GB AdS black hole solutions to the Einstein equations correspond to the local extremals of the free energy function.

In general, we can quantify the free energy landscape by plotting the Gibbs free energy as a function of the black hole radius $r_h$. The shape of the free energy landscape depends on the ensemble temperature. It can be observed that there is a specific temperature range where the free energy landscape has the shape of the double well. The minimal temperature $T_{min}$ and the maximum temperature $T_{max}$ are determined by Eq.(\ref{pGibbs}) and the following equation
\begin{eqnarray}\label{ppGibbs}
\frac{\partial^2 G}{\partial r_h^2}=8\pi P r_h+\frac{\alpha+Q^2}{r_h^3}-2\pi T+\frac{4\pi \alpha T}{r_h^2}=0\;.
\end{eqnarray}
The equation $\frac{\partial^2 G}{\partial r_h^2}=0$ indicates that there is an inflection point in the free energy landscape topography, which implies the appearance or the disappearance of the double well shape.

Combining Eq.(\ref{pGibbs}) and Eq.(\ref{ppGibbs}) and eliminating $T$, we can get the equation
\begin{eqnarray}
8\pi P r_h^6+(48\pi\alpha P-1)r_h^4+(3Q^2+5\alpha)r_h^2&&
\nonumber\\
+2\alpha Q^2 + +2\alpha^2=0&&\;.
\end{eqnarray}
When the minimal and the maximum temperatures coincide, the system lies at the critical point. This implies that, at the critical point, the above equation has two equal roots. This condition gives us the critical pressure, the critical temperature, as well we the critical black hole radius as follows
\begin{eqnarray}
T_c&=&\frac{\left(\sqrt{48\alpha^2+9Q^4+48\alpha Q^2}-3Q^2\right)}
{24\pi\alpha\sqrt{6\alpha+3Q^2+\sqrt{48\alpha^2+9Q^4+48\alpha Q^2}}}\;,\\
P_c&=&\frac{9\alpha+6Q^2+\sqrt{48\alpha^2+9Q^4+48\alpha Q^2}}{24\pi \left(6\alpha+3Q^2+\sqrt{48\alpha^2+9Q^4+48\alpha Q^2}\right)^{2}}\;,\\
r_{hc}&=&\sqrt{6\alpha+3Q^2+\sqrt{48\alpha^2+9Q^4+48\alpha Q^2}}\;,
\end{eqnarray}
which coincides with the result found in \cite{Wei:2020poh}. Therefore, we have derived the critical point of the GB AdS black hole phase transition from the generalized Gibbs free energy function.

\begin{figure}
  \centering
  \includegraphics[width=6cm]{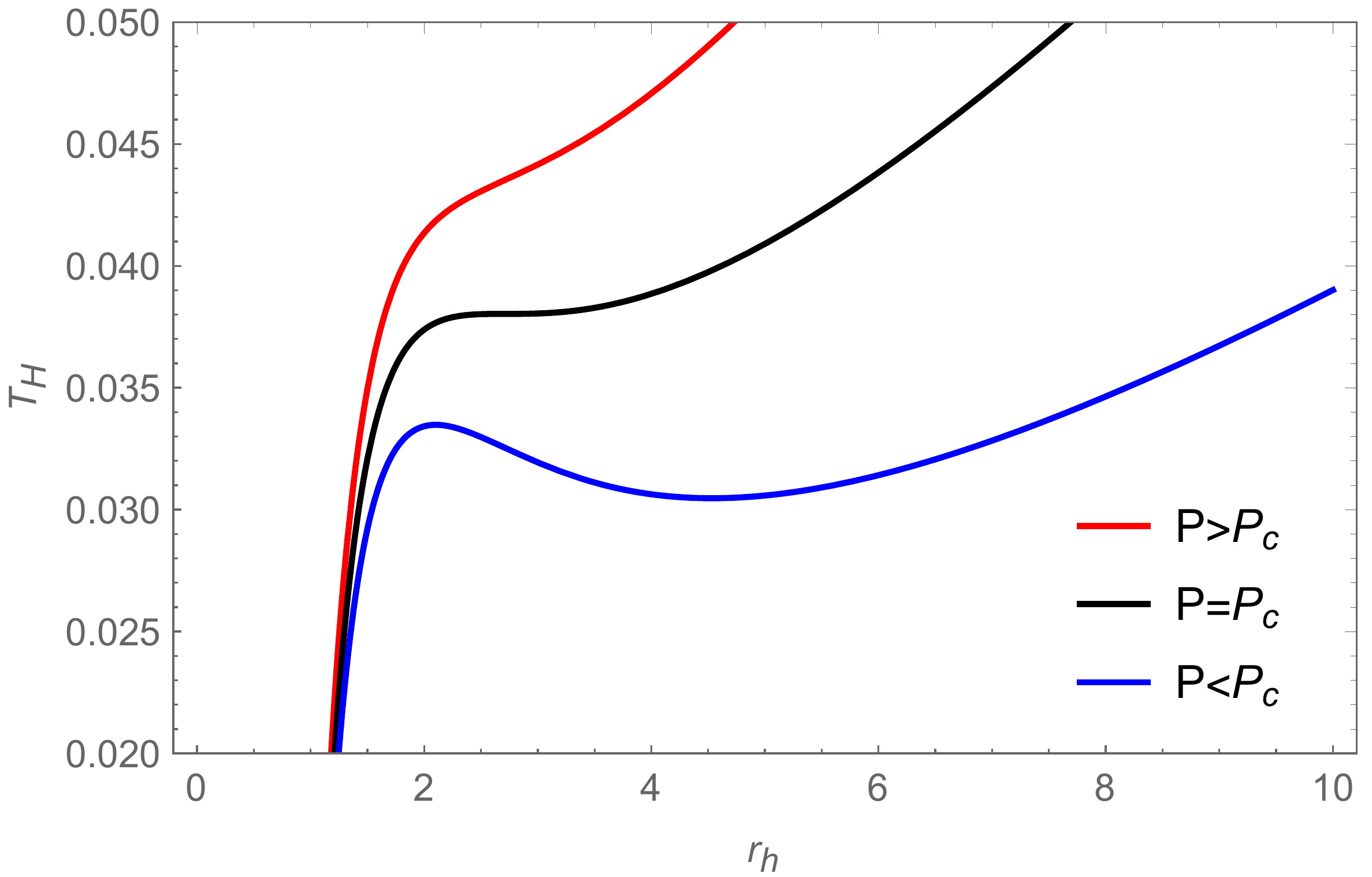}
  \caption{Black hole Temperature $T_H$ as a function of event horizon radius $r_h$ when $P<P_c$, $P=P_c$, and $P>P_c$. The electric charge $Q$ is $1$ and the coupling constant $\alpha$ is $0.1$.}
  \label{tempCE}
\end{figure}

In Fig.\ref{tempCE}, the Hawking temperature is depicted as a function of the black hole radius $r_h$ when $P<P_c$, $P=P_c$, and $P>P_c$. Above the critical pressure $P_c$ the black hole temperature $T_H$ is a monotonic function of the black hole radius $r_h$, while below the critical pressure the black hole temperature $T_H$ has the local extremal points. In our case, the local minimal and maximum values of the Hawking temperature are given by $T_{min}=0.03047$ and $T_{max}=0.03348$ for $P=0.6P_c$ and $Q=1$. When $T_{min}<T_H<T_{max}$, there exist three branches of the black hole solution (i.e. the small, the intermediate, and the large black hole of different horizon radius at the same Hawking temperature). In \cite{Wei:2020poh}, it is shown that there is a small-large black hole phase transition, which is similar to the van der Walls fluid.

\begin{figure}
  \centering
  \includegraphics[width=6cm]{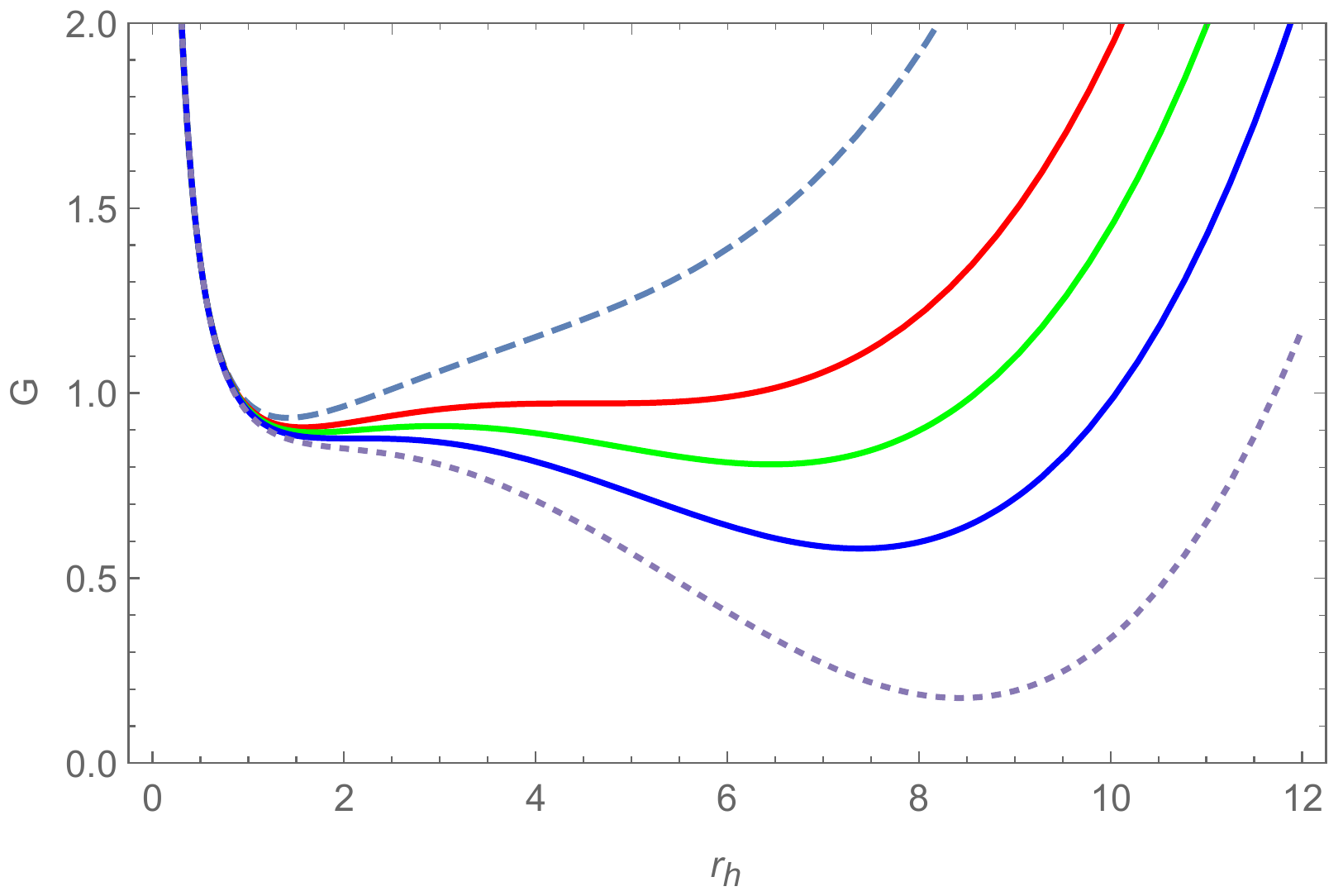}
  \caption{Gibbs free energy as a function of $r_h$ for $P=0.6P_c$ with different temperatures. The electric charge $Q$ is $1$ and the coupling constant $\alpha$ is $0.1$. The dashed, red, green, blue, and dotted curves correspond $T<T_{min}$, $T=T_{min}$, $T_{min}<T<T_{mas}$, $T=T_{max}$, and $T>T_{max}$, respectively. }
  \label{GibbsCE}
\end{figure}

\begin{figure}
  \centering
  \includegraphics[width=6cm]{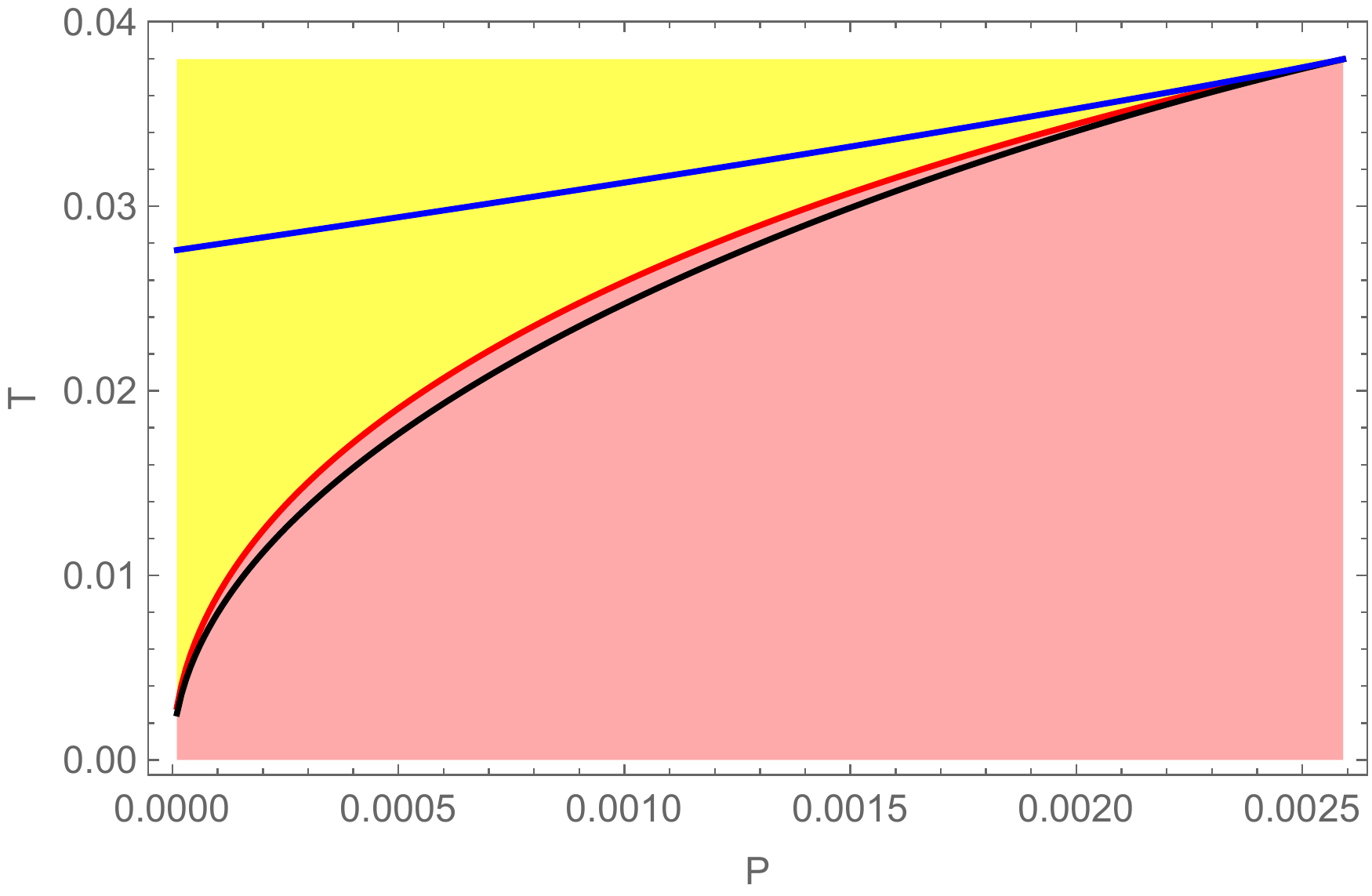}
  \caption{Phase diagram of black holes in canonical ensemble. The blue, black, and red lines are the plots of $T_{max}$, $T_{min}$, and $T_{trans}$ as a function of pressure $P$. In these plots, the range of $P$ is from $0$ to the critical pressure $P_c$. The small and the large black holes are thermodynamically stable in pink and yellow regions, respectively. }
  \label{PhaseDiagramCE}
\end{figure}

When $T<T_{min}$ or $T>T_{max}$, the free energy landscape topography has only one well or basin. The minimum point represents the small or the large black hole solution which is always thermodynamic stable. When $T_{min}<T<T_{max}$, the free energy topography has the shape of double wells (basins), which has the characteristic of the first order phase transition. In this case, there are three local extremal points corresponding to the three branches of black hole solutions. The intermediate black hole which corresponds to the local maximum point is unstable while the small black hole and large black hole are all locally stable. There exist a phase transition temperature $T_{trans}$, at which the free energies of the small and the large black hole are equal. Below/above $T_{trans}$, the basin of the small black hole is deeper/shallower than the basin of the large black hole. The thermodynamically stable black hole is then determined by the global minimum point in the free energy landscape topography. The phase diagram is presented in Fig.\ref{PhaseDiagramCE}. Note that the red curve represents the coexistence curve of the small and the large black holes with equal free energies.

\subsection{Probabilistic evolution on the free energy landscape}

In this subsection, we use the symbol $r$ to denote the black hole radius $r_h$ for simplicity. The free energy, which is a function of the black hole radius, can be denoted as $G(r)$. We now study the time evolution of the probability distribution in the canonical ensemble under the thermal fluctuations. As discussed, the canonical ensemble consists of a series of black hole spacetimes with arbitrary radius. The probability distribution of these black hole states $\rho(r, t)$, which is a function of the black hole radius $r$ and time $t$, satisfies the Fokker-Planck equation on the free energy landscape \cite{NSM,WangPRE,WangJCP,JW,BW}
\begin{eqnarray}\label{FPequation}
\frac{\partial \rho(r,t)}{\partial t}=D \frac{\partial}{\partial r}\left\{
e^{-\beta G(r)}\frac{\partial}{\partial r}\left[e^{\beta G(r)}\rho(r,t)\right]
\right\}\;,
\end{eqnarray}
where $\beta=1/kT$ is the inverse temperature and $D=kT/\zeta$ is the diffusion coefficient with $k$ being the Boltzman constant and $\zeta$ being dissipation coefficient. We take $k=\zeta=1$ without loss of generality.

In order to solve the Fokker-Planck equation, we impose the reflecting boundary condition
\begin{eqnarray}\label{bc1}
\left.
e^{-\beta G(r)}\frac{\partial}{\partial r}\left[e^{\beta G(r)}\rho(r,t)\right]\right|_{r=r_0}=0\;,
\end{eqnarray}
at the points $r_0$ where the free energy function $G(r)$ is divergent (e.g. at $r=0$ and $r=\infty$), and take the initial condition as
\begin{eqnarray}\label{initial}
\rho(r,0)=\frac{1}{\sqrt{\pi}a} e^{-(r-r_i)^2/a^2}\;,
\end{eqnarray}
with the parameter $a=1/7$. In practical numerical calculation, we typically set the reflecting boundary condition at $r=0.1$ and $r=10$ in order to avoid the numerical instability.

\begin{figure}
  \centering
  \includegraphics[width=6cm]{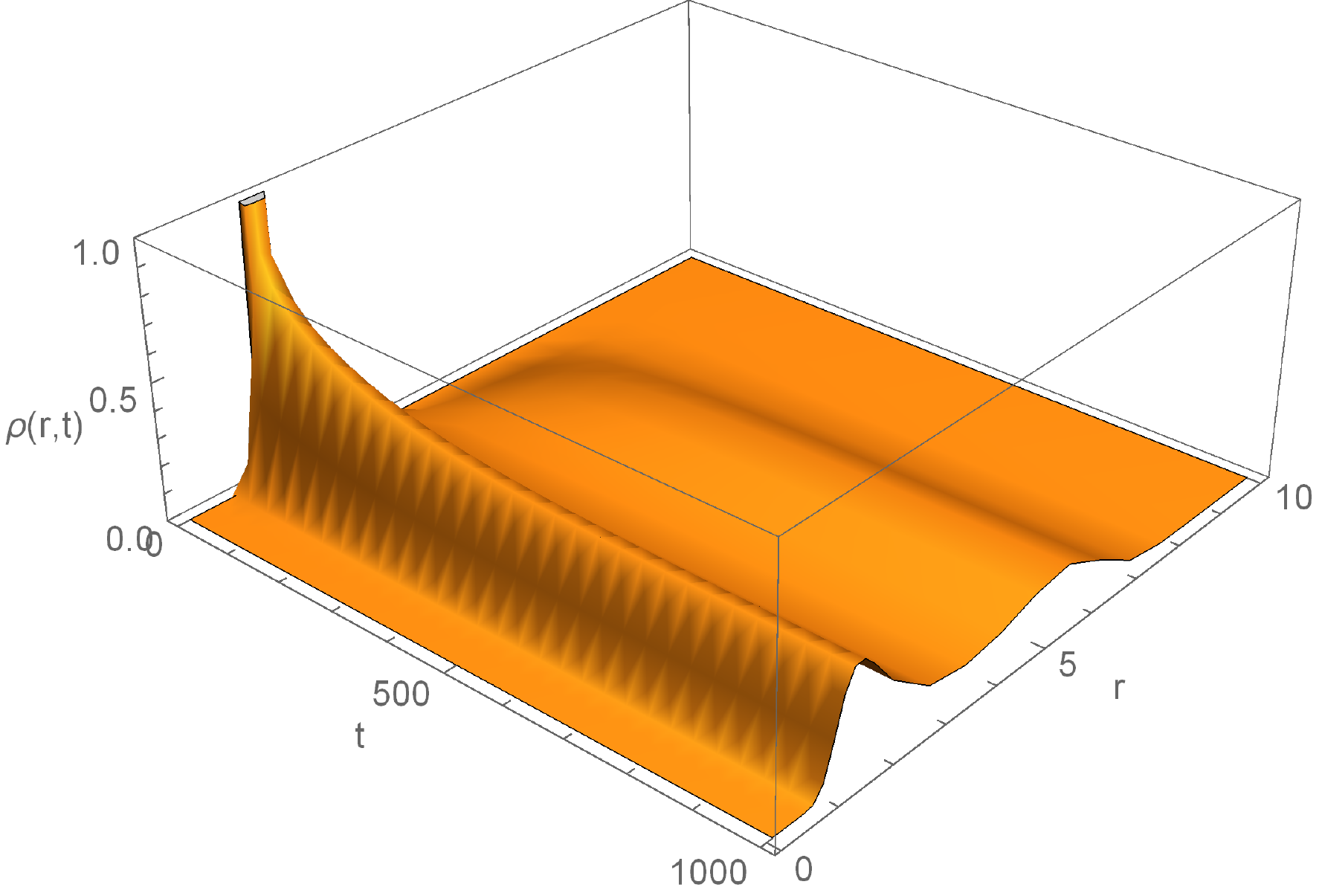}
  \includegraphics[width=6cm]{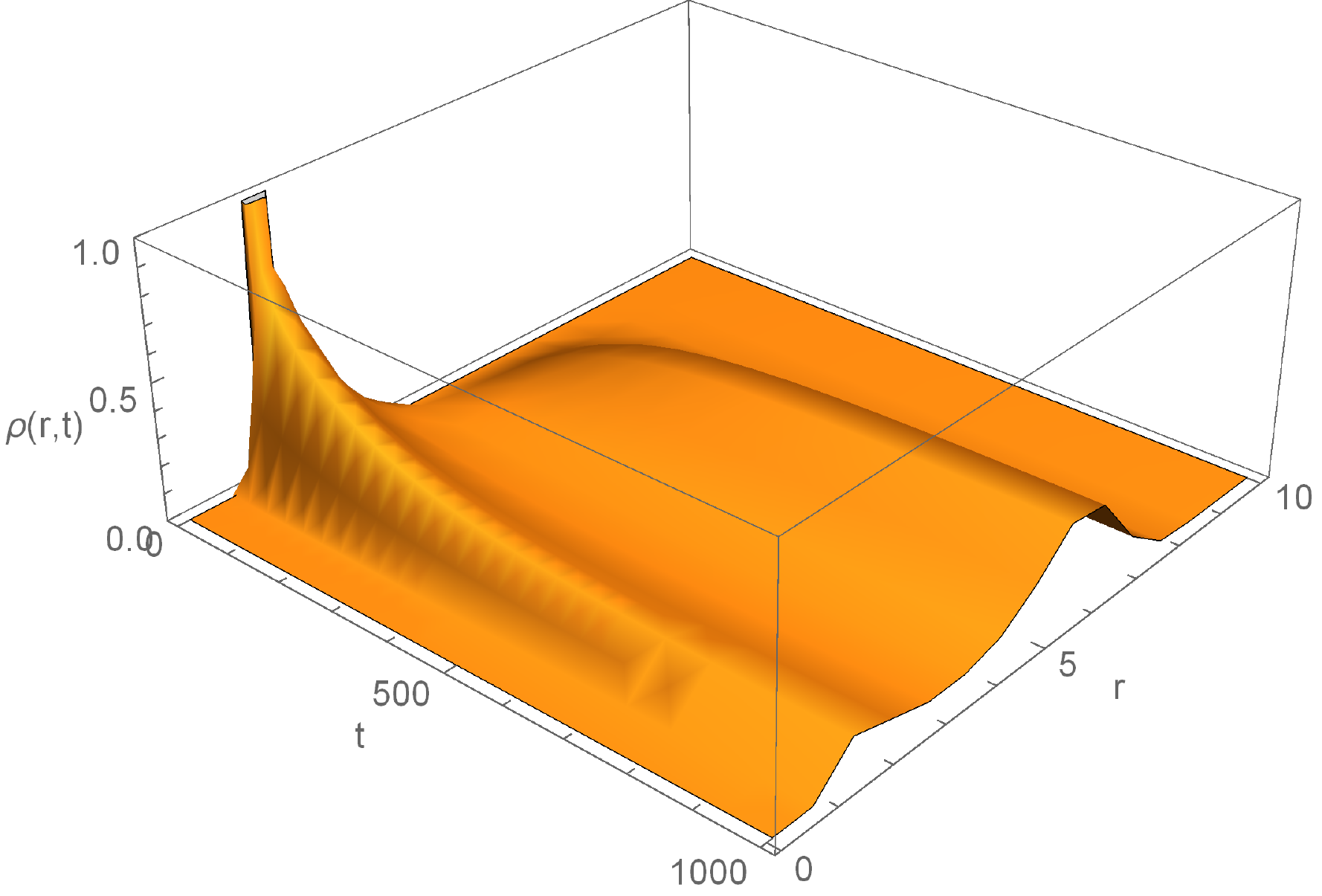}
  \caption{The time dependent behaviors of probability distribution $\rho(r, t)$ at the temperatures $T=0.031$ and $T=0.0315$. The initial wave packet is located at the small black hole state.}
  \label{FPsolutionstl}
\end{figure}

In Fig.\ref{FPsolutionstl}, we present the time dependent behaviors of the probability distribution of the black hole states in the canonical ensemble at different temperatures. The initial probability distribution is located at the small black hole state. The time evolution process consists of two stages. In the first stage, the initial gaussian type distributions reache the smooth quasi-stationary distributions in the left basin very quickly, which is not affected by the right basin. This describes the intra-basin dynamics. Then the smooth quasi-stationary distributions will spread out to the right basin and the final stationary distribution will be saturated at the long time limit. This describes the inter-basin dynamics. It can be checked that the final stationary distribution is given by $\rho_{st}(r)\propto e^{-G(r)/T}$, which is just the Boltzmann law. The similar conclusion can be obtained when choosing the large black hole state as the initial condition.

\subsection{Kinetics and its fluctuations of the black hole state switching and phase transitions}

From the ensemble viewpoint, the spacetime state can have the chance to escape from one black hole state represented by one basin in topography to the other basin under the thermal fluctuations. We now discuss the rate of this escaping, which is typically represented by the mean first passage time (MFPT). Because the underlying kinetics is stochastic due to the thermal fluctuations, the first passage time (FPT) for a spacetime state to reach the top of the barrier is also a stochastic variable. The MFPT is the statistical average of the first passage time. In \cite{Li:2020khm}, using the Fokker-Planck equation, the analytical expressions for the MFPT and 2nd moment or the fluctuation of PFT for a state from the small black hole basin to the large black hole basin are derived and given by
\begin{eqnarray}
\langle t \rangle&=&\frac{1}{D}\int_{r_s}^{r_m}dr \int_{0}^{r}dr'  e^{\beta \left(G(r)-G(r')\right)}\;,\\
\langle t^2 \rangle&=&\frac{2}{D^2}\int_{r_s}^{r_m}dr \int_{0}^{r}dr'
\int_{r'}^{r_m}dr'' \int_{0}^{r''}dr'''\nonumber\\ &&
e^{\beta \left(G(r)-G(r')+G(r'')-G(r''')\right)}\;,
\end{eqnarray}
where $r_{s/m/l}$ is the horizon radius of the small/intermediate/large black hole respectively.

There is an approximate formula for the MFPT. Note that the Gibbs free energy near the minimum and the maximum can be approximated by the quadratic expansion as
\begin{eqnarray}
G(r)&=&G(r_s)+\frac{1}{2}\omega_{min}\left(r-r_s\right)^2+\cdots\;,\nonumber\\
G(r)&=&G(r_m)-\frac{1}{2}\omega_{max}\left(r-r_m\right)^2+\cdots\;.
\end{eqnarray}
By performing the Gaussian integral, the approximate expression of the MFPT in the high barrier limit $(kT\ll G(r_m)-G(r_s))$ can be obtained
\begin{eqnarray}\label{approx}
\langle t \rangle\approxeq \frac{\pi}{\omega_{min}\omega_{max}}
e^{\beta \left(G(r_m)-G(r_s)\right)}\;.
\end{eqnarray}
This indicates that the MFPT is very dependent on the barrier hight since it is on the exponential. The prefactor also contributes to the kinetics through the fluctuation $\omega_{min}$ around the basin and $\omega_{max}$ at the top of the barrier, although to a much lesser degree than that from the barrier height in the exponential. Thus we now derived the analytical formula of the MFPT for the black hole thermodynamic state switching and phase transition.

\begin{figure}
\centering
\subfigure[]{\label{MFPTstl}
\includegraphics[width=4cm]{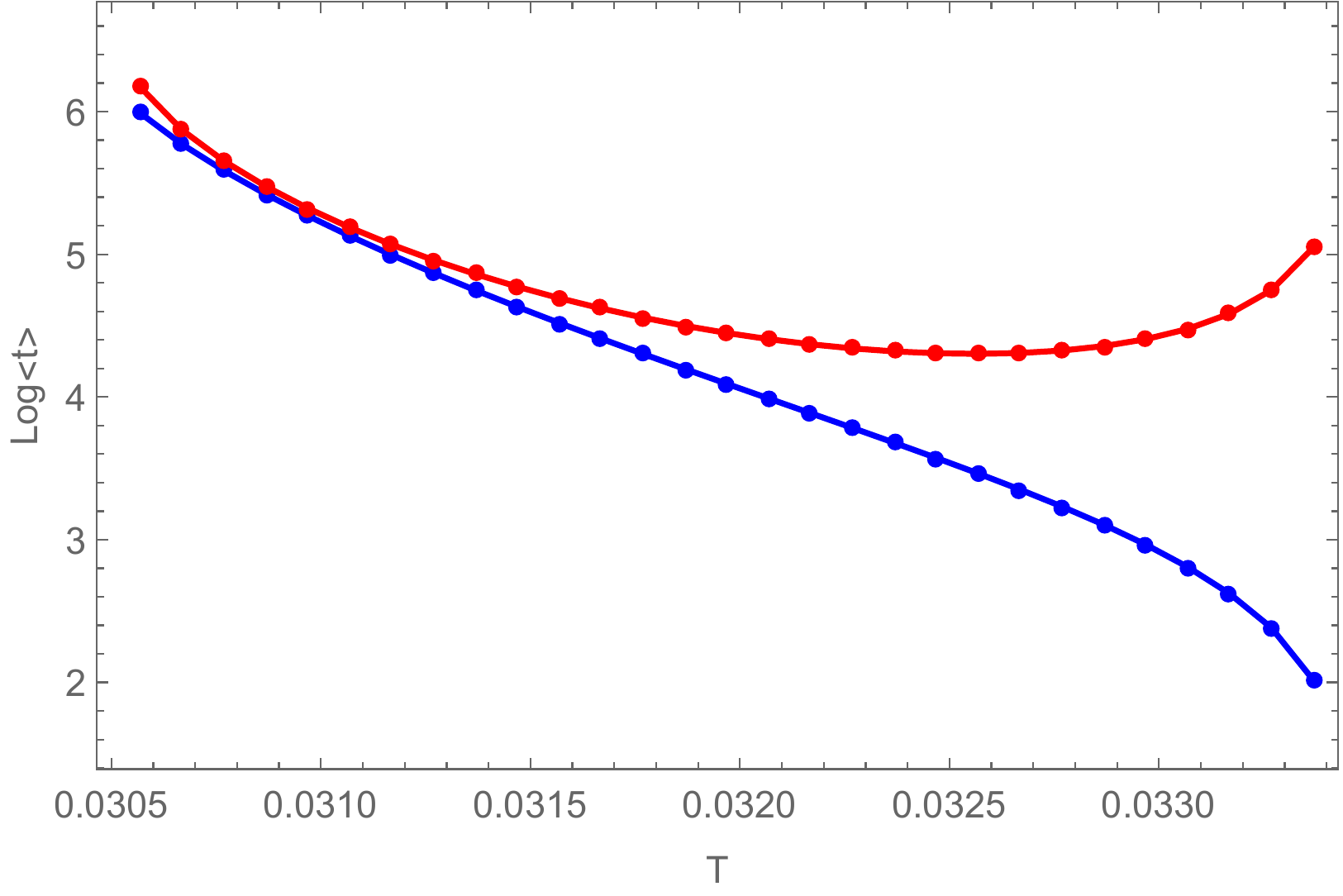}}
\subfigure[]{\label{RelFlucstl}
\includegraphics[width=4cm]{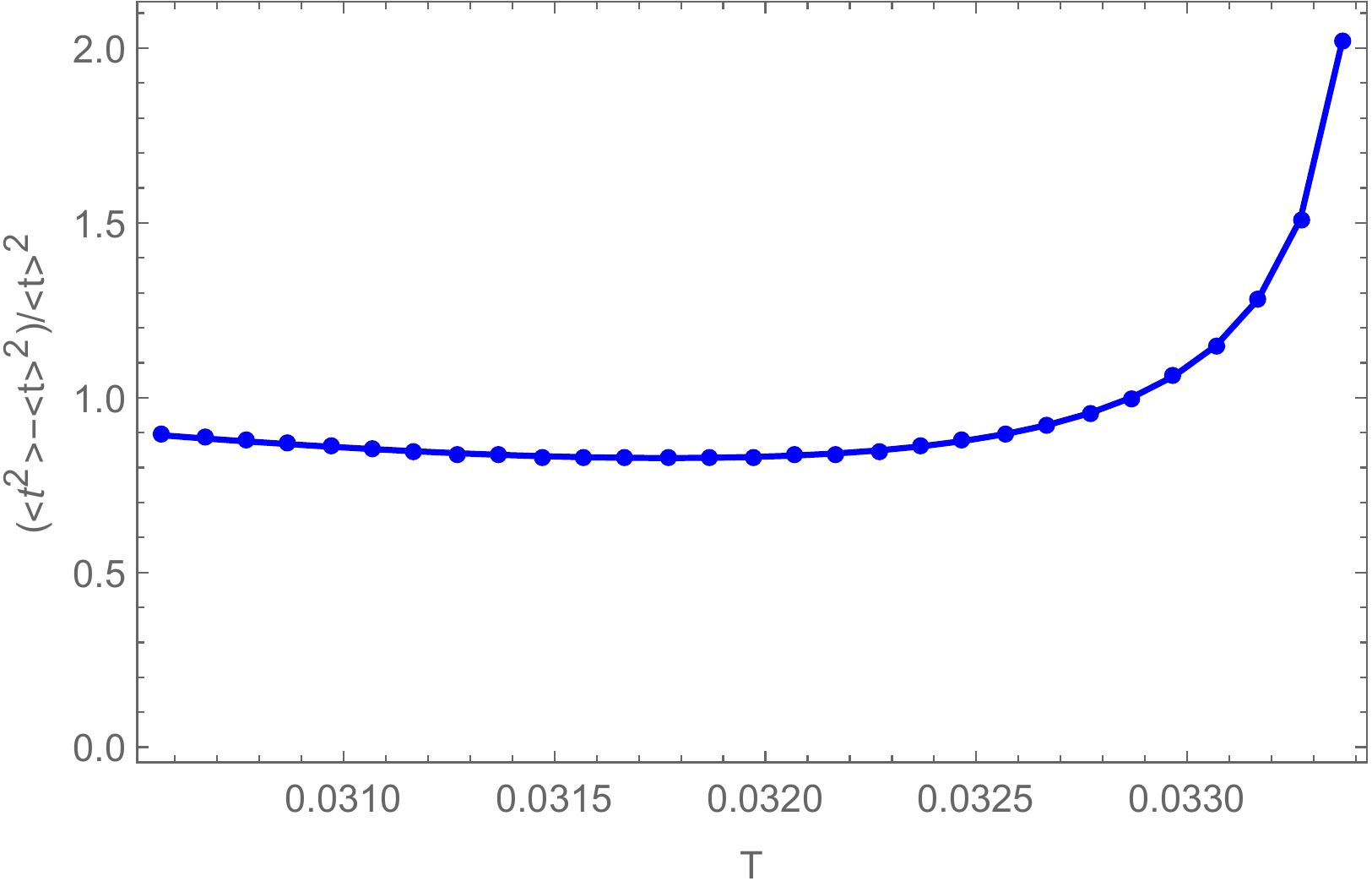}}
\subfigure[]{\label{MFPTlts}
\includegraphics[width=4cm]{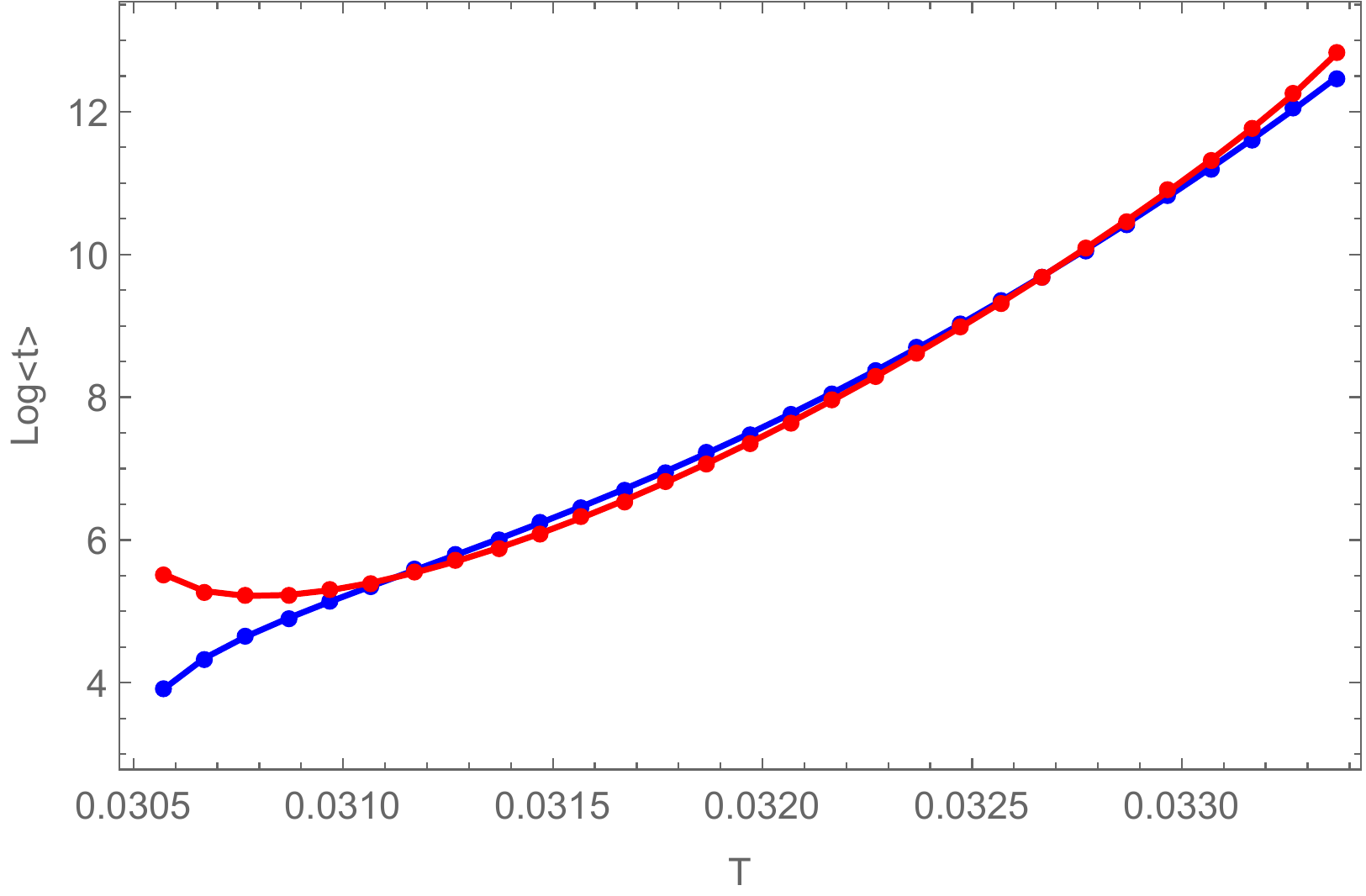}}
\subfigure[]{\label{RelFluclts}
\includegraphics[width=4cm]{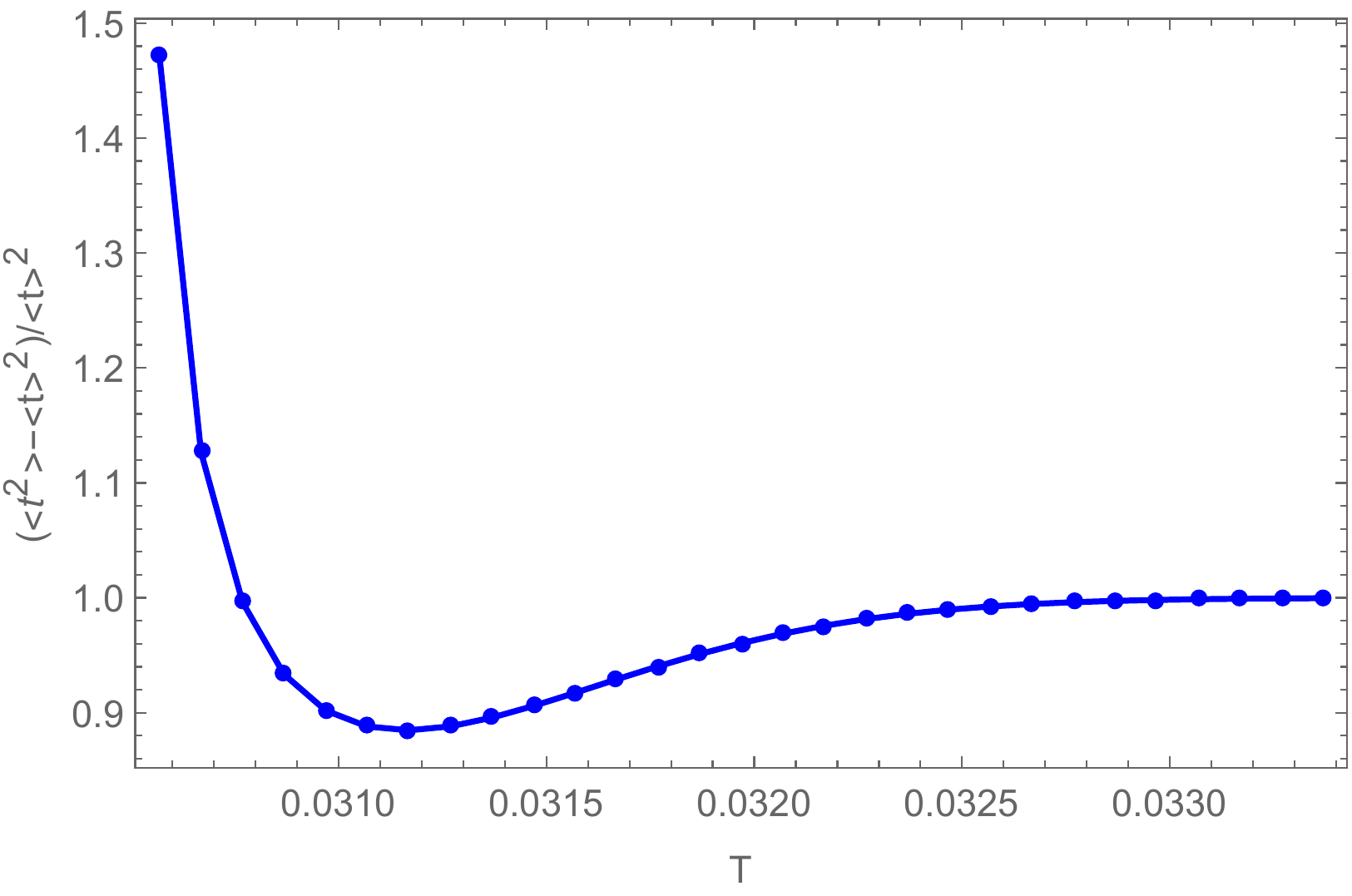}}
\caption{MFPT $\langle t \rangle$ and the relative fluctuation $(\langle t^2 \rangle-\langle t \rangle^2)/\langle t \rangle^2$ of the state transition in the canonical ensemble as a function of temperature $T$ with the parameters $\alpha=0.1$, $Q=1$, and $P=0.6P_c$.
(a): MFPT from the small to the large black hole;
(b): Relative fluctuation from the small to the large black hole;
(c): MFPT from the large to the small black hole;
(d): Relative fluctuation from the large to the small black hole.
In (a) and (c), the blue and the red curves are calculated from the integral and the approximate expressions, respectively.}
\label{MFPTCE}
\end{figure}

The numerical results are presented in Fig.\ref{MFPTCE}. We also compare the approximate analytical results of MFPT with the exact results calculated from the integral expression. In Fig.\ref{MFPTstl}, it shows that the approximate expression of the MFPT is more applicable at the low temperature and there is large deviation of the approximate results from the exact results at the high temperature. This is caused by the low barrier height at the high temperature. The exact results show that the MFPT is a monotonous decreasing function of the temperature $T$. From Fig.\ref{GibbsCE}, one can see that the barrier height between the small black hole state and the intermediate black hole state decreases when increasing the temperature. Higher temperatures tend to provide the small black hole much easier chance to go across the barrier. When the temperature approaches $T_{max}$, the barrier height gets close to zero. When $\Delta G\ll kT$, the thermal fluctuation dominates the kinetics process and the relative fluctuation of FPT is relatively large as shown in Fig.\ref{RelFlucstl}.

For the kinetics process from the large black hole to the small black hole, a similar integral expressions for the MFPT and the 2nd moment of PFT can also be derived \cite{Li:2020khm}, where the integral domains are changed. Similar to Eq.(\ref{approx}), the approximate expression of the MFPT can also be obtained. The numerical results are presented in Fig.\ref{MFPTlts} and Fig.\ref{RelFluclts}. Fig.\ref{MFPTlts} shows that the approximate expression is applicable in the high temperature region where the barrier hight is relatively large. The exact results shows that the MFPT is a monotonous increasing function of the temperature $T$ and the relative fluctuation is relatively large at $T=T_{min}$. The reason behind is that the barrier height between the large and the intermediate black holes is a monotonous increasing function of temperature and the barrier height approaches zero at $T=T_{min}$. Therefore, we can conclude that the MFPT and its fluctuations are closely related to free energy landscape topography through the barrier heights and the ensemble temperature.

\subsection{Thermodynamic driving force in the form of energy and entropy and resulting
 free energy as well as the kinetics of state transitions with different physical parameters}

For the thermodynamics and kinetics of black hole state transition in the canonical ensemble, there are four parameter that can be adjusted, the GB coupling constant $\alpha$, the electric charged $Q$, the thermodynamic pressure $P$, and the ensemble temperature $T$. We have investigated the dependence of MFPT and its relative fluctuation on the ensemble temperature in the case when the other parameters are fixed. In this subsection, we explore the effect on the kinetics and its fluctuations of black hole state transitions when varying the physical parameters $\alpha$, $Q$, and $P$.

\subsubsection{The energy, entropy and free energy landscapes at different coupling constant $\alpha$}

\begin{figure}
  \centering
  \includegraphics[width=6cm]{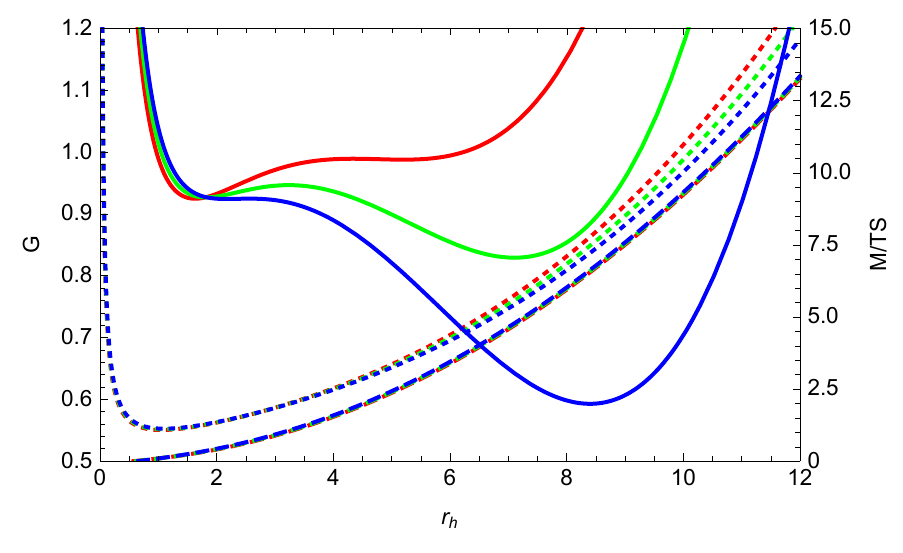}
  \caption{Gibbs free energy $G$ (solid), black hole mass $M$ (dotted) as well as the product of temperature $T$ and entropy $S$ (dashed) as the functions of $r_h$ with different GB coupling constant $\alpha$. The red, green, and blue curves correspond to $\alpha=0.15$, $0.2$, and $0.25$, respectively. The electric charge $Q=1$, the pressure $P=0.6P_c$, and the ensemble temperature $T=0.029$. }
  \label{GPlotalpha}
\end{figure}

\begin{figure}
  \centering
  \includegraphics[width=6cm]{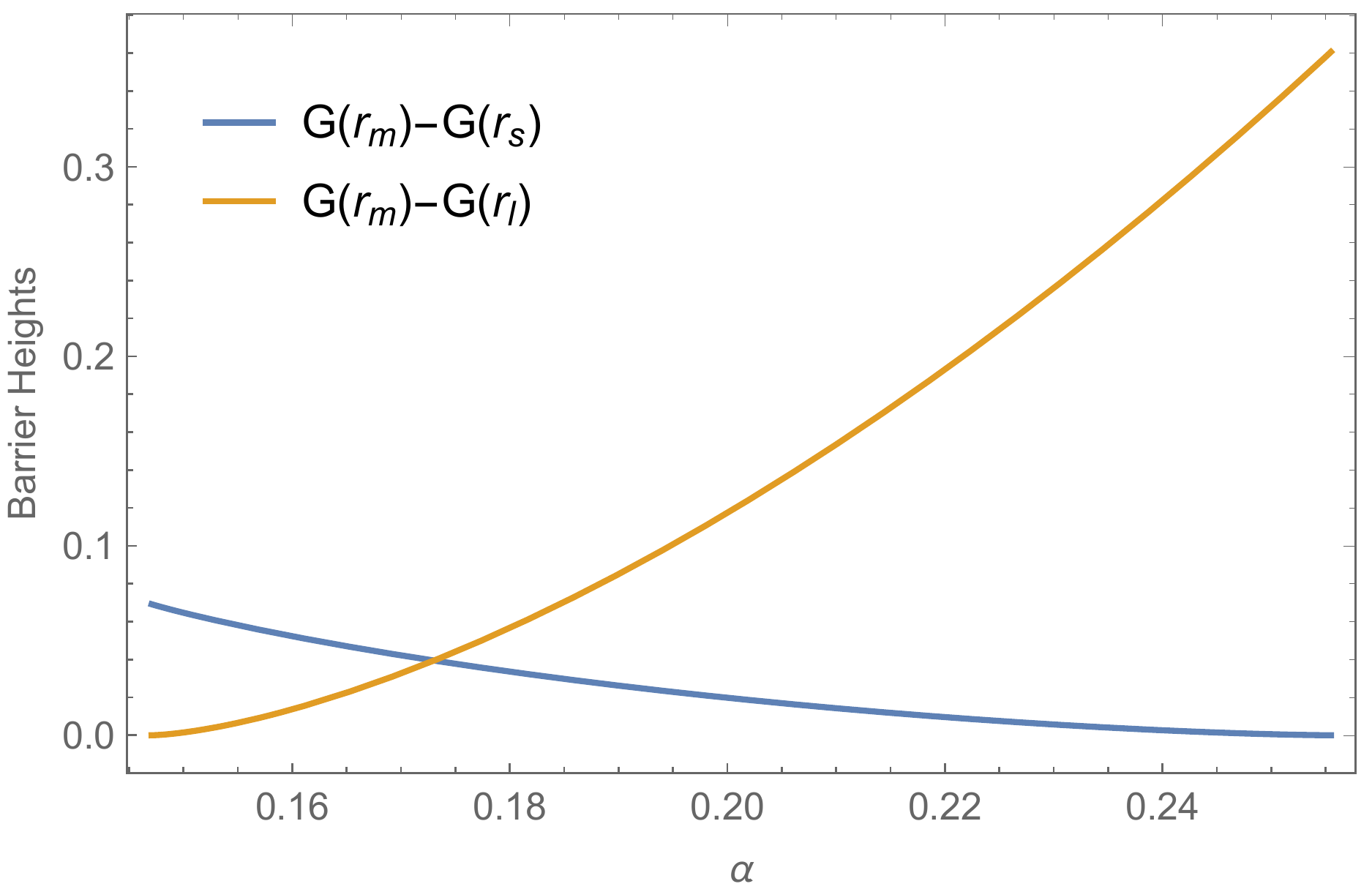}
  \caption{The barrier heights of Gibbs free energy landscape as the function of GB coupling constant. The barrier height between the small (large) black hole and the intermediate black hole monotonically decreases (increases) with $\alpha$. The electric charge $Q=1$, the pressure $P=0.6P_c$, and the ensemble temperature $T=0.029$. }
  \label{BHalpha}
\end{figure}

In Fig.\ref{GPlotalpha}, the Gibbs free energy is plotted as the function of the black hole radius for different GB coupling constant. When the ensemble temperature $T=0.029$, the three curves all have the shapes of double well. From the free energy landscape topography, we observe that, the large black hole is more sensitive to the change of $\alpha$. When $\alpha$ increases, the barrier height between the small black hole and the intermediate black hole decreases while the barrier height between the large black hole and the intermediate black hole increases. The barrier heights as the functions of $\alpha$ are shown explicitly in Fig.\ref{BHalpha}. The variation of the free energy barrier height should be caused by the variations of the mass and the entropy of the three branches of black hole solutions. We have plotted the mass and the product of temperature and entropy of black hole with different coupling constants in Fig.\ref{GPlotalpha}. We can observe that the shape and the trend of Gibbs free energy is completely determined by the black hole mass and the product of the temperature and entropy.

We can see that the black hole mass or energy is nonmonotonic with respect to $r_h$, In fact, certain small size black hole is preferred (minimum of $M$ vs $r_h$). We can also see that the entropy multiplied by the temperature monotonically increases as $r_h$ increases. In fact, low energy or mass or high entropy are always preferred when acting alone. Since both mass and entropy are nonlinear functions of the black hole size, the competition between the two can generate interesting behaviors and possible new phases unexpected from the action of the mass or the entropy alone. As shown in Fig.\ref{GPlotalpha}, when the coupling $\alpha$ is smaller, the small sized black hole is more preferred from the mass versus $r_h$ under steeper slope while when the coupling $\alpha$ becomes bigger, the small sized black hole is less stable because the mass versus $r_h$ has shallower slope. Considering the entropy alone, we see that it always prefers the large size black hole. The change of the slope of entropy versus coupling is not very significant as shown in Fig.\ref{GPlotalpha}. The Gibbs free energy as the result of the competition between the mass (energy) and the entropy leads to two possible stable black hole phases in certain parameter regimes, a small size black hole favored by energy or mass and a large size black hole favored by the entropy. In other words, the free energy has two minimum representing the two phases of the black hole, a low energy and low entropy phase of the small black hole  where the energy element of the free energy in the competition with the entropy element wins and a high energy and high entropy phase of the large black hole where the entropy element of the free energy in the competition with the energy element wins. From the changes of the slope of the mass versus $r_h$ with respect to the coupling $\alpha$, we can see from the free energy landscapes in Figure 6 that the black hole phases change from the small black hole preference (changing to shallower basin) to the large black hole preference (changing to deeper basin) as the coupling $\alpha$ increases.

Notice that the the black hole mass and entropy multiplied by $T$ are both much larger than that of the resulting Gibbs free energy as shown in Fig.\ref{GPlotalpha}. Therefore, the free energy landscape and the black hole phases are the results of the delicate balance between the two large numbers or scales, the mass and the entropy, giving rise to a much smaller number or scale of the free energy. Due to the scale difference of the free energy (small) in comparison to the energy and entropy (large), accurate quantifications of the energy and entropy are essential for obtaining the accurate information of the free energy and thermodynamics of the black hole. Small errors in estimating the energy and entropy can generate relatively larger errors in the free energy estimation. In other words, it might be difficult to switch from one state to another when considering the energy/mass or entropy barrier alone. For example, it is very difficult to switch from a small size black hole state to the large size black hole state from the energy/mass perspective since the barrier ($\Delta E$, where $\Delta E$ is the energy barrier from initial to the final state) for switching is high compared to the thermal temperature and it is difficult to realize the transition through thermal motion. On the other hand, it is very difficult to switch from a large size black hole state to the small size black hole state from the entropy perspective since the barrier for switching ($T\Delta S$,  where $T \Delta S$ is the entropy barrier from initial to the final state) is high compared to the thermal temperature and it is difficult to realize the transition through thermal motion. However, the competition between the mass/energy and the entropy due to their intricate balance can lead to the significantly reduced free energy and the associated free energy barrier between the small and large size black holes. The free energy barrier height is in the order of $T$ which makes the thermal transitions possible between the the small and large size black holes.

\begin{table}[!htbp]\caption{Thermodynamic quantities of different types of black hole at different GB coupling constant with $Q=1$, $P=0.6P_c$, and $T=0.029$.\\}
\centering
\begin{tabular}{|c|c|c|c|c|c|}
\hline
BH Type&$\alpha$&$r_h$&$M$&$S$&$G$\\
\hline
\multirow{3}*{Small}&0.15&1.62915&1.19307&9.25817&0.924583\\
\cline{2-6}
&0.2&1.81213&1.26926&11.8105&0.926752\\
\cline{2-6}
&0.25&2.17098&1.42415&17.2422&0.924124\\
\hline
\multirow{3}*{Intermediate}&0.15&4.33678&2.78293&61.8516&0.989233\\
\cline{2-6}
&0.2&3.2348&1.98542&35.8238&0.946531\\
\cline{2-6}
&0.25&2.56711&1.61095&23.6651&0.924667\\
\hline
\multirow{3}*{Large}&0.15&5.19668&3.53822&87.9467&0.987769\\
\cline{2-6}
&0.2&7.10596&5.57245&163.562&0.829148\\
\cline{2-6}
&0.25&8.40065&7.21608&228.391&0.592734\\
\hline
\end{tabular}
\label{Tablealpha}
\end{table}

We have also listed the thermodynamic quantities of these black holes with different GB coupling constants on Table \ref{Tablealpha}. The radii of the small and the large black holes increase while the radius of the intermediate black hole decreases as the GB coupling constant increases. The masses and the entropies of these black holes have the same trends while the variations of the free energy barriers are the results of the subtle competitions between the energies and entropies of these black hole states. From the variations of the barrier heights, we anticipate that the small (large) black hole becomes less (more) stable and the MFPT from the small (large) black hole to the large (small) black hole will decrease (increase) when $\alpha$ increases.

\subsubsection{Kinetics and associated fluctuations versus temperature at different coupling
constant $\alpha$}

\begin{figure}
\centering
\subfigure[]{\label{MFPTstlalpha}
\includegraphics[width=4cm]{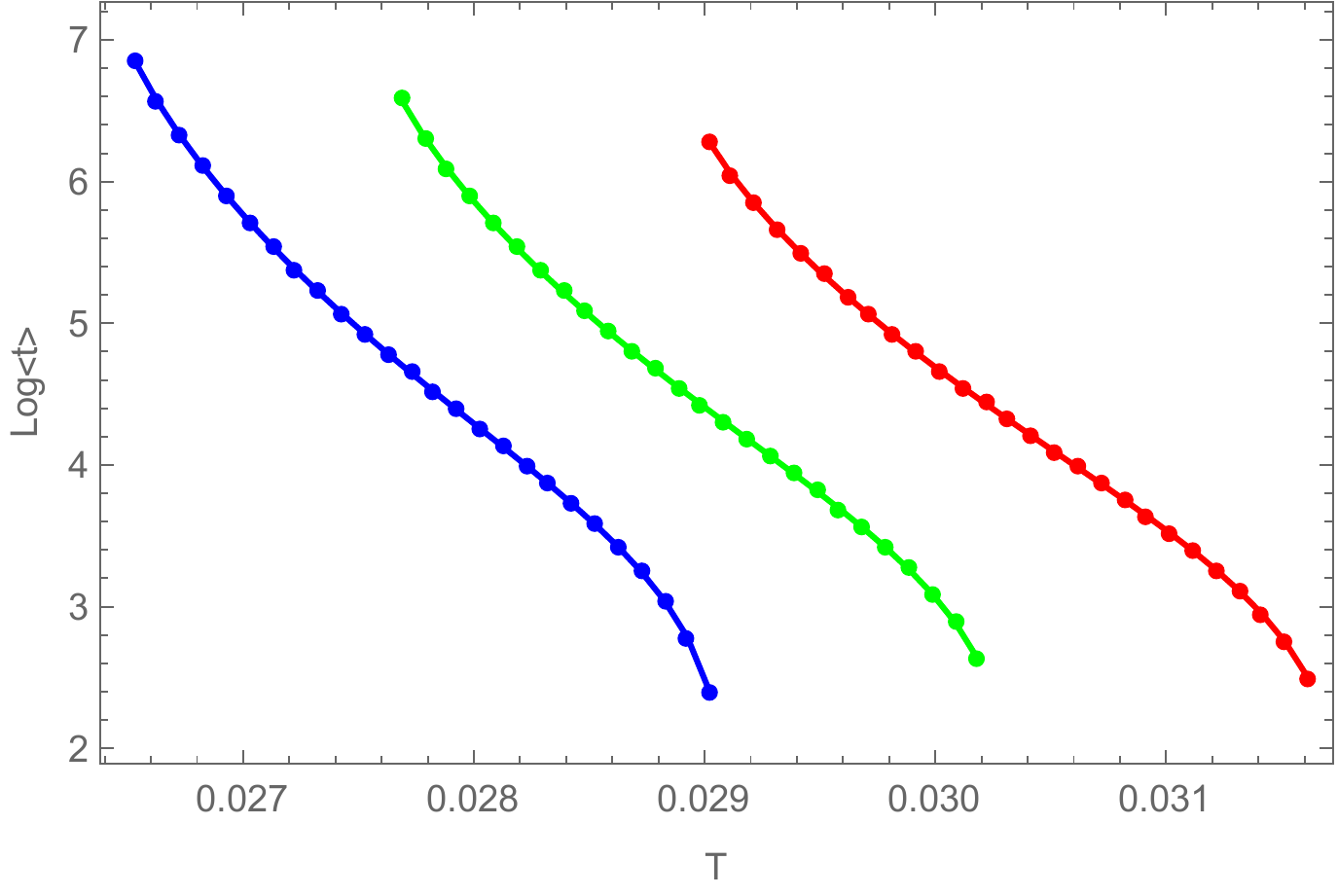}}
\subfigure[]{\label{RelFlucstlalpha}
\includegraphics[width=4cm]{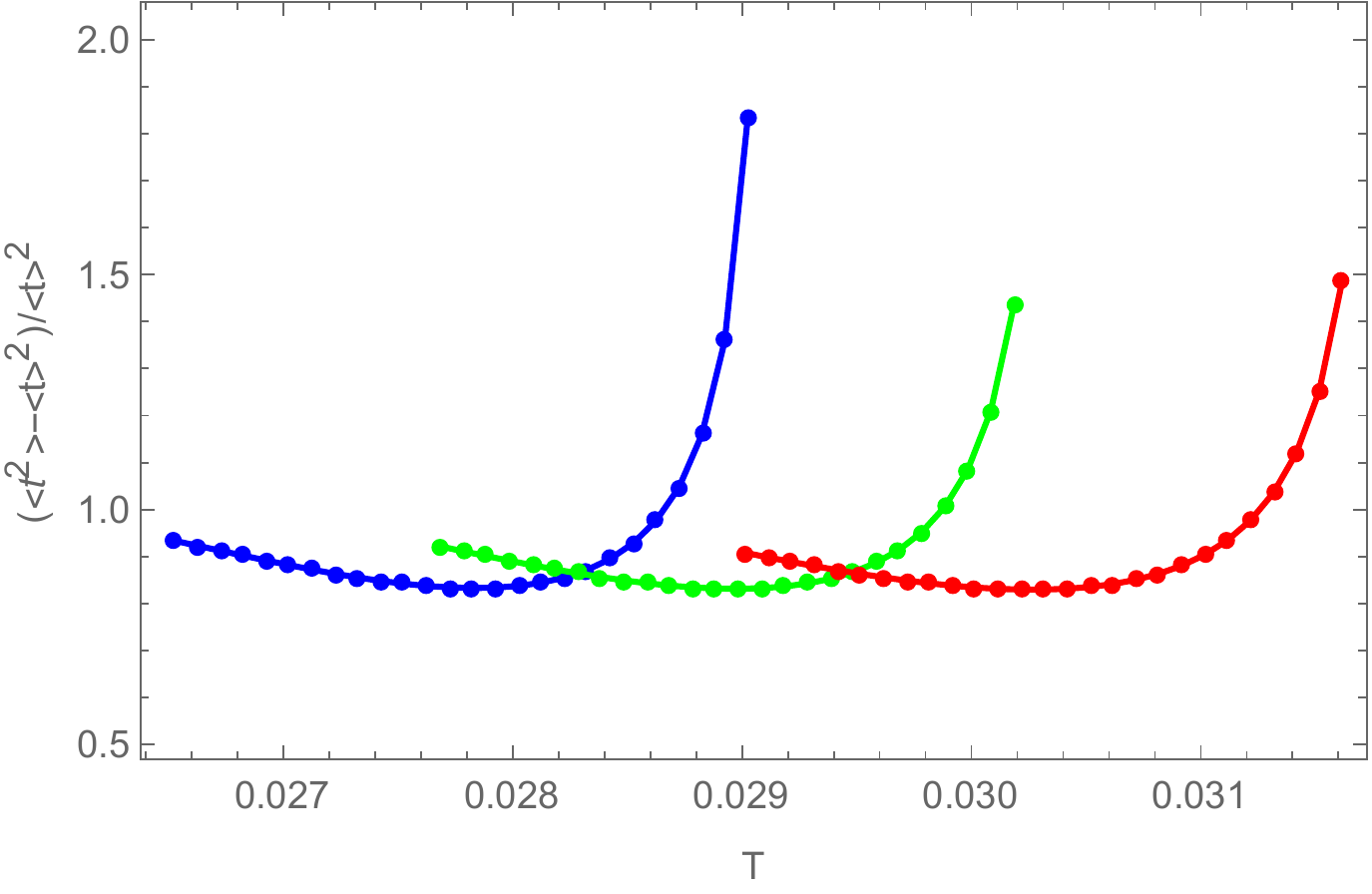}}
\subfigure[]{\label{MFPTltsalpha}
\includegraphics[width=4cm]{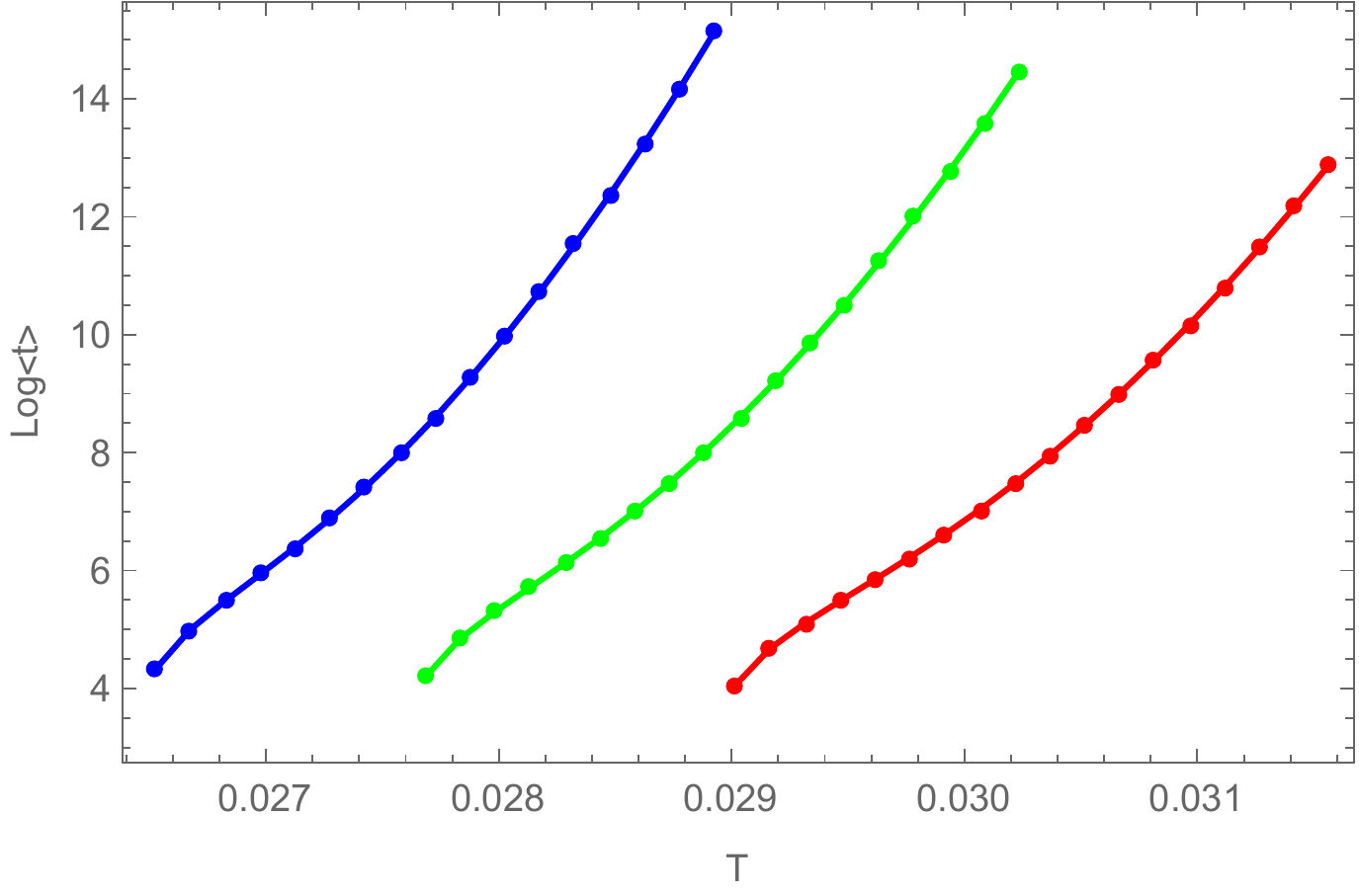}}
\subfigure[]{\label{RelFlucltsalpha}
\includegraphics[width=4cm]{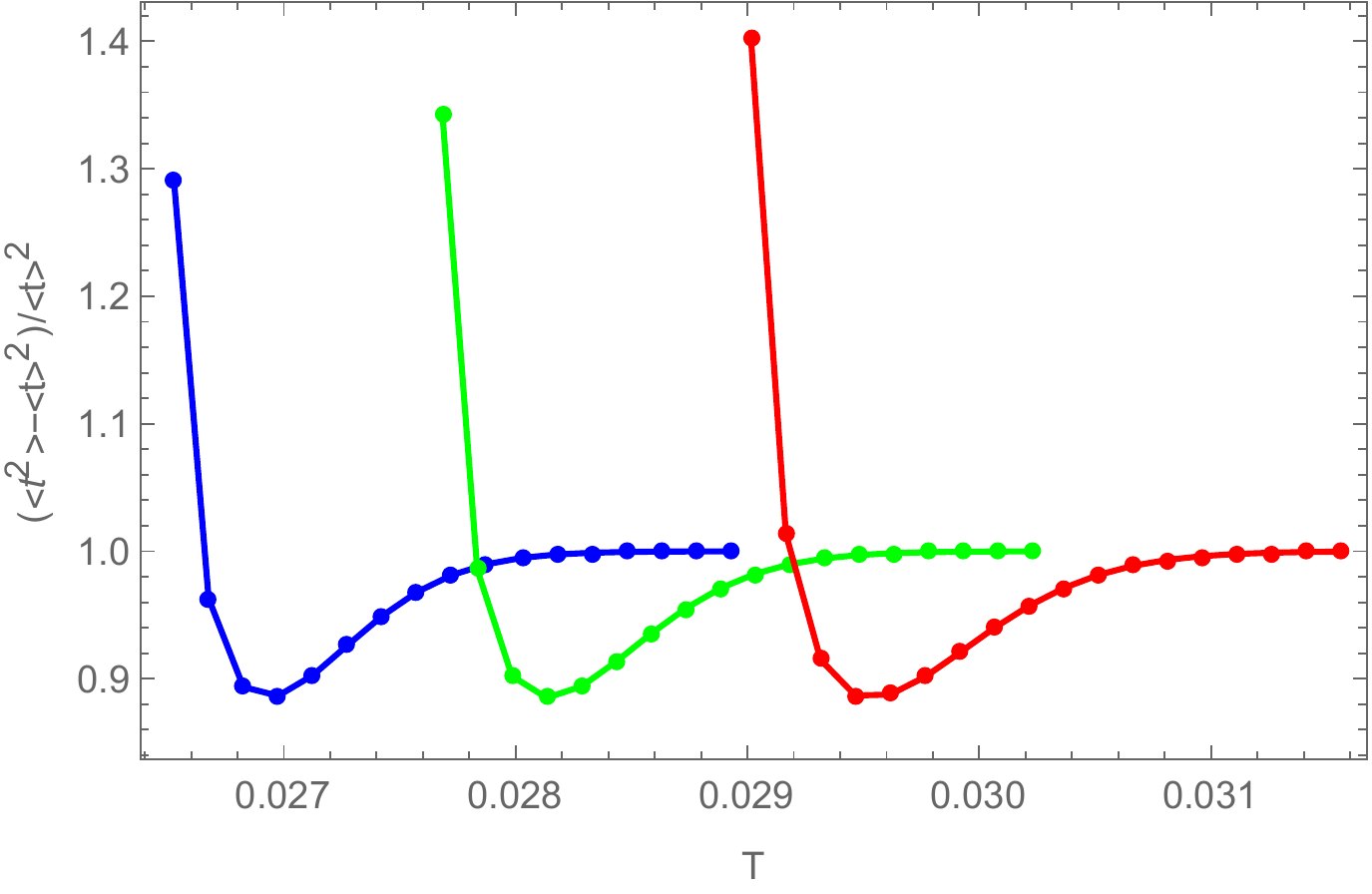}}
\caption{MFPT and the relative fluctuation of the state transition in the canonical ensemble as a function of temperature $T$ for the different GB coupling constant $\alpha$.
(a): MFPT from the small to the large black hole;
(b): Relative fluctuation from the small to the large black hole;
(c): MFPT from the large to the small black hole;
(d): Relative fluctuation from the large to the small black hole.
In each panel, the red, green, and blue curves correspond to $\alpha=0.15$, $0.2$, and $0.25$.
The electric charge $Q=1$, and the pressure $P=0.6P_c$, which are kept fixed.}
\label{MFPTCEalpha}
\end{figure}

The MFPT and the relative fluctuation as the functions of the ensemble temperature $T$ are plotted for the different GB coupling constant $\alpha$ in Fig.\ref{MFPTCEalpha}. It is observed that, when $\alpha$ increases, the range of the ensemble temperature, where the double well shaped free energy landscape topography appears, moves to the left. As mentioned, the minimal temperature $T_{min}$ and the maximum temperature $T_{max}$ are determined by Eq.(\ref{pGibbs}) and Eq.(\ref{ppGibbs}), which have nonlinear relationships between the temperature and the GB coupling constant. The variations of the barrier heights leads to the behavior that the MFPT from the small to the large black hole decreases with $\alpha$ while the MFPT from the large to the small black hole increases with $\alpha$, as shown in Fig.\ref{MFPTstlalpha} and Fig.\ref{MFPTltsalpha} respectively. Because the parameter $\alpha$ is the coupling strength of the higher curvature corrections to the pure Einstein gravity, the variation of $\alpha$ represents the effect of the higher order curvature terms on the kinetics of GB black hole phase transition. It should be noted that, when varying $\alpha$, the higher order curvature terms will influence the mass and entropy of black hole, and indirectly lead to the variation of Gibbs free energy landscape and the barrier heights. The numerical results indicate that the higher order curvature terms become more important for the black hole state switching process because it is more easier (harder) to switch from the small (large) black hole to the (large) small black hole when increasing the GB coupling constant.

\subsubsection{The energy, entropy and the resulting free energy landscapes at different charge $Q$}

\begin{figure}
  \centering
  \includegraphics[width=6cm]{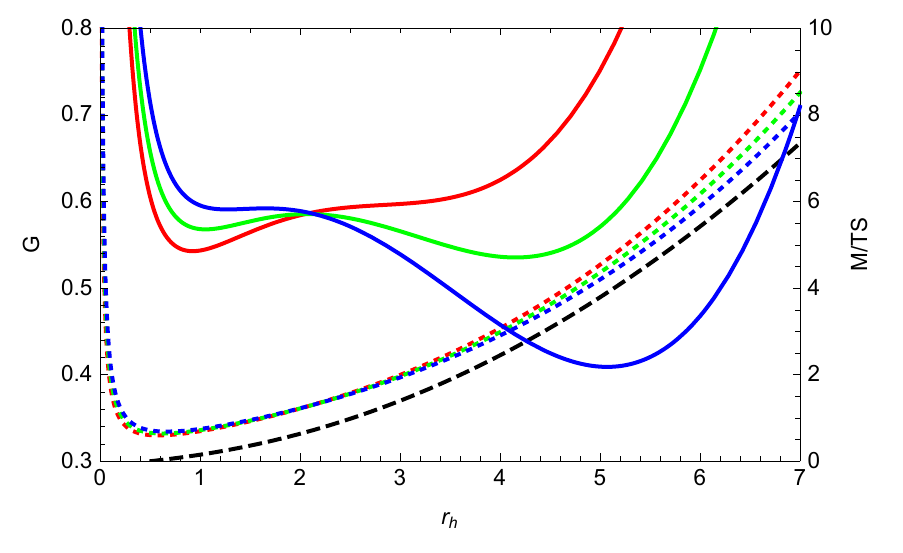}
  \caption{Gibbs free energy $G$ (solid), black hole mass $M$ (dotted) as well as the product of temperature $T$ and entropy $S$ (dashed black line) as the functions of $r_h$ with different electric charge $Q$. The red, green, and blue curves correspond to $Q=0.5$, $0.55$, and $0.6$, respectively. The coupling constant $\alpha=0.1$, the pressure $P=0.6P_c$, and the ensemble temperature $T=0.047$.}
  \label{GPlotQ}
\end{figure}

\begin{table}[!htbp]\caption{Thermodynamic quantities of different types of black hole at different electric charge with $\alpha=0.1$, $P=0.6P_c$, and $T=0.047$.\\}
\centering
\begin{tabular}{|c|c|c|c|c|c|}
\hline
BH Type&$Q$&$r_h$&$M$&$S$&$G$\\
\hline
\multirow{2}*{Small}&0.55&1.05403&0.734971&3.5564&0.56782\\
\cline{2-6}
&0.6&1.27449&0.84503&5.40778&0.590865\\
\hline
\multirow{2}*{Intermediate}&0.55&2.07103&1.2618&14.3896&0.58549\\
\cline{2-6}
&0.6&1.65606&1.02685&9.24982&0.59211\\
\hline
\multirow{2}*{Large}&0.55&4.15588&3.16968&56.0497&0.535347\\
\cline{2-6}
&0.6&5.07128&4.30217&82.8354&0.408902\\
\hline
\end{tabular}
\label{TableQ}
\end{table}

\begin{figure}
  \centering
  \includegraphics[width=6cm]{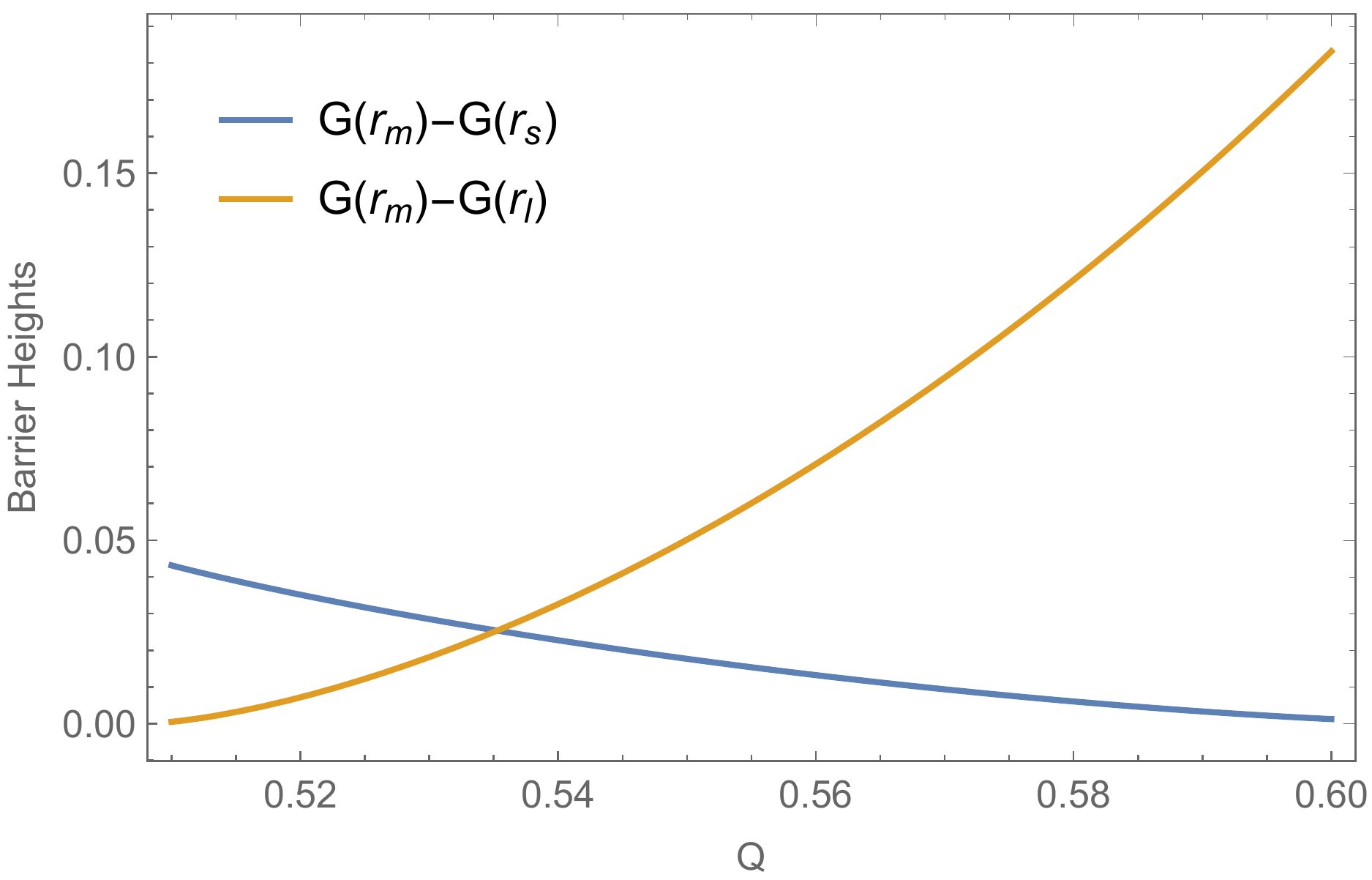}
  \caption{The barrier heights of Gibbs free energy landscape as the function of the electric charge. The barrier height between the small (large) black hole and the intermediate black hole monotonically decreases (increases) with $Q$. The coupling constant $\alpha=0.1$, the pressure $P=0.6P_c$, and the ensemble temperature $T=0.047$.}
  \label{BHQ}
\end{figure}

In Fig.\ref{GPlotQ}, the Gibbs free energy is plotted for different electric charge $Q$. It is shown that, at the ensemble temperature $T=0.047$, the red curve ($Q=0.5$) has the shape of one well while the other two curves have the shapes of double well. Therefore, there is no small/large black hole phase transition in the case of the red curve and the only one black hole state emerged as the global minimum of the free energy landscape is always stable. From Fig.\ref{GPlotQ}, one observe that increasing $Q$ has larger impacts to the large black hole than the smaller black hole. In Fig.\ref{BHQ}, the barrier heights in the Gibbs free energy landscape are plotted as the functions of $Q$. It is observed that, when $Q$ increases, the barrier height between the small black hole and the intermediate black hole decreases while the barrier height between the large black hole and the intermediate black hole increases. We also present the thermodynamic quantities of different types of black hole with $Q=0.55$ and $0.6$ in Table \ref{TableQ}, from which we can observe the variation trends of these quantities. These variation trends are also caused by the competitions of the masses and the entropies of the three branches of black hole, as explicitly shown in Fig.\ref{GPlotQ}, where the black hole mass and the product of temperature and entropy are plotted for different electric charge.

In Fig.\ref{GPlotQ}, we can see that the black hole mass or energy is nonmonotonic with respect to $r_h$, In fact, certain small size black hole is preferred (minimum of $M$ vs $r_h$). We can also see that the entropy multiplied by the temperature monotonically increases as $r_h$ increases. In fact, low energy or mass or high entropy are always preferred when acting alone. Since both mass and entropy are nonlinear functions of the black hole size, the competition between the two can generate interesting behaviors and possible new phases unexpected from the action of the mass or the entropy alone. As shown in Fig.\ref{GPlotQ}, when the charge $Q$ is smaller, the small sized black hole is more preferred from the mass versus $r_h$ under steeper slope while when the charge $Q$ becomes bigger, the small sized black hole is less stable because the mass versus $r_h$ has shallower slope. Considering the entropy alone, we see that it always prefers the large size black hole. The entropy is not explicitly dependent on charge $Q$ from the analytical expression and as shown in Fig.\ref{GPlotQ}. The Gibbs free energy as the result of the competition between the mass (energy) and the entropy leads to two possible stable black hole phases in certain parameter regimes, a small size black hole favored by energy or mass and a large size black hole favored by the entropy. In other words, the free energy has two minimum representing the two phases of the black hole, a low energy and low entropy phase of the small black hole  where the energy element of the free energy in the competition with the entropy element wins and a high energy and high entropy phase of the large black hole where the entropy element of the free energy in the competition with the energy element wins. From the changes of the slope of the mass versus $r_h$ with respect to the charge $Q$, we can see from the free energy landscapes in Fig.\ref{GPlotQ} that the black hole phases change from the small black hole preference (changing to shallower basin) to the large black hole preference (changing to deeper basin) as the charge $Q$ increases.

Notice that the the black hole mass and entropy multiplied by $T$ are both much larger than that of the resulting Gibbs free energy as shown in Fig.\ref{GPlotQ}. Therefore, the free energy landscape and the black hole phases are the results of the delicate balance between the two large numbers or scales, the mass and the entropy,  giving rise to a much smaller number or scale of the free energy. Due to the scale difference of the free energy (small) in comparison to the energy and entropy (large), accurate quantifications of the energy and entropy are essential for obtaining the accurate information of the free energy and thermodynamics of the black hole. Small errors in estimating the energy and entropy can generate relatively larger errors in the free energy estimation. In other words, it might be difficult to switch from one state to another when considering the energy/mass or entropy barrier alone. For example, it is very difficult to switch from a small size black hole state to the large size black hole state from the energy/mass perspective since the barrier ($\Delta E$, where $\Delta E$ is the energy barrier from initial to the final state) for switching is high compared to the thermal temperature and it is difficult to realize the transition through thermal motion. On the other hand, it is very difficult to switch from a large size black hole state to the small size black hole state from the entropy perspective since the barrier for switching  ($T\Delta S$,  where $T \Delta S$ is the entropy barrier from initial to the final state) is high compared to the thermal temperature and it is difficult to realize the transition through thermal motion. However the competition between the mass/energy and the entropy due to their intricate balance can lead to the significantly reduced free energy and the associated free energy barrier between the small and large size black holes. The free energy barrier height is in the order of $T$ which makes the thermal transitions possible between the the small and large size black holes.

\subsubsection{Kinetics and associated fluctuations versus temperatures at different charge $Q$}

\begin{figure}
\centering
\subfigure[]{\label{MFPTstlQ}
\includegraphics[width=4cm]{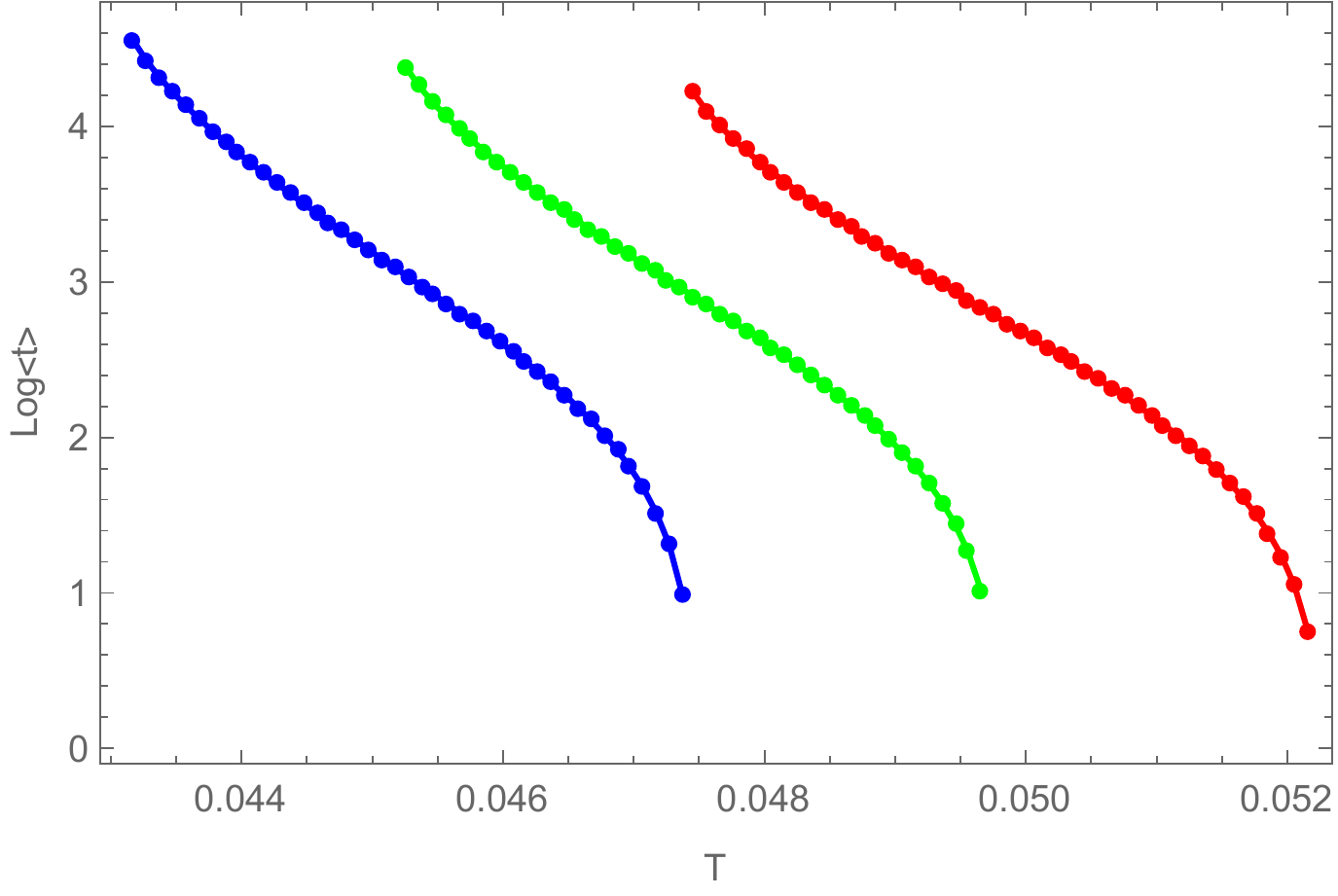}}
\subfigure[]{\label{RelFlucstlQ}
\includegraphics[width=4cm]{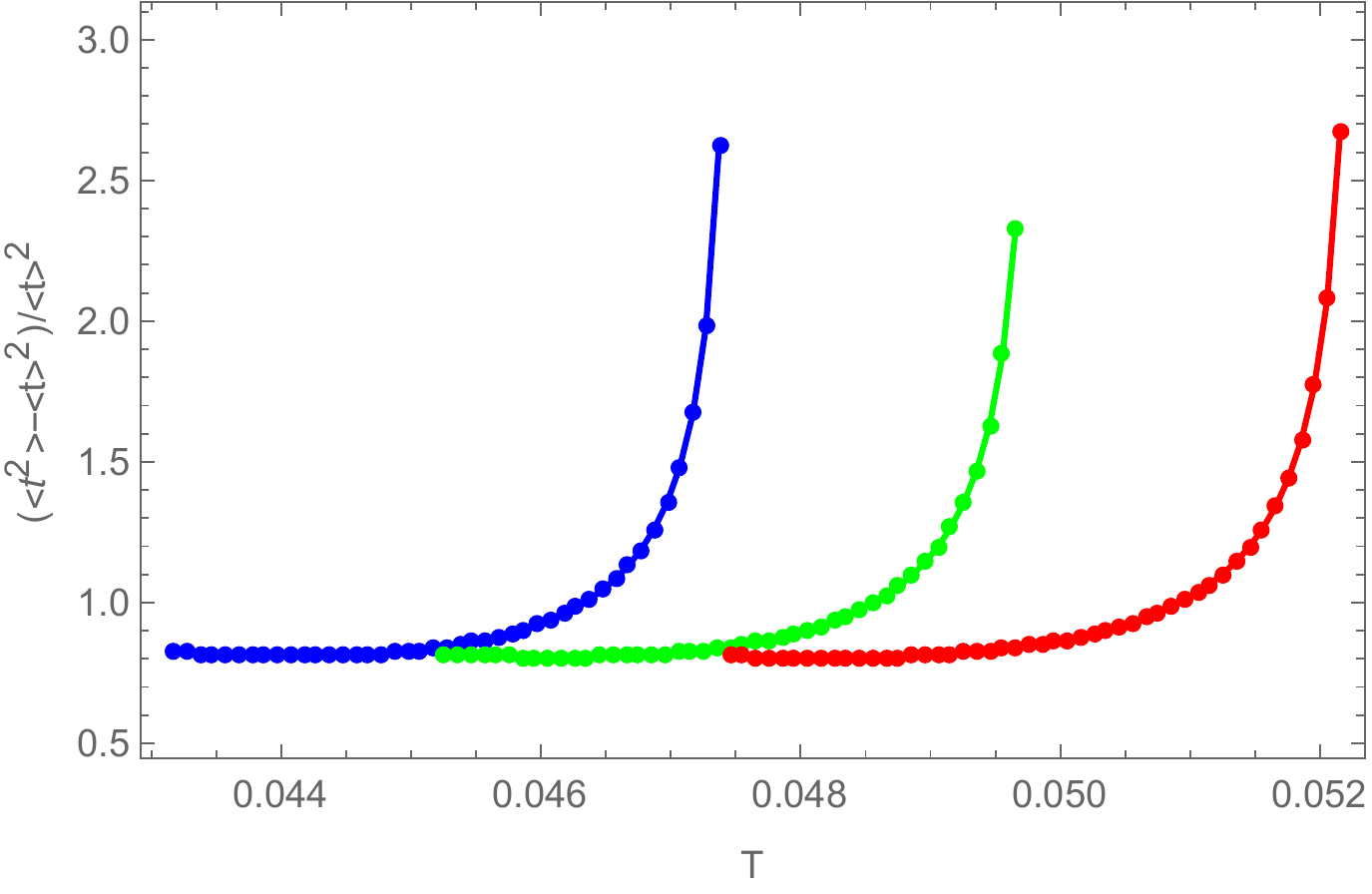}}
\subfigure[]{\label{MFPTltsQ}
\includegraphics[width=4cm]{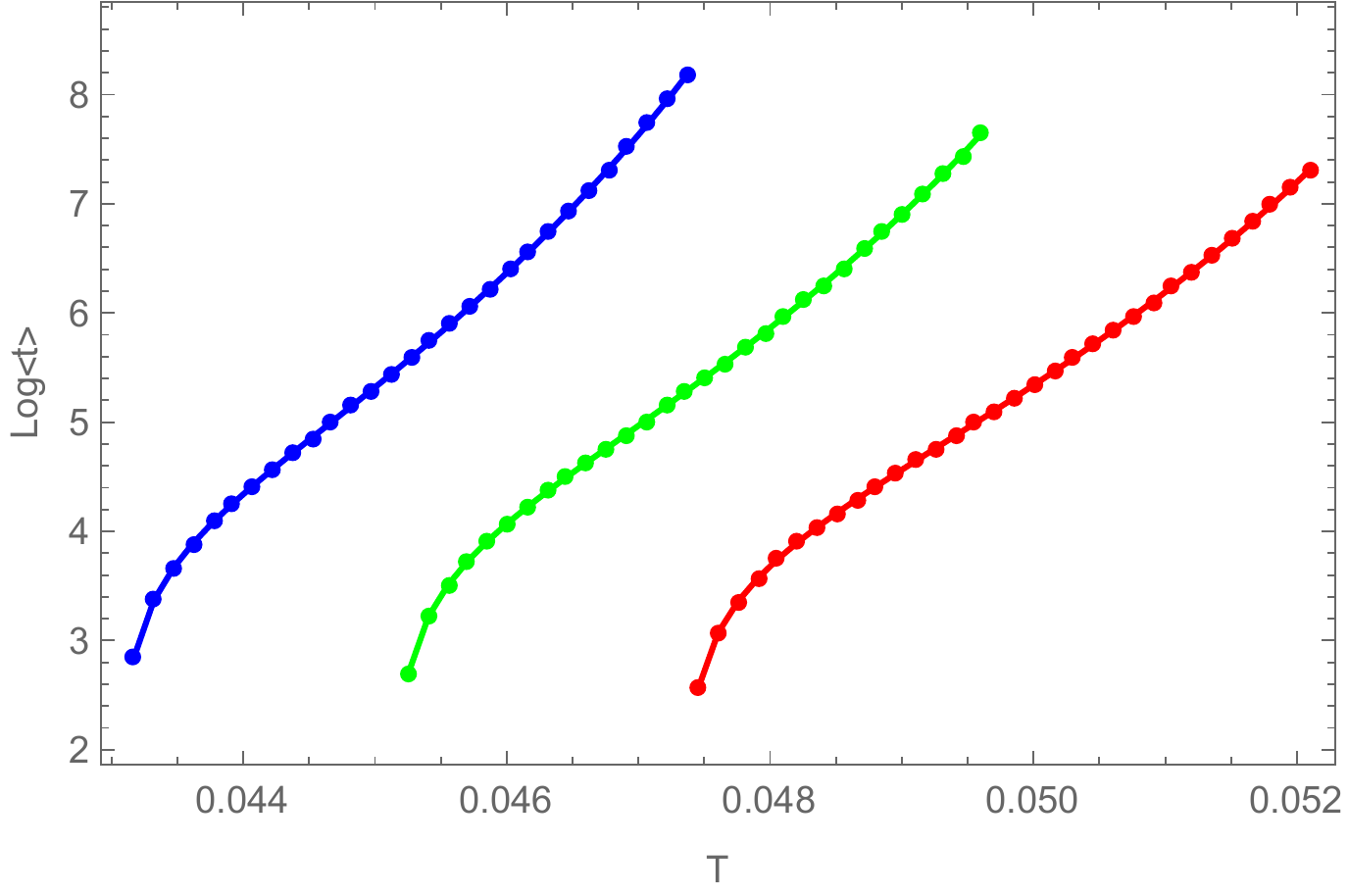}}
\subfigure[]{\label{RelFlucltsQ}
\includegraphics[width=4cm]{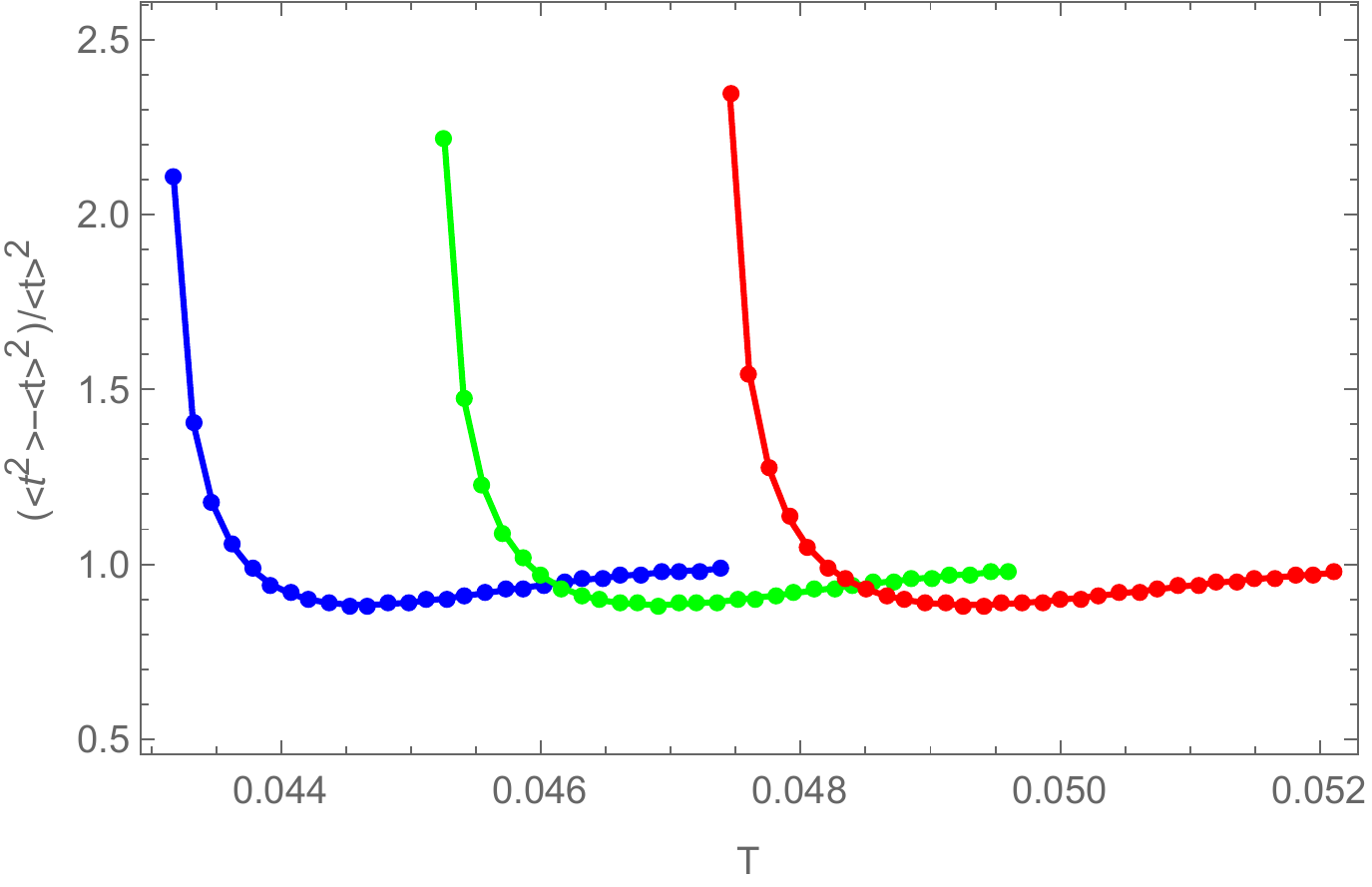}}
\caption{MFPT and the relative fluctuation of the state transition in the canonical ensemble as a function of temperature $T$ for the different electric charge $Q$.
The panels (a)-(d) shows the same quantities as in fig.\ref{MFPTCEalpha}.
In each panel, the red, green, and blue curves correspond to $Q=0.5$, $0.55$, and $0.6$.
The parameters $\alpha=0.1$ and $P=0.6P_c$ are kept fixed.}
\label{MFPTCEQ}
\end{figure}

The MFPT and the relative fluctuation as the functions of $T$ are plotted for different $Q$ in Fig.\ref{MFPTCEQ}. When $Q$ increases, the range of the ensemble temperature, where the double well shaped free energy landscape topography appears, also moves to the left. As shown in Fig.\ref{MFPTstlQ} and Fig.\ref{MFPTltsQ}, the MFPT from the small to the large black hole decreases with $Q$ while the MFPT from the large to the small black hole increases when increasing $Q$. This behavior can also interpreted by the variations of free energy barrier heights. The results indicate that it is easier (harder) to switch from the small (large) black hole to the large (small) black hole when increasing the black hole charge, i.e. the small (large) black hole becomes less (more) stable.

\subsubsection{The energy, entropy and the resulting free energy landscapes at different pressures or cosmological constants}

\begin{figure}
  \centering
  \includegraphics[width=6cm]{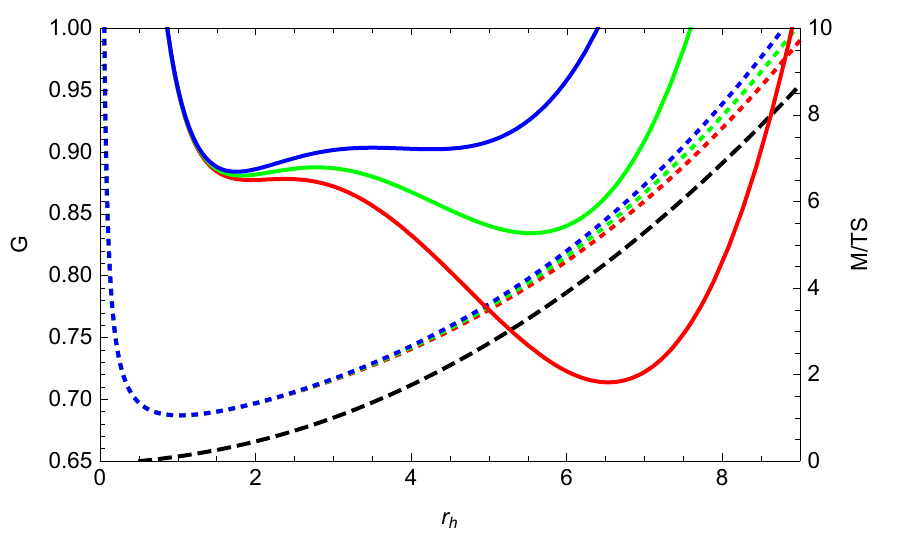}
  \caption{Gibbs free energy $G$ (solid), black hole mass $M$ (dotted) as well as the product of temperature $T$ and entropy $S$ (dashed black line) as a function of $r_h$ with different pressure $P$. The electric charge $Q=1$, the coupling constant $\alpha=0.1$, and the ensemble temperature $T=0.0338$. The red, green, and blue curves correspond to $P=0.65P_c$, $0.7P_c$, and $0.75P_c$, respectively.}
  \label{GPlotP}
\end{figure}

\begin{table}[!htbp]\caption{Thermodynamic quantities of different types of black hole at different pressure with $\alpha=0.1$, $Q=1$, and $T=0.0338$.\\}
\centering
\begin{tabular}{|c|c|c|c|c|c|}
\hline
BH Type&$P/P_c$&$r_h$&$M$&$S$&$G$\\
\hline
\multirow{3}*{Small}&0.65&1.94204&1.3061&12.6827&0.877429\\
\cline{2-6}
&0.7&1.82156&1.25882&11.1776&0.881014\\
\cline{2-6}
&0.75&1.75471&1.23495&10.3796&0.884119\\
\hline
\multirow{3}*{Intermediate}&0.65&2.37843&1.51575&18.8606&0.878265\\
\cline{2-6}
&0.7&2.78008&1.75176&25.5657&0.887639\\
\cline{2-6}
&0.75&3.49913&2.25686&40.0393&0.903534\\
\hline
\multirow{3}*{Large}&0.65&6.53258&5.3249&136.425&0.713748\\
\cline{2-6}
&0.7&5.53685&4.16243&98.04615&0.834426\\
\cline{2-6}
&0.75&4.26528&2.89572&58.9766&0.902307\\
\hline
\end{tabular}
\label{TableP}
\end{table}

\begin{figure}
  \centering
  \includegraphics[width=6cm]{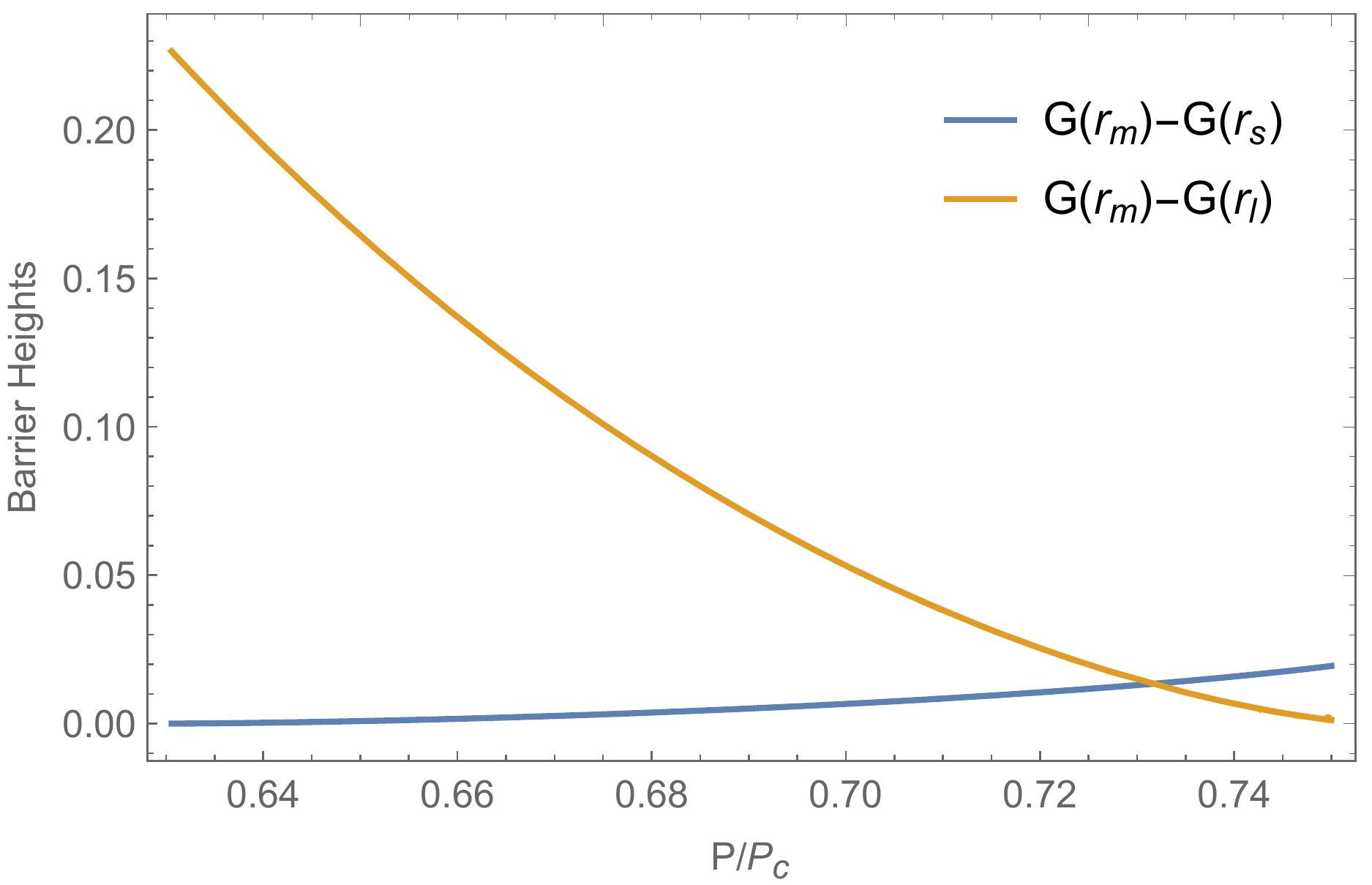}
  \caption{The barrier heights of Gibbs free energy landscape as the function of the pressure. The barrier height between the small (large) black hole and the intermediate black hole monotonically increases (decreases) with $P$. The electric charge $Q=1$, the coupling constant $\alpha=0.1$, and the ensemble temperature $T=0.0338$.}
  \label{BHP}
\end{figure}

The Gibbs free energy for different pressure $P$ is plotted in Fig.\ref{GPlotP}. At the ensemble temperature $T=0.0338$, all the three cases have the shapes of double well. Increasing the pressure (the absolute value of the cosmological constant) leads to more significant changes in the large black hole than the small black hole. From the plots of barrier heights in Fig.\ref{BHP}, it is observed that, when $P$ increases, the barrier height between the small black hole and the intermediate black hole increases while the barrier height between the large black hole and the intermediate black hole decreases. In Table \ref{TableP}, the thermodynamic quantities of three branches of black hole state are presented explicitly. The radii of the small and the large black hole decrease while the radius of the intermediate black hole increases with the pressure.

Furthermore, in Fig.\ref{GPlotP}, we have shown the plots of the black hole mass and the product of temperature and entropy for different pressures. The masses and the entropies have the same trends while the variations of the free energies depend on the competitions of the masses and entropies. These behavior lead to the variations of the free energy barrier heights, which can be considered as the underlying reason of the trends of the MFPTs.

From Fig.\ref{GPlotP}, we can see that the black hole mass or energy is nonmonotonic with respect to $r_h$, In fact, certain small size black hole is preferred (minimum of $M$ vs $r_h$). We can also see that the entropy multiplied by the temperature monotonically increases as $r_h$ increases. In fact, low energy or mass or high entropy are always preferred when acting alone. Since both mass and entropy are nonlinear functions of the black hole size, the competition between the two can generate interesting behaviors and possible new phases unexpected from the action of the mass or the entropy alone. As shown in Fig.\ref{GPlotP}, when the effective pressure $P$ or the equivalently absolute value of the cosmological constant $\Lambda$ is small, the small sized black hole is less preferred from the mass versus $r_h$ under shallower slope while when the effective pressure $P$ or the equivalently absolute value of the cosmological constant $\Lambda$ becomes bigger, the small sized black hole is more stable because the mass versus $r_h$ has steeper slope. Considering the entropy alone, we see that it always prefers the large size black hole.The entropy is not explicitly dependent on the effective pressure $P$ or the equivalently absolute value of the cosmological constant $\Lambda$  from the analytical expression and as shown in Fig.\ref{GPlotP}. The Gibbs free energy as the result of the competition between the mass (energy) and the entropy leads to two possible stable black hole phases in certain parameter regimes, a small size black hole favored by energy or mass and a large size black hole favored by the entropy. In other words, the free energy has two minimum representing the two phases of the black hole, a low energy and low entropy phase of the small black hole  where the energy element of the free energy in the competition with the entropy element wins and a high energy and high entropy phase of the large black hole where the entropy element of the free energy in the competition with the energy element wins. From the changes of the slope of the mass versus $r_h$ with respect to the effective pressure $P$ or the equivalently absolute value of the cosmological constant $\Lambda$, we can see from the free energy landscapes in Fig.\ref{GPlotP} that the black hole phases change from the large black hole preference (changing to shallower basin) to the small black hole preference (changing to deeper basin) as the the effective pressure $P$ or the equivalently absolute value of the cosmological constant $\Lambda$ increases.

Notice that the the black hole mass and entropy multiplied by T are both much larger than that of the resulting Gibbs free energy as shown in Fig.\ref{GPlotP}. Therefore, the free energy landscape and the black hole phases are the results of the delicate balance between the two large numbers or scales, the mass and the entropy  giving rise to a much smaller number or scale of the free energy. Due to the scale difference of the free energy (small) in comparison to the energy and entropy (large), accurate quantifications of the energy and entropy are essential for obtaining the accurate information of the free energy and thermodynamics of the black hole. Small errors in estimating the energy and entropy can generate relatively larger errors in the free energy estimation. In other words, it is difficult to switch from one state to another when considering the energy/mass or entropy  barrier alone. For example, it is very difficult to switch from a small size black hole state to the large size black hole state from the energy/mass perspective since the barrier ($\Delta E$, where $\Delta E$ is the energy barrier from initial to the final state) for switching is high compared to the thermal temperature and it is difficult to realize the transition through thermal motion. On the other hand, it is very difficult to switch from a large size black hole state to the small size black hole state from the entropy perspective since the barrier ($T\Delta S$,  where $T \Delta S$ is the entropy barrier from initial to the final state) for switching is high compared to the thermal temperature and it is difficult to realize the transition through thermal motion. However the competition between the mass/energy and the entropy due to their intricate balance can lead to the significantly reduced free energy and the associated free energy barrier between the small and large size black holes. The free energy barrier height is in the order of $T$ which makes the thermal transitions possible between the the small and large size black holes.

\subsubsection{Kinetics and associated fluctuations versus temperature at different pressures or cosmological constants}

\begin{figure}
\centering
\subfigure[]{\label{MFPTstlP}
\includegraphics[width=4cm]{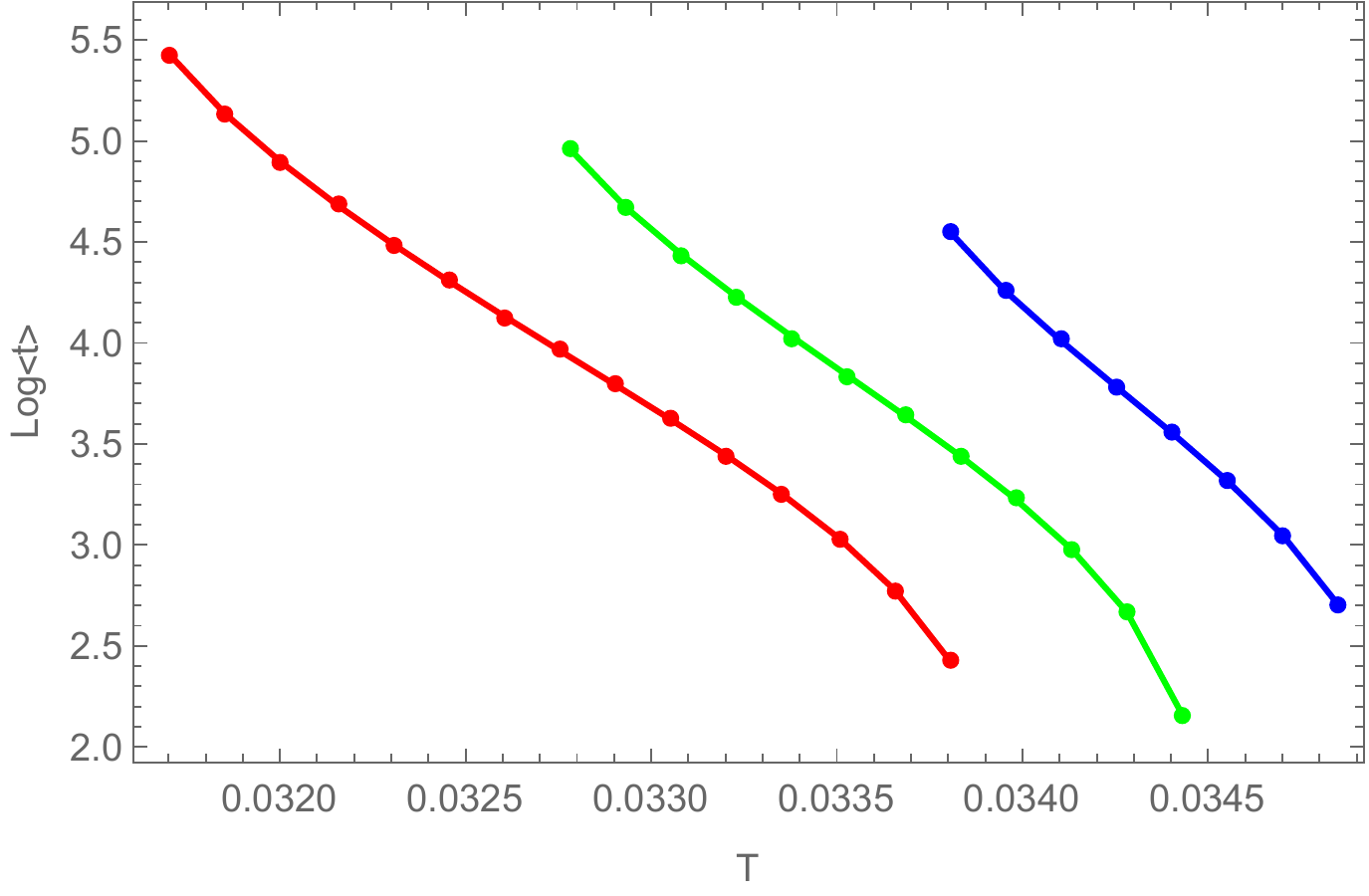}}
\subfigure[]{\label{RelFlucstlP}
\includegraphics[width=4cm]{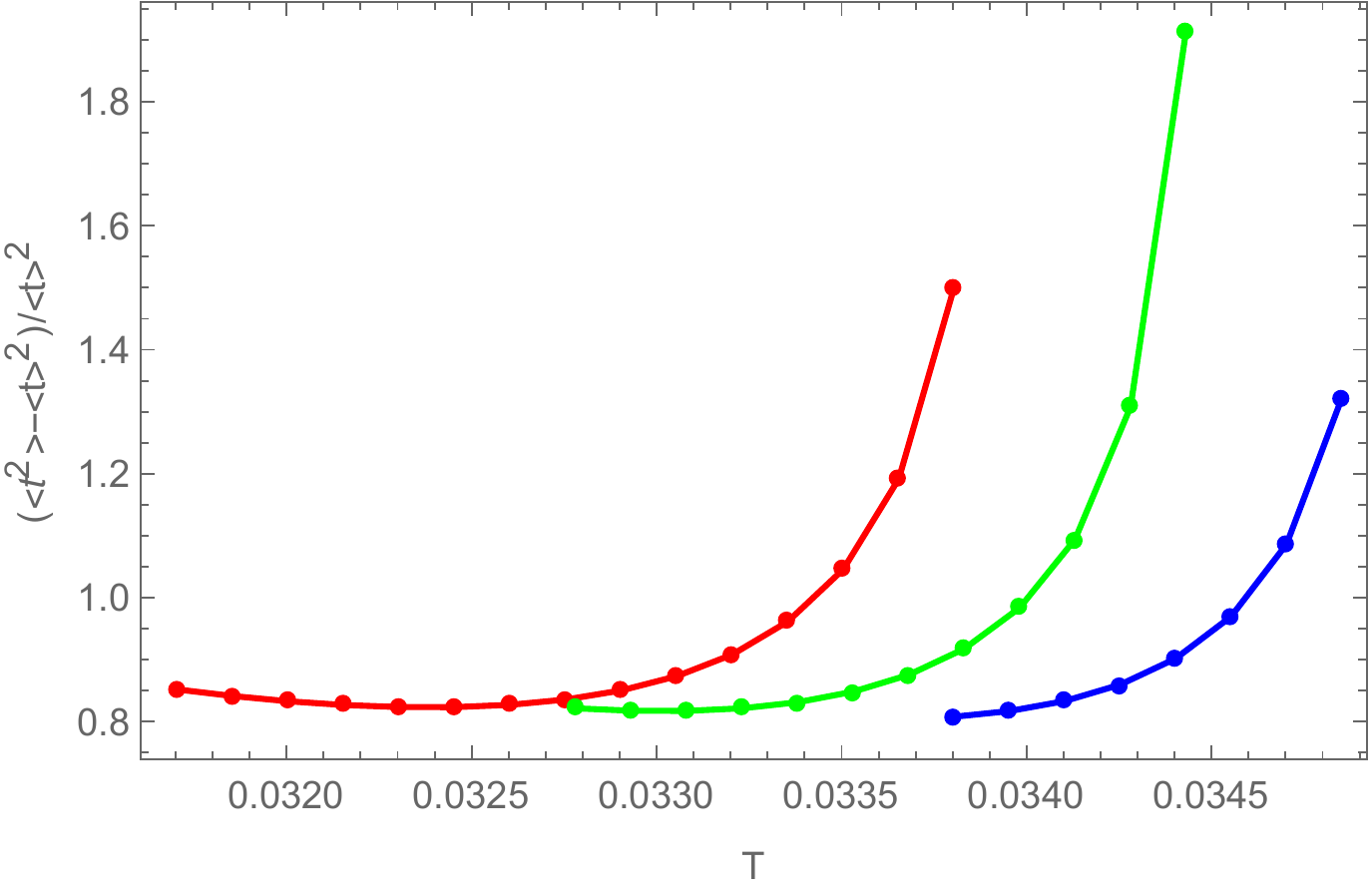}}
\subfigure[]{\label{MFPTltsP}
\includegraphics[width=4cm]{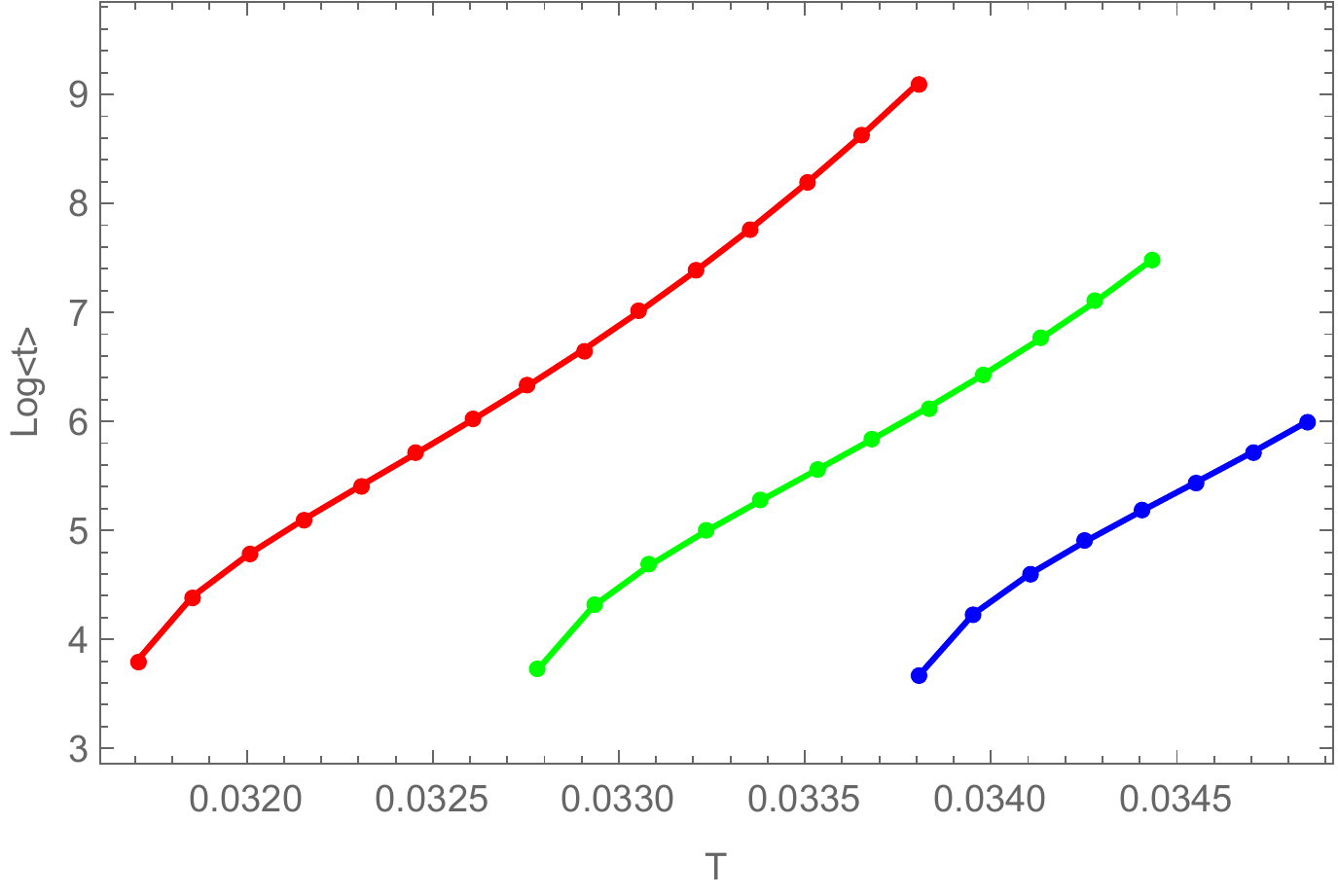}}
\subfigure[]{\label{RelFlucltsP}
\includegraphics[width=4cm]{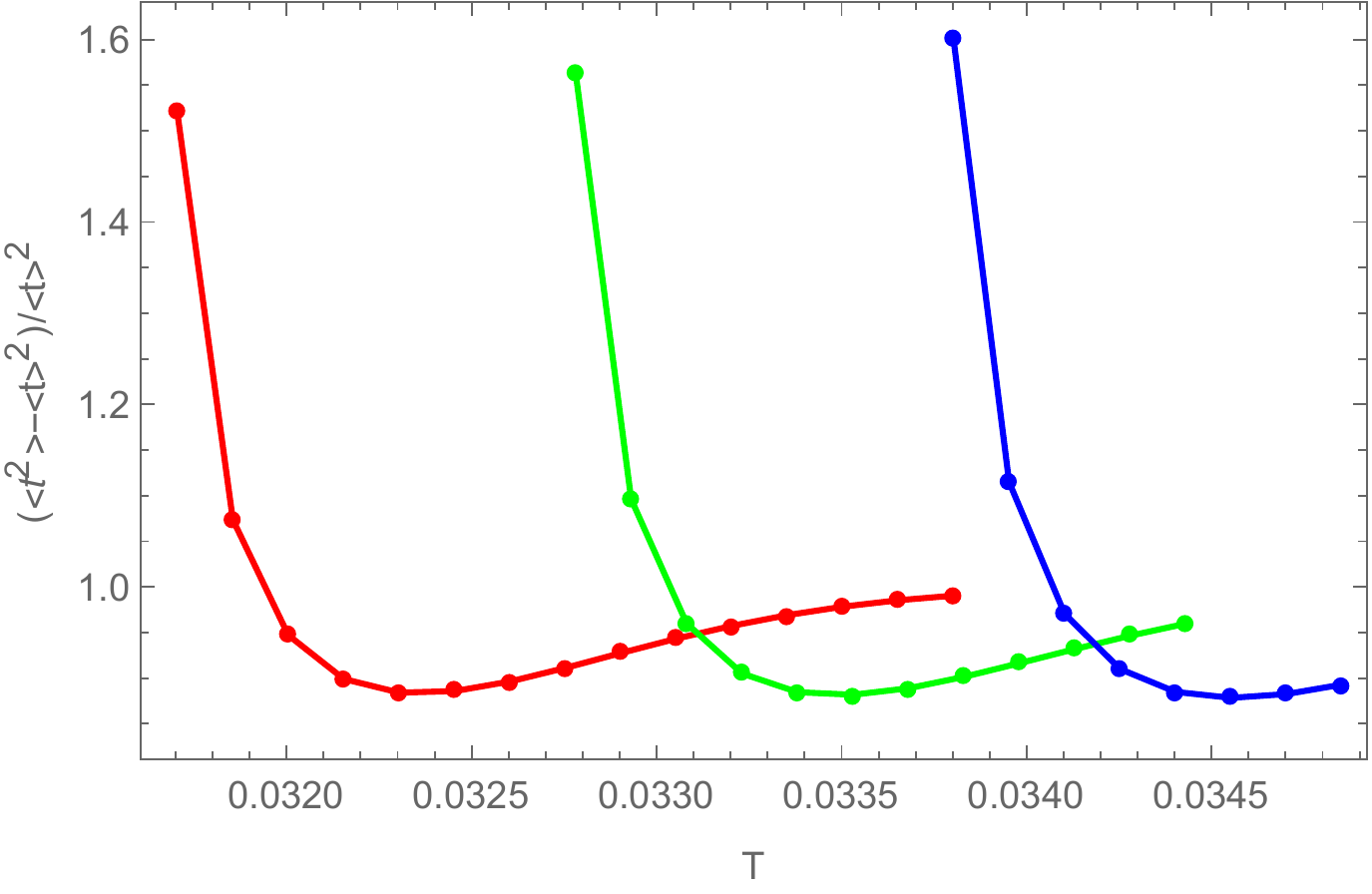}}
\caption{MFPT and the relative fluctuation of the state transition in the canonical ensemble as a function of temperature $T$ for the different pressure $P$.
The panels (a)-(d) shows the same quantities as in fig.\ref{MFPTCEalpha}.
In each panel, the red, green, and blue curves correspond to $P=0.65P_c$, $0.7P_c$, and $0.75P_c$.
The electric charge $Q=1$, and the GB coupling constant $\alpha=0.1$, which are kept fixed.}
\label{MFPTCEP}
\end{figure}

The MFPT and the relative fluctuation as the functions of $T$ for different pressure $P$ are plotted in Fig.\ref{MFPTCEP}. When $P$ increases, the range of the ensemble temperature, where the double well shaped free energy landscape topography appears, moves from left to the right.
The MFPT from the small to the large black hole increases with $P$ while the MFPT from the large to the small black hole decreases with $P$. It is more harder (easier) to switch from the small (large) black hole to the (large) small black hole when increasing the pressure. Because the thermodynamic pressure is related to the cosmological constant, this result indicates that varying the cosmological constant will make the small (larger) black hole more (less) stable.

Therefore, we can conclude that, in the canonical ensemble, when the GB coupling coupling constant increases, or the electric charge increases, or the pressure decreases, the small black hole state is more easier to escape to the large black hole state while the inverse process becomes more harder, i.e. the small (large) black hole state will becomes less (more) stable. The reason behind is that the barrier height between the small (large) black hole state and the intermediate black hole state becomes lower (higher) when increasing the GB coupling coupling constant, or increasing the electric charge, or decreasing the thermodynamic pressure while the variations of the free energy barrier heights are caused by the competitions of the energies or the masses and the entropies of three branches of black hole states.

At last, it can be seen from Fig.\ref{MFPTCEalpha}, Fig.\ref{MFPTCEQ}, and Fig.\ref{MFPTCEP},
for different physical parameters, the MFPT of the state transition from the small to the large black hole decreases with the temperature while the MFPT from the large to the small black hole increases with the temperature. The relative fluctuations for different physical parameters are large at the higher temperature for the state switching process from the small to the large black hole while being large at the lower temperature for the process from the large to the small black hole. These behavior are similar to those of Fig.\ref{MFPTCE}. The reason behind is that these quantities are related to the barrier height in the free energy landscape topography and the ensemble temperature, just as the discussions in the last subsection.

\section{Phase transition and Kinetics in Grand canonical Ensemble}

\subsection{Free energy landscape and phase diagram in the grand canonical}

In this subsection, we discuss the phase transition and the kinetics in the grand canonical ensemble where the electric potential $\Phi$ is fixed. In the grand canonical ensemble, we can define the generalized off-shell free energy for the transient black hole state as
\begin{eqnarray}
F&=&M-TS-Q\Phi\nonumber\\
&=&\frac{r_h}{2}\left(1+\frac{8\pi P}{3}r_h^2+\frac{\alpha}{r_h^2}+\Phi^2\right)
\nonumber\\
&&-T \left(\pi r_h^2+4\pi\alpha \log(r_h)\right)-\Phi^2 r_h\;,
\end{eqnarray}
where we have used the relation $Q=\Phi /r_h$. According to the definitions of Gibbs free energy as $U-TS+pV$ and of Grand potential as $U-TS-\mu N$ (where $U$ represents the internal energy of the system), we call it Gibbs grand potential. In the following, without the confusion, we also call it free energy of the grand canonical ensemble. In this way, we can then formulate the free energy landscape in the grand canonical ensemble. All the information about the phase transition of GB black holes in the grand canonical ensemble are included in the generalized free energy function.

It is also shown in \cite{Wei:2020poh} that there is a small-large black hole phase transition in the grand canonical ensemble. The critical pressure of the phase transition is given by
\begin{eqnarray}
P_c=\frac{\left(\Phi^2-1\right)^2\left(9-3\Phi^2+\lambda\right)}
{24\pi \alpha\left(6-3\Phi^2+\lambda\right)^2}\;,
\end{eqnarray}
with $\lambda=\sqrt{9\Phi^4-48\Phi^2+48}$. In principle, we can also derive this result from the generalized free energy function by employing the same procedure in the last section.

\begin{figure}
  \centering
  \includegraphics[width=6cm]{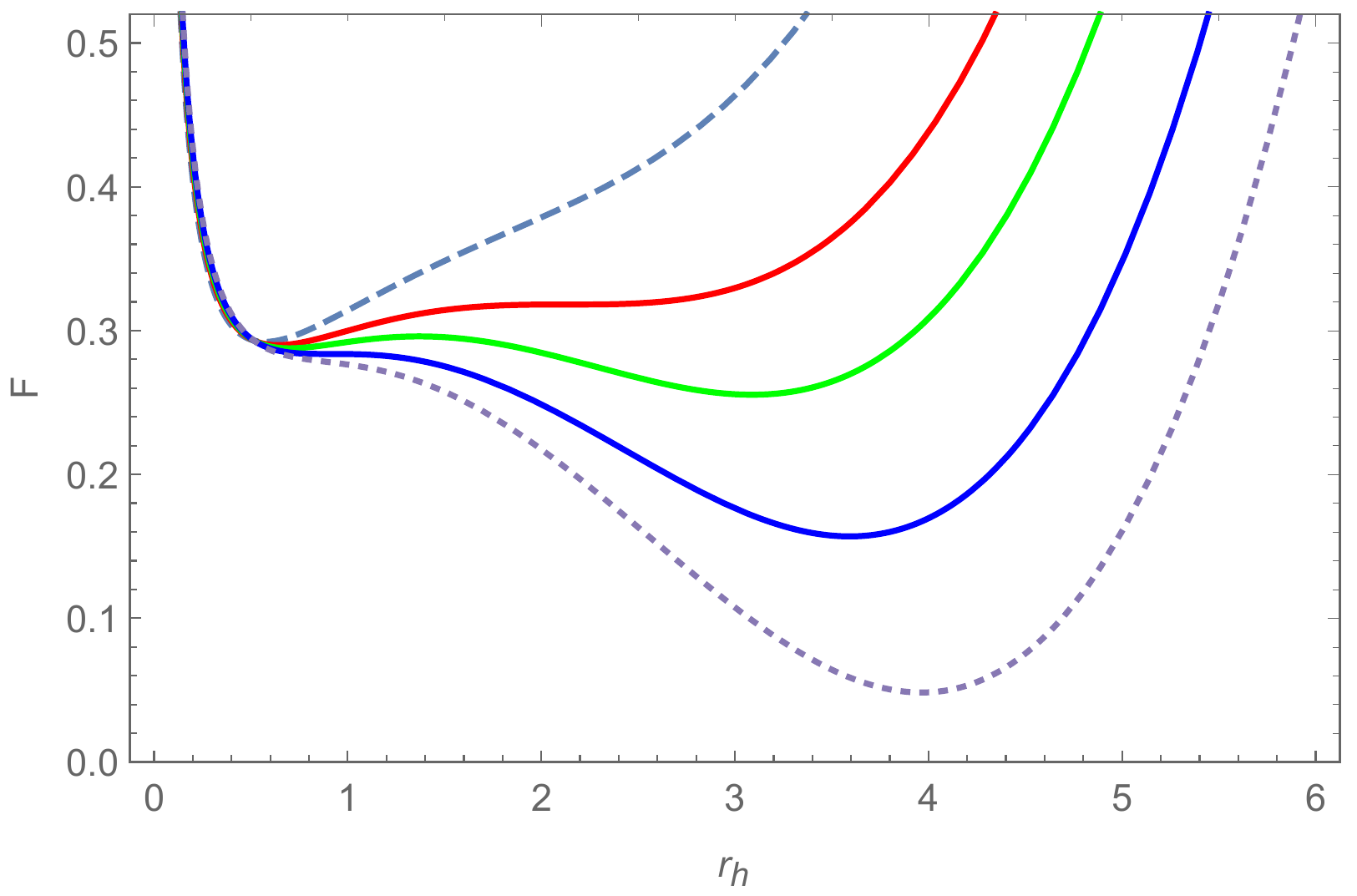}
  \caption{The free energy landscape in grand canonical ensemble for $P=0.6P_c$ with $\Phi=0.5$ and $\alpha=0.1$. In this case, $T_{min}=0.04651$ and $T_{max}=0.05167$. The dashed, red, green, blue, and dotted curves correspond $T<T_{min}$, $T=T_{min}$, $T_{min}<T<T_{mas}$, $T=T_{max}$, and $T>T_{max}$, respectively. }
  \label{GibbsGCE}
\end{figure}

Similar to the case in the canonical ensemble, the black hole temperature is a monotonic function of the black hole radius when $P>P_c$, and has the local minimum $T_{min}$ and the local maximum $T_{max}$ when $P<P_c$. When $T_{min}<T_H<T_{max}$, there also exist three branches of black hole solution (i.e. small, intermediate, and large black hole).

\begin{figure}
  \centering
  \includegraphics[width=6cm]{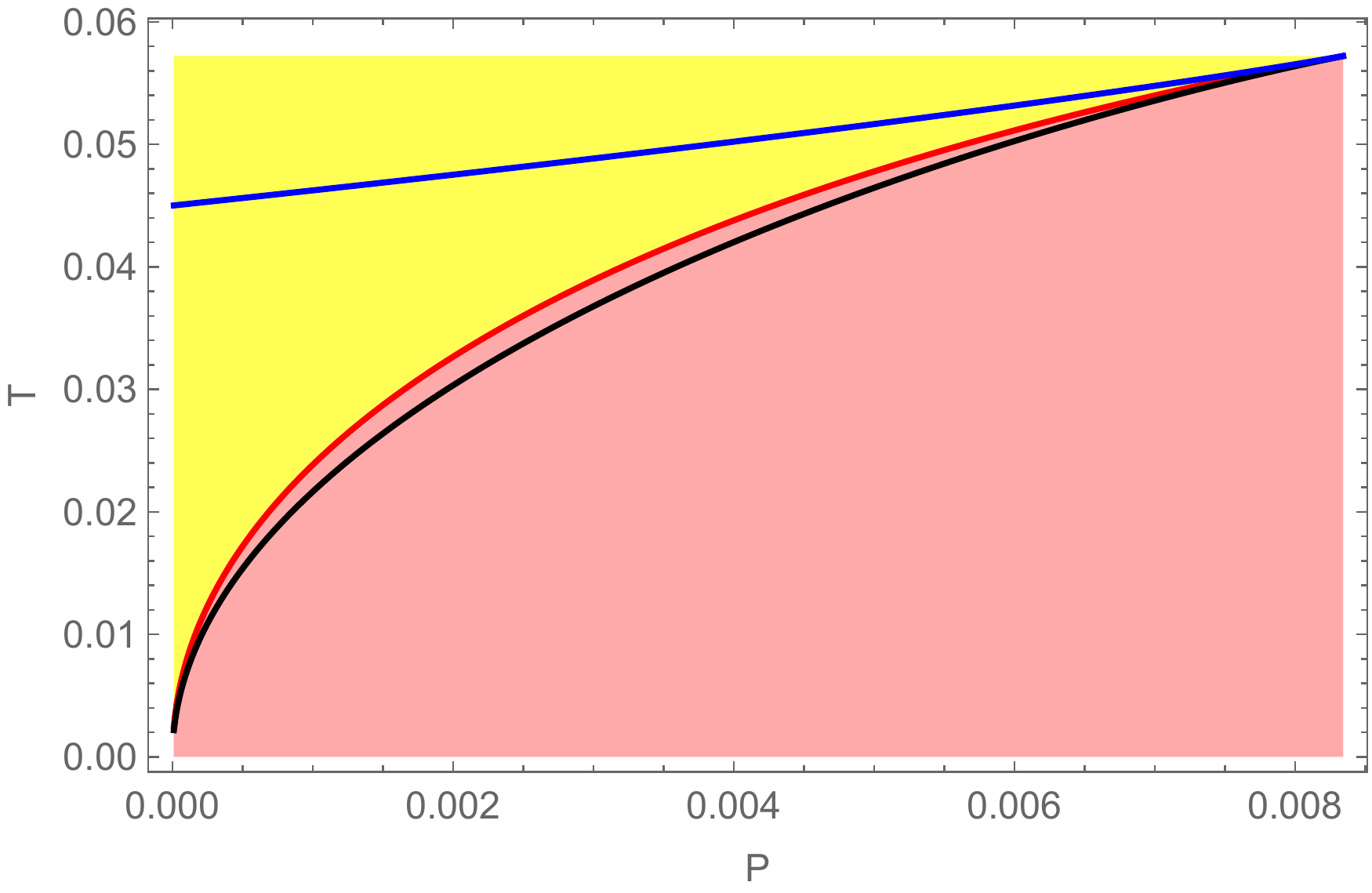}
  \caption{Phase diagram in grand canonical ensemble. The blue, black, and red lines are the plots of $T_{max}$, $T_{min}$, and $T_{trans}$ as a function of pressure $P$. In these plots, the range of $P$ is from $0$ to the critical pressure $P_c$. The small and large black holes are thermodynamically stable in pink and yellow regions, respectively. }
  \label{PhaseDiagramGCE}
\end{figure}

The free energy landscape at different temperatures are presented in Fig.\ref{GibbsGCE}. It can be seen that the free energy landscape topographies in the grand canonical ensemble and in the canonical ensemble have the similar shapes. When $T<T_{min}$ or $T>T_{max}$, there is only one basin in the landscape topography. The small or the large black hole which is represented by the minimum point in the free energy landscape topography is always thermodynamically stable. When $T_{min}<T<T_{max}$, there are two basins in the landscape topography. The three local extremal points correspond to the three branches of black hole solutions. Two of them correspond to the locally stable small or large black hole states while the other one corresponds to an unstable intermediate black hole state. The thermodynamically stable phase is also determined by the global minimum point in the free energy landscape topography. By adjusting the ensemble temperature, there is a first order phase transition between the small black hole and the large black hole where the free energies of the two are equal. The phase diagram is summarized in Fig.\ref{PhaseDiagramGCE}.

\subsection{Free energy landscapes and the kinetics of state transitions under different parameters}

\subsubsection{varying the coupling constant $\alpha$}

\begin{figure}
  \centering
  \includegraphics[width=6cm]{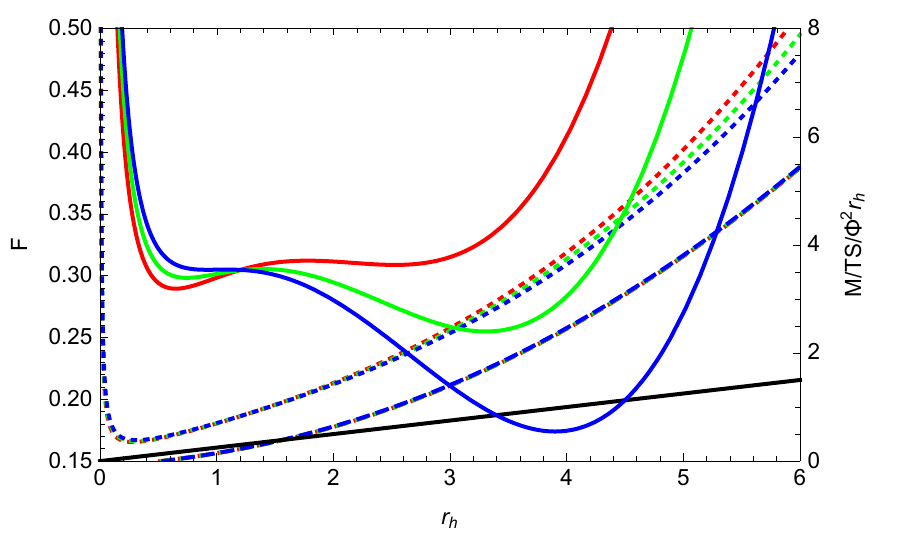}
  \caption{The free energy $F$ (solid), black hole mass $M$ (dotted), the product $TS$ of entropy and temperature (dashed) and the electric contribution term $\Phi^2 r_h$ to the free energy (black solid) as the functions of $r_h$ with different coupling constant $\alpha$. The red, green, and blue curves correspond to $\alpha=0.1$, $0.11$, and $0.12$, respectively. The electric potential $\Phi=0.5$, the pressure $P=0.6P_c$, and the ensemble temperature $T=0.047$. }
  \label{FPlotalpha}
\end{figure}

\begin{table}[!htbp]\caption{Thermodynamic quantities of different types of black hole at different GB coupling constant with $\Phi=0.5$, $P=0.6P_c$, and $T=0.047$.\\}
\centering
\begin{tabular}{|c|c|c|c|c|c|}
\hline
BH Type&$\alpha$&$r_h$&$M$&$S$&$F$\\
\hline
\multirow{3}*{Small}&0.1&0.648406&1.17816&0.7764&0.2895\\
\cline{2-6}
&0.11&0.742197&1.12669&1.31845&0.298268\\
\cline{2-6}
&0.12&0.924297&1.08184&2.56523&0.304782\\
\hline
\multirow{3}*{Intermediate}&0.1&1.78142&1.3182&10.6953&0.312171\\
\cline{2-6}
&0.11&1.38999&1.14556&6.52496&0.305425\\
\cline{2-6}
&0.12&1.08933&1.08137&3.85699&0.304928\\
\hline
\multirow{3}*{Large}&0.1&2.5281&1.82101&21.2443&0.308736\\
\cline{2-6}
&0.11&3.30429&2.50904&35.9531&0.254889\\
\cline{2-6}
&0.12&3.90074&3.13288&49.8543&0.173953\\
\hline
\end{tabular}
\label{TablealphaGCE}
\end{table}

\begin{figure}
  \centering
  \includegraphics[width=6cm]{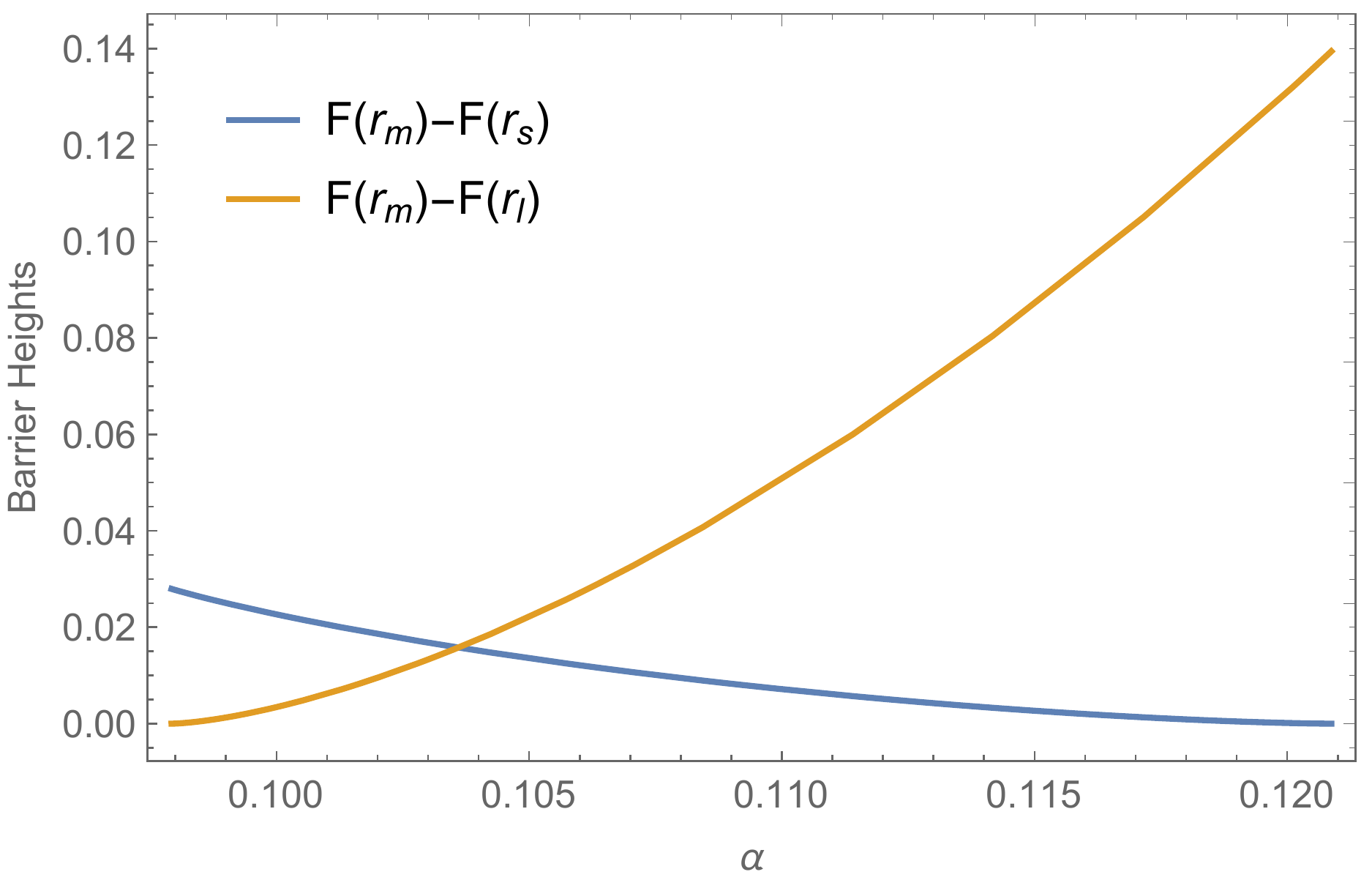}
  \caption{The barrier heights of free energy landscape as the function of GB coupling constant. The barrier height between the small (large) black hole and the intermediate black hole monotonically decreases (increases) with $\alpha$. The electric potential $\Phi=0.5$, the pressure $P=0.6P_c$, and the ensemble temperature $T=0.047$. }
  \label{BHalphaGCE}
\end{figure}

\begin{figure}
\centering
\subfigure[]{\label{MFPTstlalphaGCE}
\includegraphics[width=4cm]{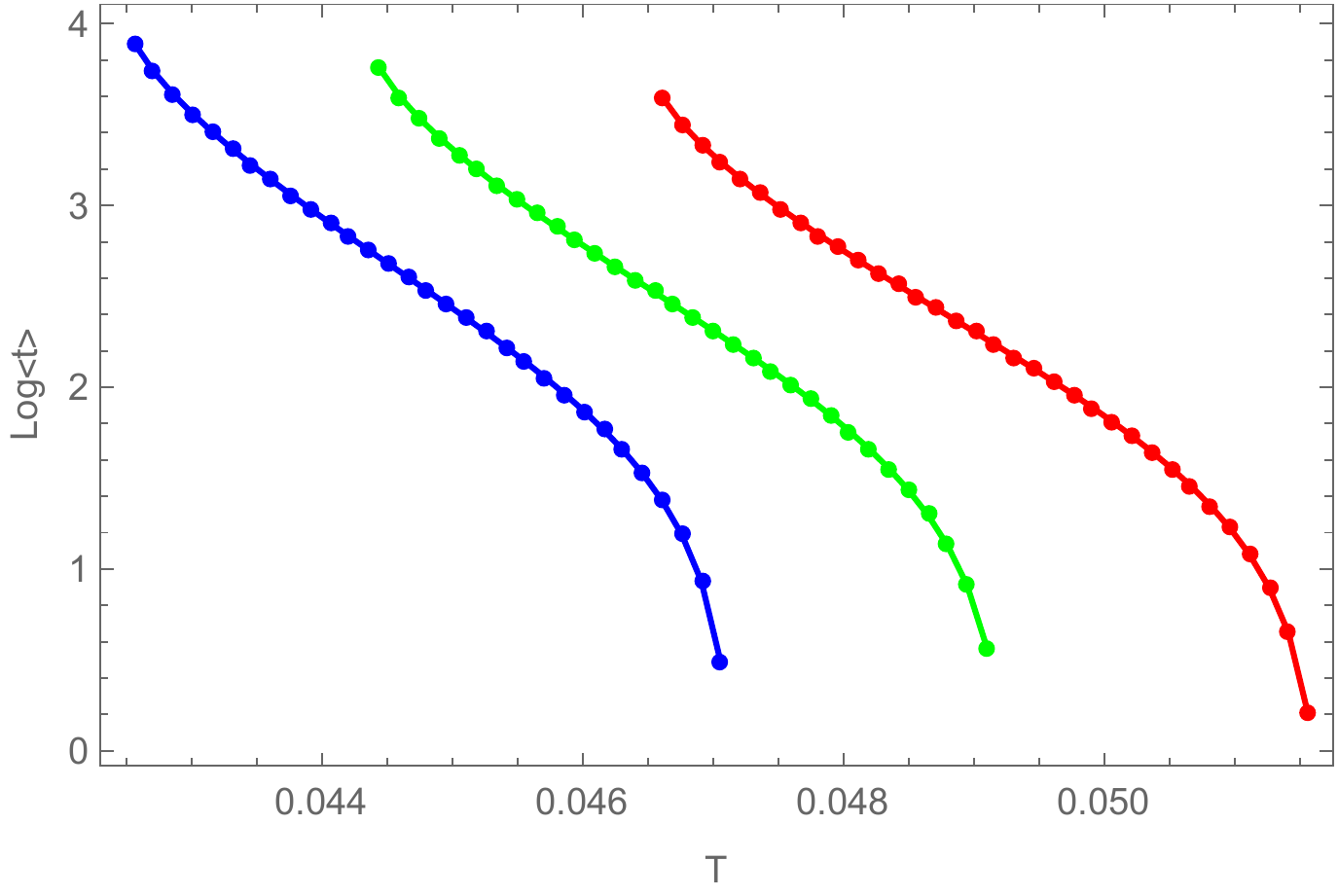}}
\subfigure[]{\label{RelFlucstlalphaGCE}
\includegraphics[width=4cm]{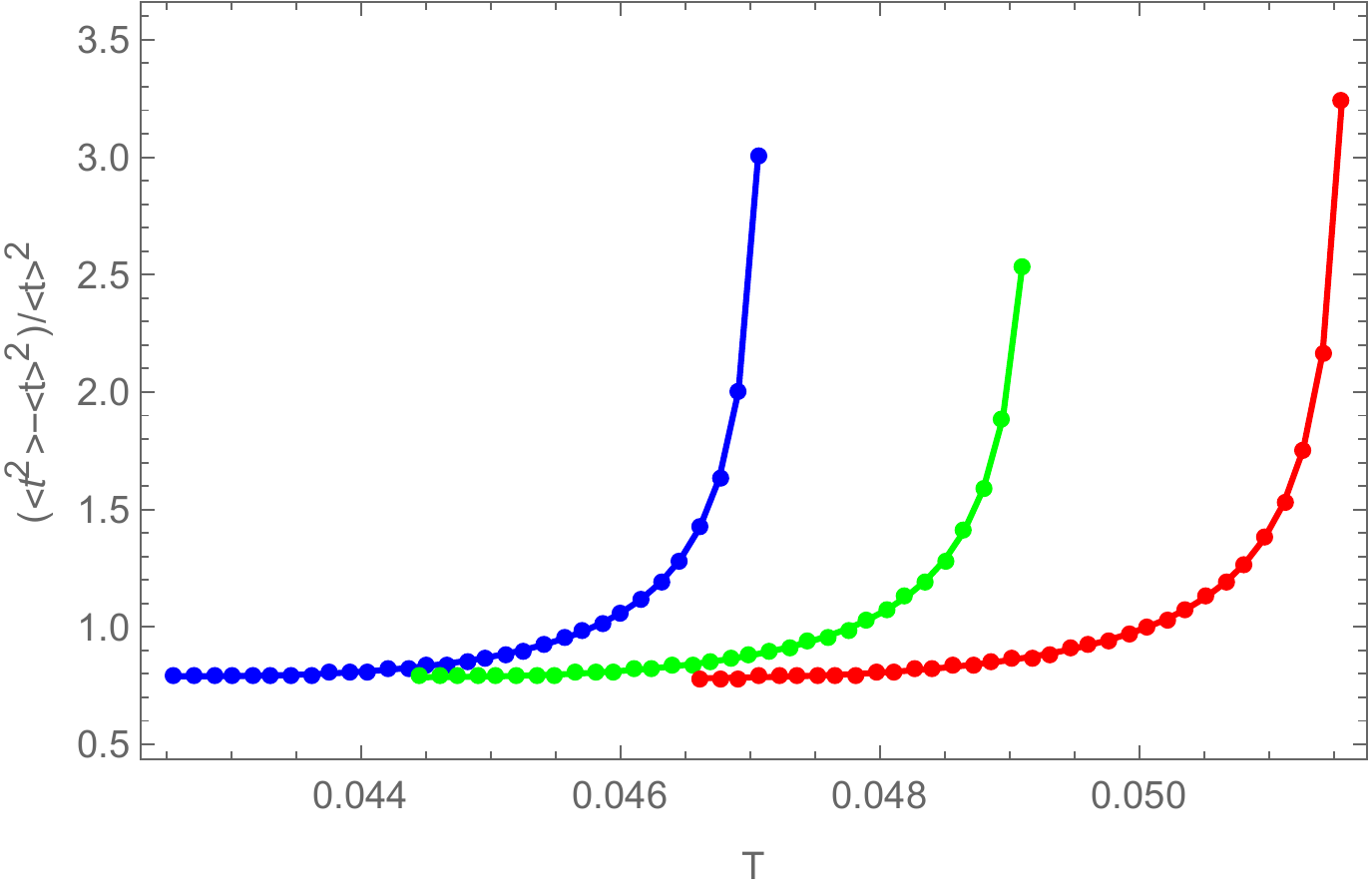}}
\subfigure[]{\label{MFPTltsalphaGCE}
\includegraphics[width=4cm]{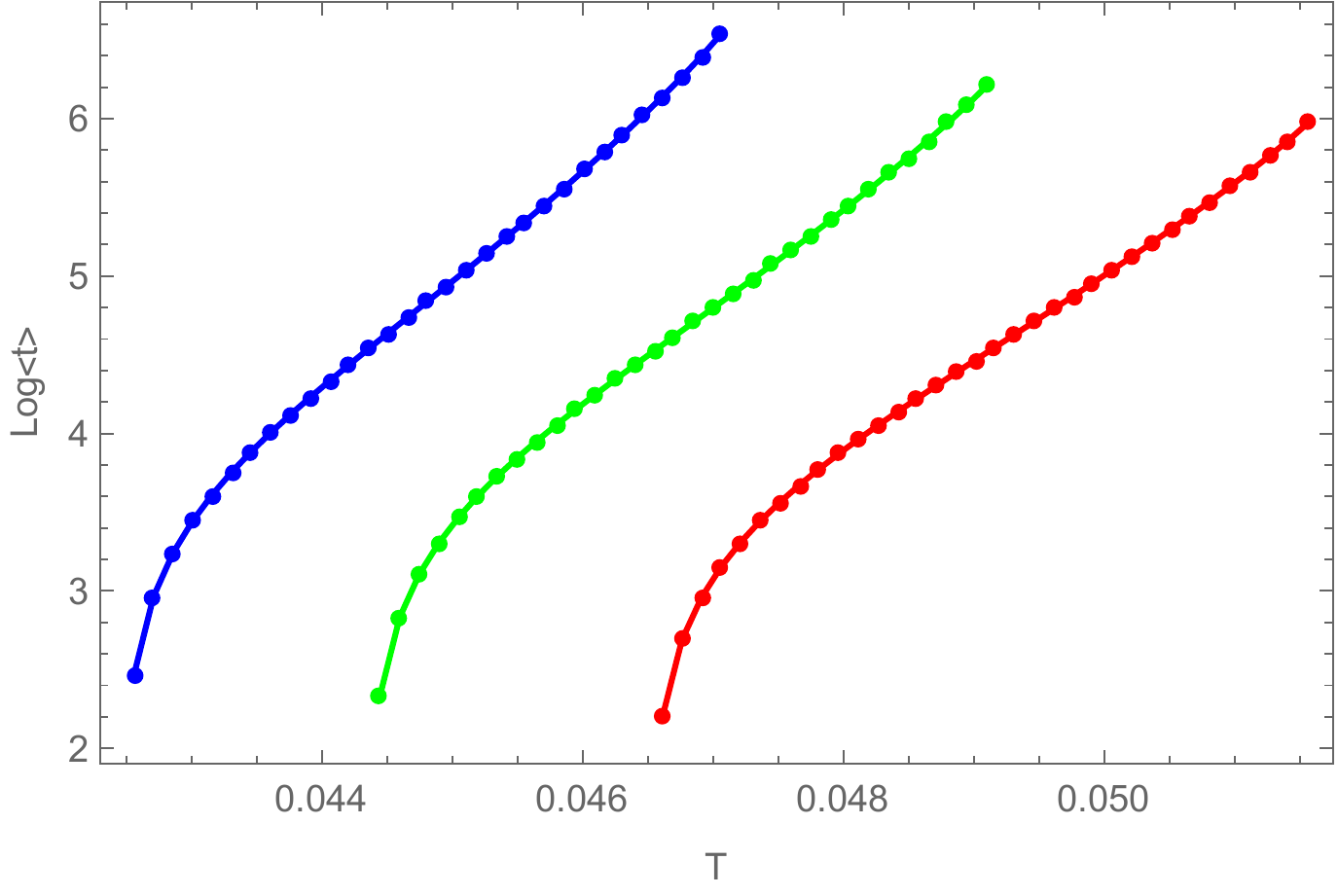}}
\subfigure[]{\label{RelFlucltsalphaGCE}
\includegraphics[width=4cm]{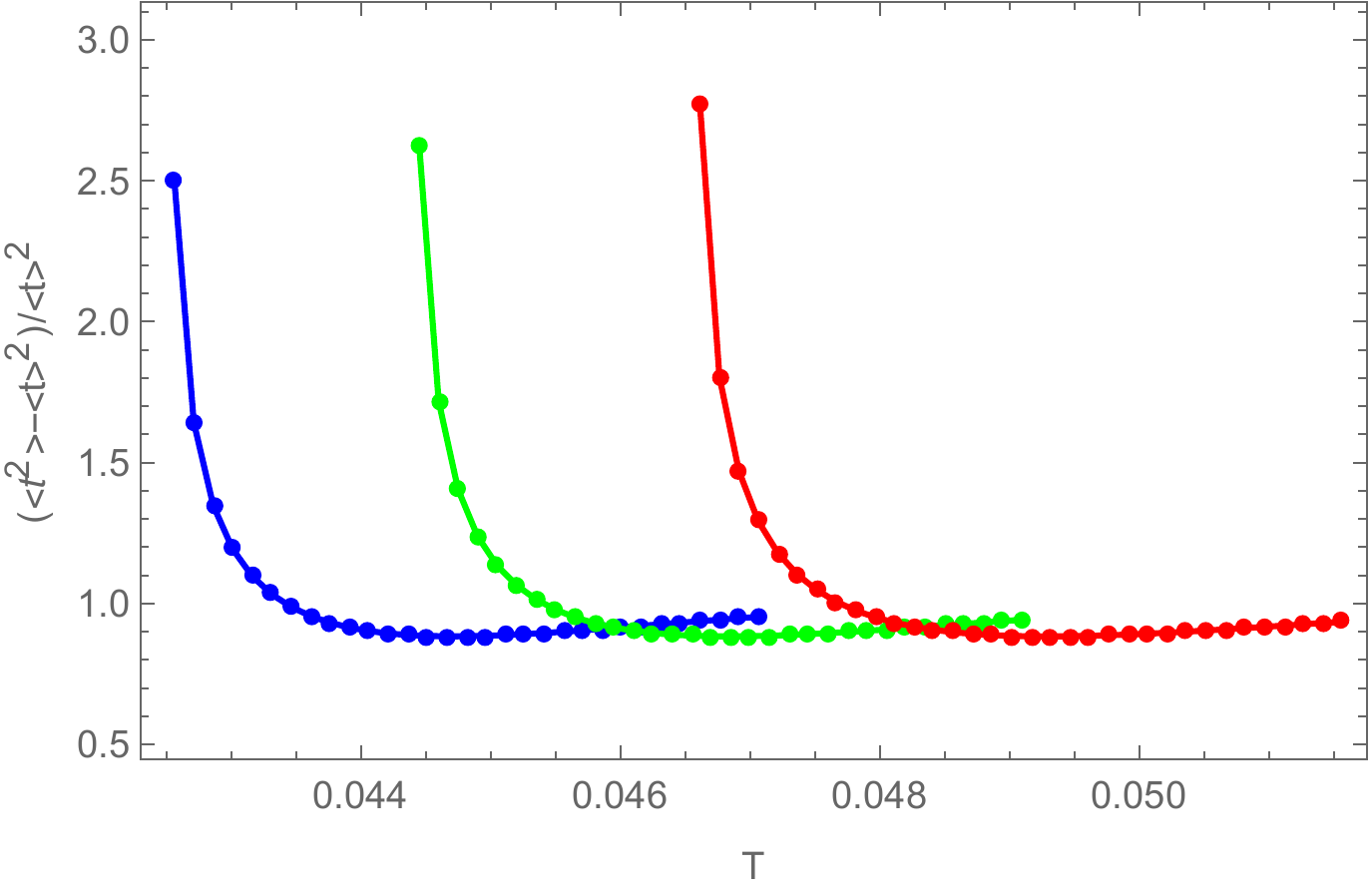}}
\caption{MFPT and the relative fluctuation of the state transition in the grand canonical ensemble as a function of temperature $T$ for the different GB coupling constant $\alpha$.
(a): MFPT from the small to the large black hole;
(b): Relative fluctuation from the small to the large black hole;
(c): MFPT from the large to the small black hole;
(d): Relative fluctuation from the large to the small black hole.
In each panel, the red, green, and blue curves correspond to $\alpha=0.1$, $0.11$, and $0.12$.
The electric potential $\Phi=0.5$, and the pressure $P=0.6P_c$, which are kept fixed.}
\label{MFPTGCEalpha}
\end{figure}

In Fig.\ref{FPlotalpha}, the free energy $F$, black hole mass $M$, the product $TS$ of entropy and temperature and the electric contribution term $\Phi^2 r_h$ as the functions of $r_h$ with different coupling constant are plotted. We can see that the black hole mass or energy is nonmonotonic with respect to $r_h$, In fact, certain small size black hole is preferred (minimum of $M$ vs $r_h$). We can also see that the entropy multiplied by the temperature and electric potential related term $Q\Phi$ monotonically increase as $r_h$ increases. In fact, low energy or mass or high entropy or low electric contribution term $Q\Phi$ are always preferred when acting alone. Since both mass and entropy are nonlinear functions of the black hole size and the electric potential related term $Q\Phi$ is linear in the size of the black hole, the competition among the three can generate interesting behaviors and possible new phases unexpected from the action of the mass or the entropy or the electric potential related term $Q\Phi$ alone. As shown in Fig.\ref{FPlotalpha}, when the coupling $\alpha$ is smaller, the small sized black hole is more preferred from the mass versus $r_h$ under steeper slope while when the coupling $\alpha$ becomes bigger, the small sized black hole is less stable because the mass versus $r_h$ has shallower slope. Considering the entropy or the electric potential related term $Q\Phi$ alone, we see that the large size black hole is always preferred. The change of the slope of entropy and electric potential related term versus coupling is not significant as shown in Fig.\ref{FPlotalpha}. The grand thermodynamic potential as the result of the competition among the mass (energy),  the entropy and the electric potential related term leads to two possible stable black hole phases in certain parameter regimes, a small size black hole favored by the energy or mass and a large size black hole favored by the entropy or electric potential related term $Q\Phi$. In other words, the grand thermodynamic potential has two minimum representing the two phases of the black hole, a low energy, low entropy phase and low electric potential related term of the small black hole where the energy element of the grand thermodynamic potential in the competition with the entropy and electric potential related term elements wins and a high energy, high entropy and high electric potential related contribution phase of the large black hole where the entropy element and the electric potential related term of the grand thermodynamic potential in the competition with the energy element wins. From the changes of the slope of the mass versus $r_h$ with respect to the coupling $\alpha$, we can see that from the grand thermodynamic potential landscapes in Fig.\ref{FPlotalpha} that the black hole phases change from the small black hole preference (changing to shallower basin) to the large black hole preference (changing to deeper basin) as the coupling $\alpha$ increases.

Notice that the the black hole mass, the entropy multiplied by $T$ and electric potential related term are all much larger than that of the resulting grand thermodynamic potential as shown in Fig.\ref{FPlotalpha}. Therefore, the Gibbs grand potential landscape and the black hole phases are the results of the delicate balance among the three large numbers or scales, the mass, the entropy and the electric potential related contribution, giving rise to a much smaller number or scale of the Gibbs grand potential. Due to the scale difference of the Gibbs grand potential (small) in comparison to the energy, entropy and electric potential related contribution (large), accurate quantifications of the energy, entropy and electric potential related contribution are essential for obtaining the accurate information of the Gibbs grand potential and thermodynamics of the black hole. Small errors in estimating the energy, entropy and electric potential related contribution can generate relatively larger errors in the Gibbs grand potential estimation. In other words, it might be difficult to switch from one state to another when considering the energy/mass or entropy or the electric potential related contribution barrier alone. For example, it is very difficult to switch from a small size black hole state to the large size black hole state from the energy/mass perspective since the barrier ($\Delta E$, where $\Delta E$ is the energy barrier from initial to the final state) for switching is high compared to the thermal temperature and it is difficult to realize the transition through thermal motion. On the other hand, it is very difficult to switch from a large size black hole state to the small size black hole state from the entropy and electric potential related contribution perspective since the barrier for switching ($T\Delta S$ or $-Q \Delta \Phi$,  where $T \Delta S$ is the entropy barrier from initial to the final state and $-Q \Delta \Phi$ is the electric potential related term barrier from initial to the final state) is high compared to the thermal temperature and it is difficult to realize the transition through thermal motion. However, the competition among the mass/energy, the entropy and electric potential related contribution due to their intricate balance can lead to the significantly reduced Gibbs grand potential and the associated Gibbs grand potential barrier between the small and large size black holes. The Gibbs grand potential barrier height is in the order of $T$ which makes the thermal transitions possible between the the small and large size black holes.

\subsubsection{Varying the electric potential $\Phi$}

\begin{figure}
  \centering
  \includegraphics[width=6cm]{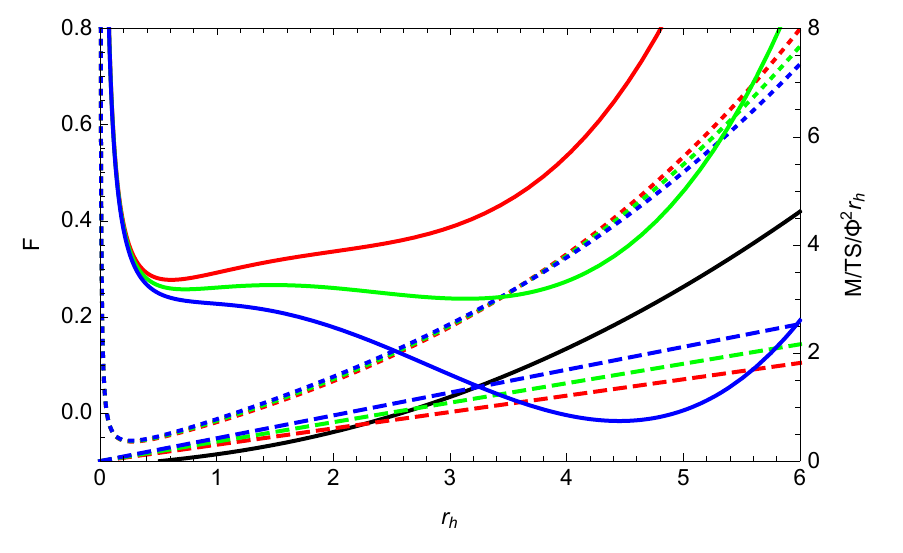}
  \caption{The free energy $F$ (solid), black hole mass $M$ (dotted), the product $TS$ of entropy and temperature (black solid) and the electric potential related term $\Phi^2 r_h$ (dashed) as the functions of $r_h$ with different electric potential $\Phi$. The red, green, and blue curves correspond to $\Phi=0.55$, $0.6$, and $0.65$, respectively. The coupling constant $\alpha=0.1$, the pressure $P=0.6P_c$, and the ensemble temperature $T=0.04$. }
  \label{FPlotPhi}
\end{figure}

\begin{table}[!htbp]\caption{Thermodynamic quantities of different types of black hole with $\alpha=0.1$, $\Phi=0.6$, $P=0.6P_c$, and $T=0.04$.\\}
\centering
\begin{tabular}{|c|c|c|c|c|}
\hline
BH Type&$r_h$&$M$&$S$&$F$\\
\hline
\multirow{1}*{Small}&0.727878&1.12596&1.2653&0.257399\\
\hline
\multirow{1}*{Intermediate}&1.48512&1.16724&7.42607&0.266204\\
\hline
\multirow{1}*{Large}&3.13357&2.25275&32.2834&0.237805\\
\hline
\end{tabular}
\label{TablePhiGCE}
\end{table}

\begin{figure}
  \centering
  \includegraphics[width=6cm]{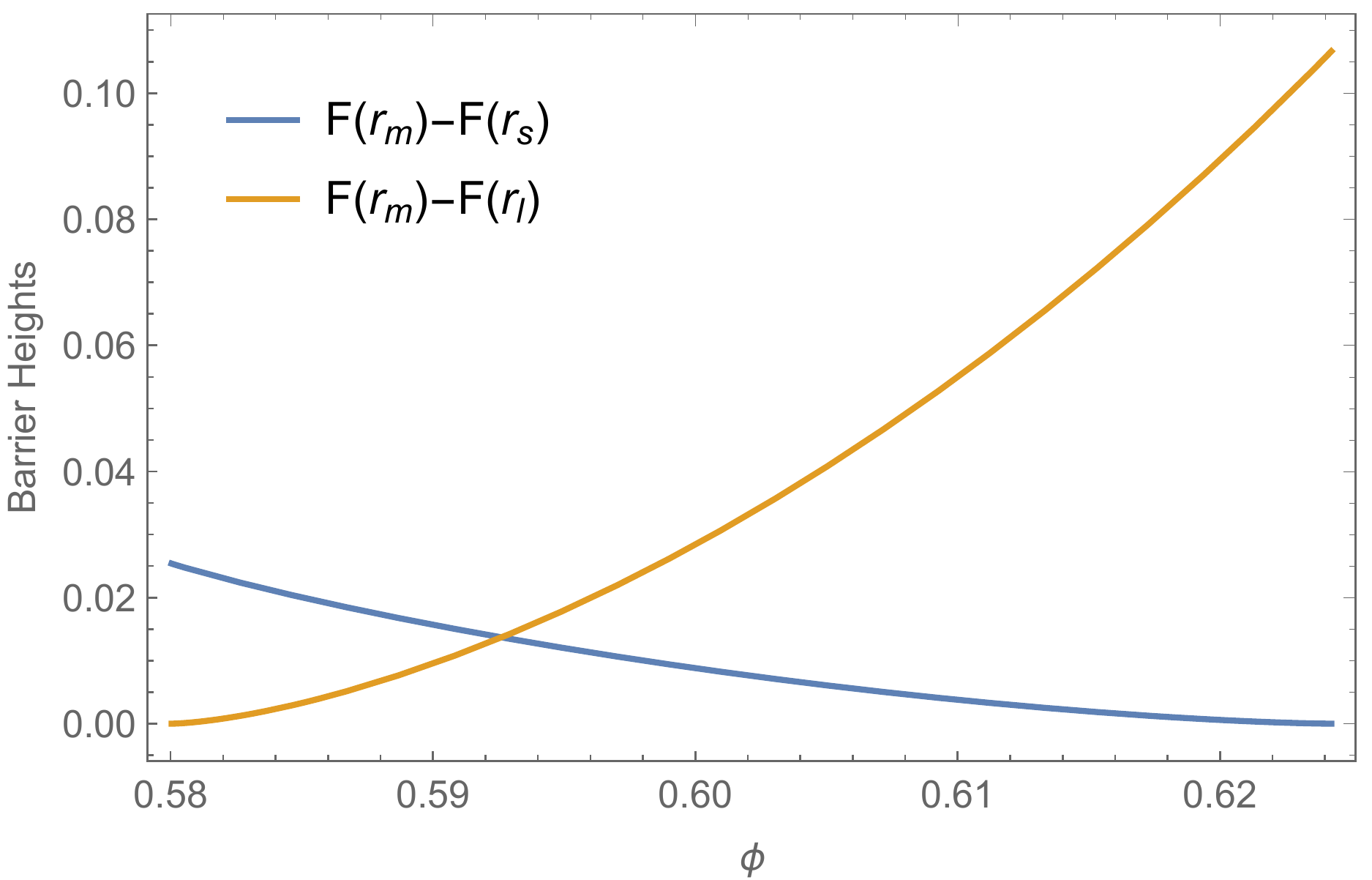}
  \caption{The barrier heights of free energy landscape as the function of the electric potential. The barrier height between the small (large) black hole and the intermediate black hole monotonically decreases (increases) with $\Phi$. The coupling constant $\alpha=0.1$, the pressure $P=0.6P_c$, and the ensemble temperature $T=0.04$. }
  \label{BHPhiGCE}
\end{figure}

\begin{figure}
\centering
\subfigure[]{\label{MFPTstlPhiGCE}
\includegraphics[width=4cm]{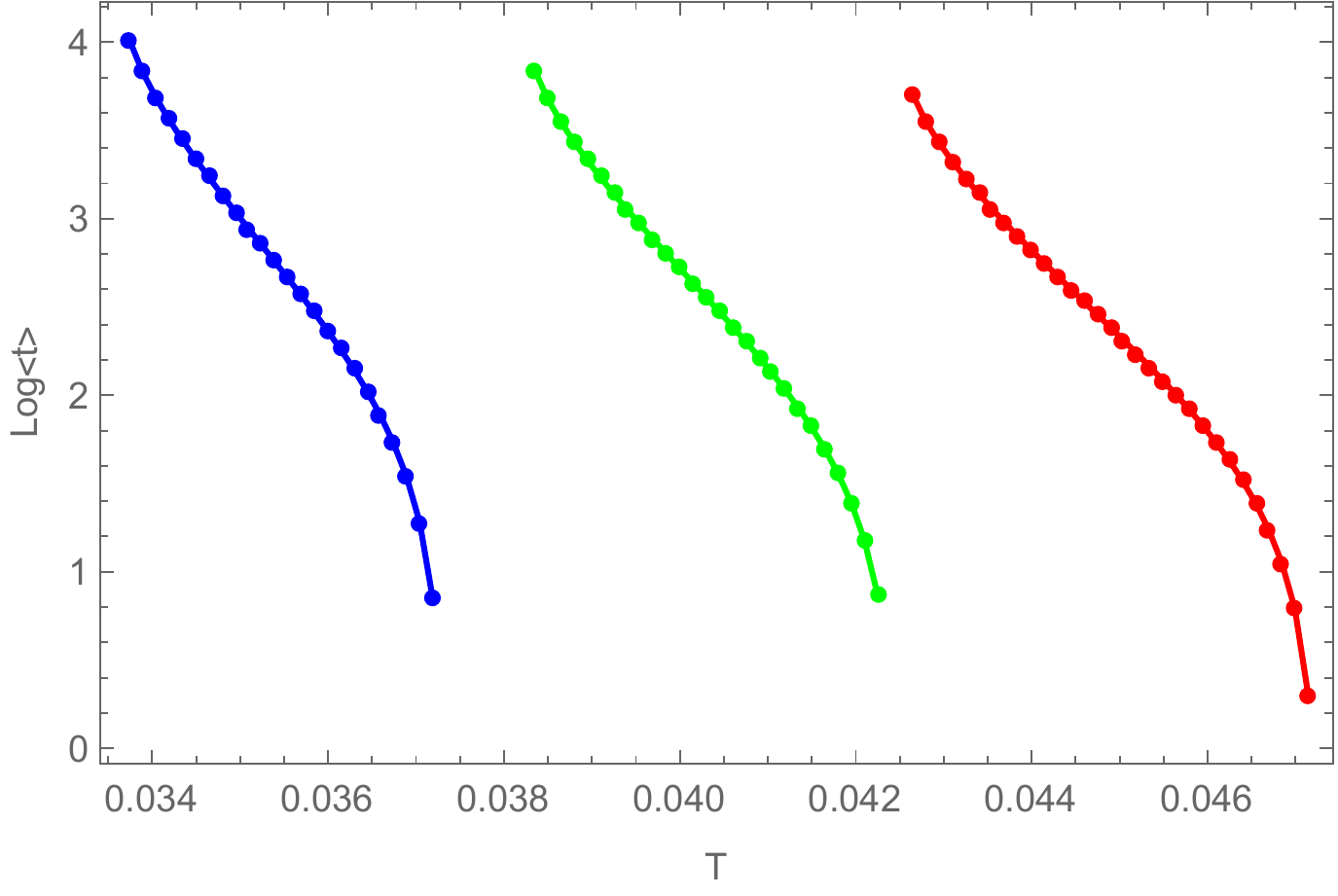}}
\subfigure[]{\label{RelFlucstlPhiGCE}
\includegraphics[width=4cm]{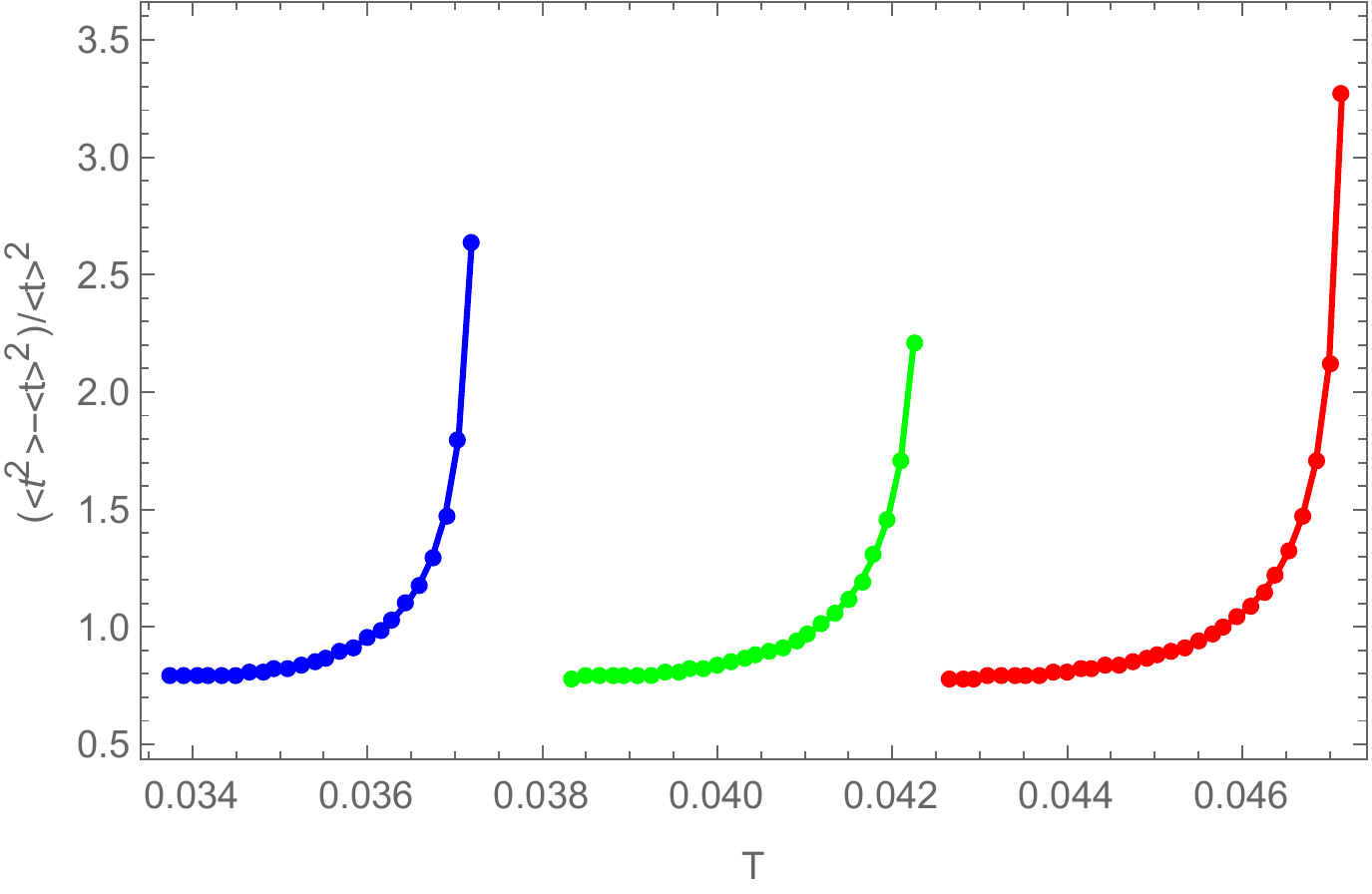}}
\subfigure[]{\label{MFPTltsPhiGCE}
\includegraphics[width=4cm]{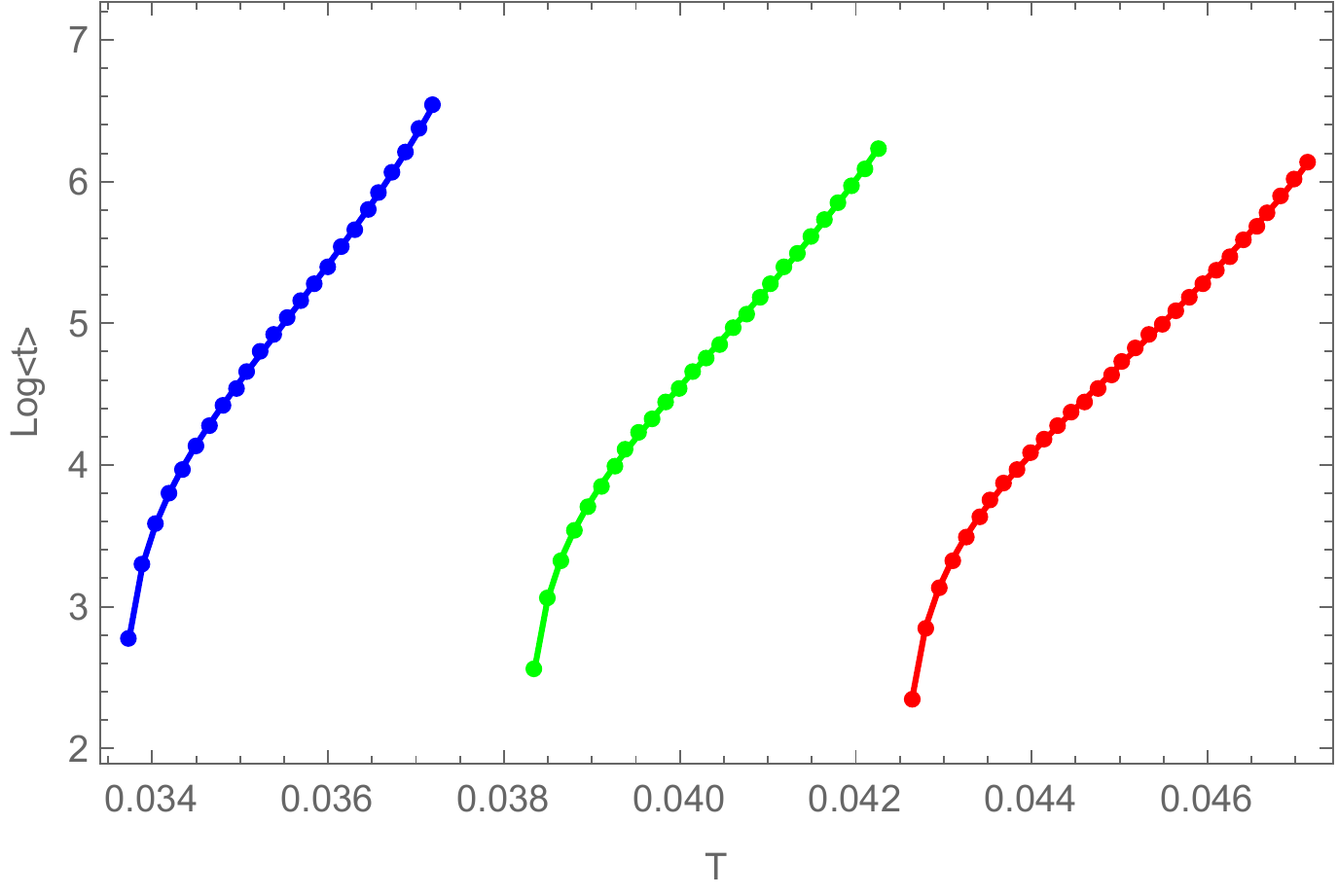}}
\subfigure[]{\label{RelFlucltsPhiGCE}
\includegraphics[width=4cm]{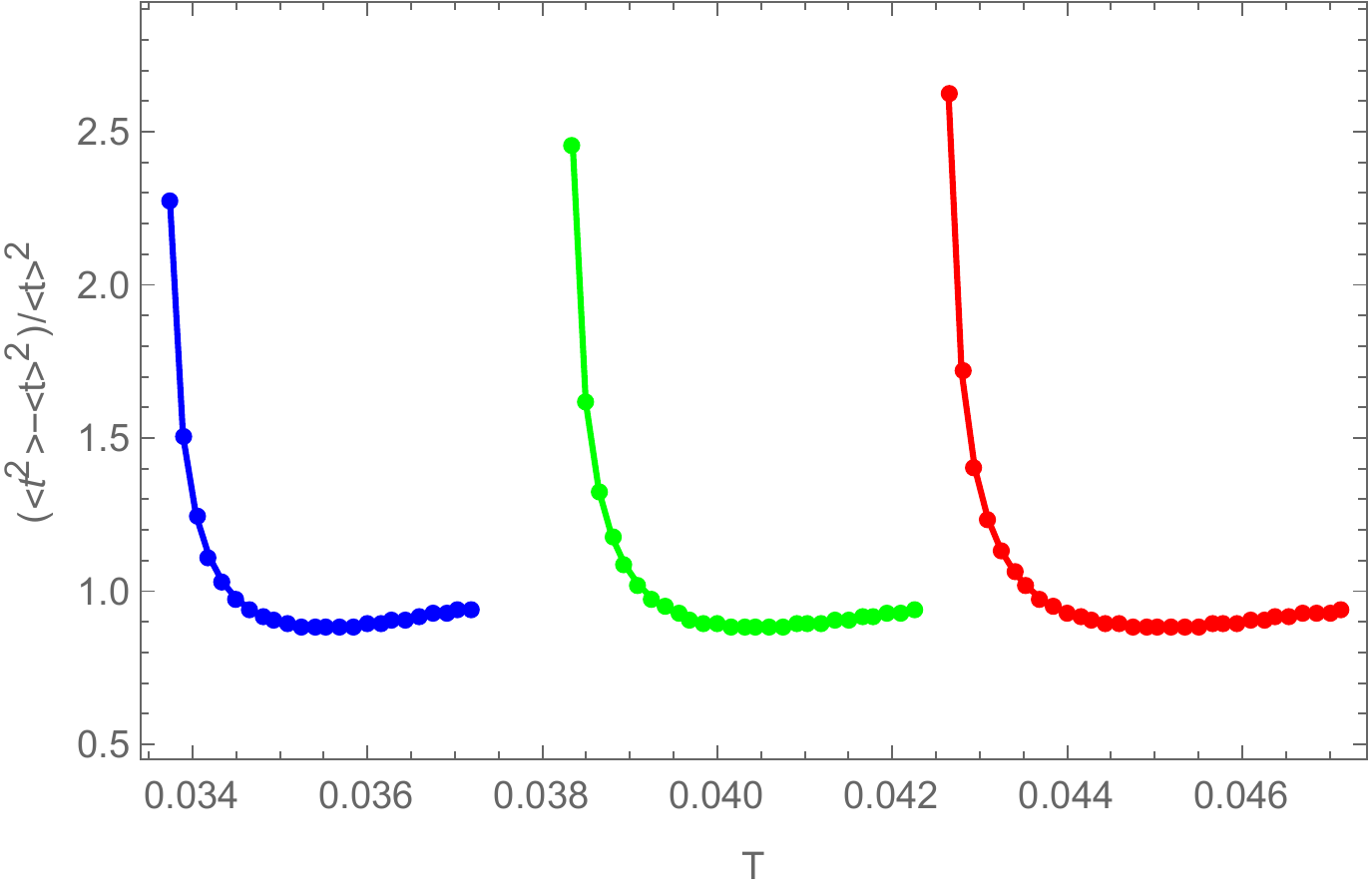}}
\caption{MFPT and the relative fluctuation of the state transition in the grand canonical ensemble as a function of temperature $T$ for the different electric potential $\Phi$.
The panels (a)-(d) shows the same quantities as in fig.\ref{MFPTGCEalpha}.
In each panel, the red, green, and blue curves correspond to $\Phi=0.55$, $0.6$, and $0.65$.
The GB coupling constant $\alpha=0.1$, and the pressure $P=0.6P_c$, which are kept fixed.}
\label{MFPTGCEPhi}
\end{figure}

In Fig.\ref{FPlotPhi}, the free energy $F$, black hole mass $M$, the product $TS$ of entropy and temperature and the electric potential related term $\Phi^2 r_h$ as the functions of $r_h$ with different electric potential are plotted. We can see that the black hole mass or energy is nonmonotonic with respect to $r_h$, In fact, certain small size black hole is preferred (minimum of $M$ vs $r_h$). We can also see that the entropy multiplied by the temperature and electric potential related term $Q \Phi$ monotonically increases as $r_h$ increases. In fact, low energy or mass or high entropy or low electric potential related contribution ($-Q \Phi$) are always preferred when acting alone. Since both mass and entropy are nonlinear functions of the black hole size, and the electric potential related contribution is linear in size of the black hole, the competition among the three can generate interesting behaviors and possible new phases unexpected from the action of the mass or the entropy or the electric potential related contribution alone. As shown in Fig.\ref{FPlotPhi}, when the electric potential $\Phi$ is smaller, the small sized black hole is more preferred from the mass versus $r_h$ under steeper slope while when the electric potential $\Phi$ becomes bigger, the small sized black hole is less stable because the mass versus $r_h$ has shallower slope. Considering the entropy or the electric potential related contribution alone, we see that it always prefers the large size black hole. The entropy is not explicitly dependent on electric potential $\Phi$ from the analytical expression and as shown in Fig.\ref{FPlotPhi}. The Gibbs grand thermodynamic potential as the result of the competition among the mass (energy), the entropy and electric potential related contribution leads to two possible stable black hole phases in certain parameter regimes, a small size black hole favored by the energy or mass and a large size black hole favored by the entropy and electric potential related contribution. In other words, the grand thermodynamic potential has two minimum representing the two phases of the black hole, a low energy, low entropy and low electric potential related contribution phase of the small black hole where the energy element of the free energy in the competition with the entropy and electric potential related element wins and a high energy, high entropy and high electric potential related contribution phase of the large black hole where the entropy and electric potential related contribution element of the grand thermodynamic potential in the competition with the energy element wins. From the changes of the slope of the mass versus $r_h$ with respect to the electric potential $\Phi$, we can see from the grand thermodynamic potential landscapes in Fig.\ref{FPlotPhi} that the black hole phases change from the small black hole preference (changing to shallower basin) to the large black hole preference (changing to deeper basin) as the charge $Q$ increases. On the other hand, from the changes of the slope of the electric potential related term ($Q \Phi$) versus $r_h$ with respect to the increase of the electric potential $\Phi$, large black hole phase is more preferred. We can also see from the grand thermodynamic potential landscapes in Fig.\ref{FPlotPhi} that the black hole phases change from the small black hole preference to the large black hole preference (changing to deeper and steeper basin) as the electric potential $\Phi$ increases.

Notice that the the black hole mass, entropy multiplied by $T$ and electric potential related contribution are all much larger than that of the resulting grand thermodynamic potential as shown in Fig.\ref{FPlotPhi}. Therefore, the grand thermodynamic potential landscape and the black hole phases are the results of the delicate balance among the three large numbers or scales, the mass, the entropy and the electric potential related contribution, giving rise to a much smaller number or scale of the grand thermodynamic potential. Due to the scale difference of the grand thermodynamic potential energy (small) in comparison to the energy, entropy and the electric potential related contribution (large), accurate quantifications of the energy, entropy and electric potential related contribution are essential for obtaining the accurate information of the grand thermodynamic potential and the thermodynamics of the black hole. Small errors in estimating the energy, entropy and electric potential can generate relatively larger errors in the grand thermodynamic potential estimation.  In other words, it might be difficult to switch from one state to another when considering the energy/mass or entropy or electric potential barrier alone. For example, it is very difficult to switch from a small size black hole state to the large size black hole state from the energy/mass perspective since the barrier ($\Delta E$, where $\Delta E$ is the energy barrier from initial to the final state) for switching is high compared to the thermal temperature and it is difficult to realize the transition through thermal motion. On the other hand, it is very difficult to switch from a large size black hole state to the small size black hole state from the entropy or electric potential related contribution perspective since the barrier for switching ($T\Delta S$ or $Q \Delta \Phi$,  where $T \Delta S$ is the entropy barrier from initial to the final state and $Q \Delta \Phi$ is the electric potential barrier from the initial to the final state) is high compared to the thermal temperature and it is difficult to realize the transition through thermal motion. However the competition among the mass/energy, the entropy and the electric potential related contribution due to their intricate balance can lead to the significantly reduced grand thermodynamic potential and the associated grand thermodynamic potential barrier between the small and large size black holes. The grand thermodynamic potential barrier height is in the order of $T$ which makes the thermal transitions possible between the the small and large size black holes.

\subsubsection{Varying the pressure or the cosmological constant}

\begin{figure}
  \centering
  \includegraphics[width=6cm]{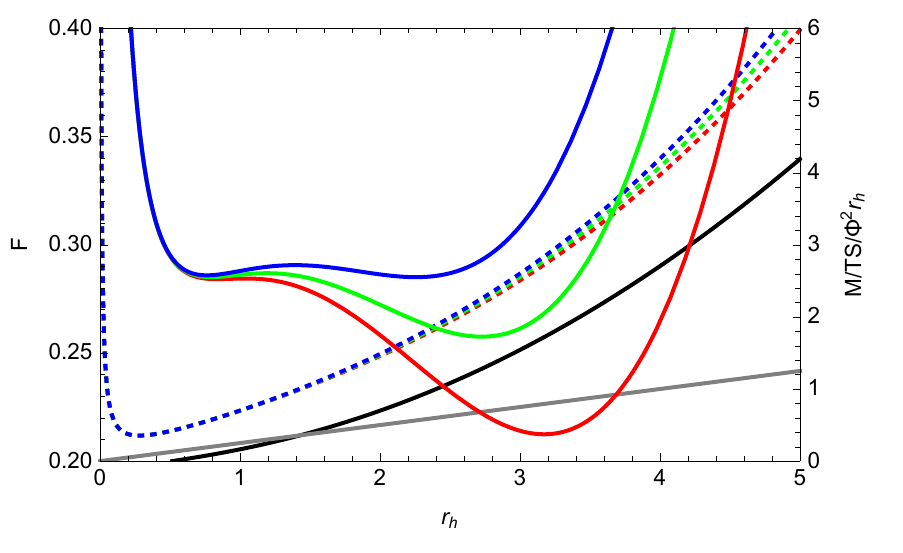}
  \caption{The free energy $F$ (solid), black hole mass $M$ (dotted), the product $TS$ of entropy and temperature (black solid) and the electric potential related term $\Phi^2 r_h$ (gray solid) as the functions of $r_h$ with different pressure $P$. The electric potential $\Phi=0.5$, the coupling constant $\alpha=0.1$, and the ensemble temperature $T=0.052$. The red, green, and blue curves correspond to $P=0.65P_c$, $0.7P_c$, and $0.75P_c$, respectively.}
  \label{FPlotP}
\end{figure}

\begin{table}[!htbp]\caption{Thermodynamic quantities of different types of black hole at different pressure with $\alpha=0.1$, $\Phi=0.5$, and $T=0.052$.\\}
\centering
\begin{tabular}{|c|c|c|c|c|c|}
\hline
BH Type&$P/P_c$&$r_h$&$M$&$S$&$F$\\
\hline
\multirow{3}*{Small}&0.65&0.838655&1.08856&1.9885&0.284136\\
\cline{2-6}
&0.7&0.790626&1.10308&1.66855&0.285072\\
\cline{2-6}
&0.75&0.762708&1.11412&1.48714&0.28589\\
\hline
\multirow{3}*{Intermediate}&0.65&1.03956&1.07441&3.44382&0.284418\\
\cline{2-6}
&0.7&1.19188&1.09889&4.68349&0.286859\\
\cline{2-6}
&0.75&1.39423&1.16276&6.52453&0.290586\\
\hline
\multirow{3}*{Large}&0.65&3.17251&2.48623&33.0703&0.212408\\
\cline{2-6}
&0.7&2.72629&2.06147&24.6106&0.257531\\
\cline{2-6}
&0.75&2.2555&1.67288&17.0043&0.28504\\
\hline
\end{tabular}
\label{TablePGCE}
\end{table}

\begin{figure}
  \centering
  \includegraphics[width=6cm]{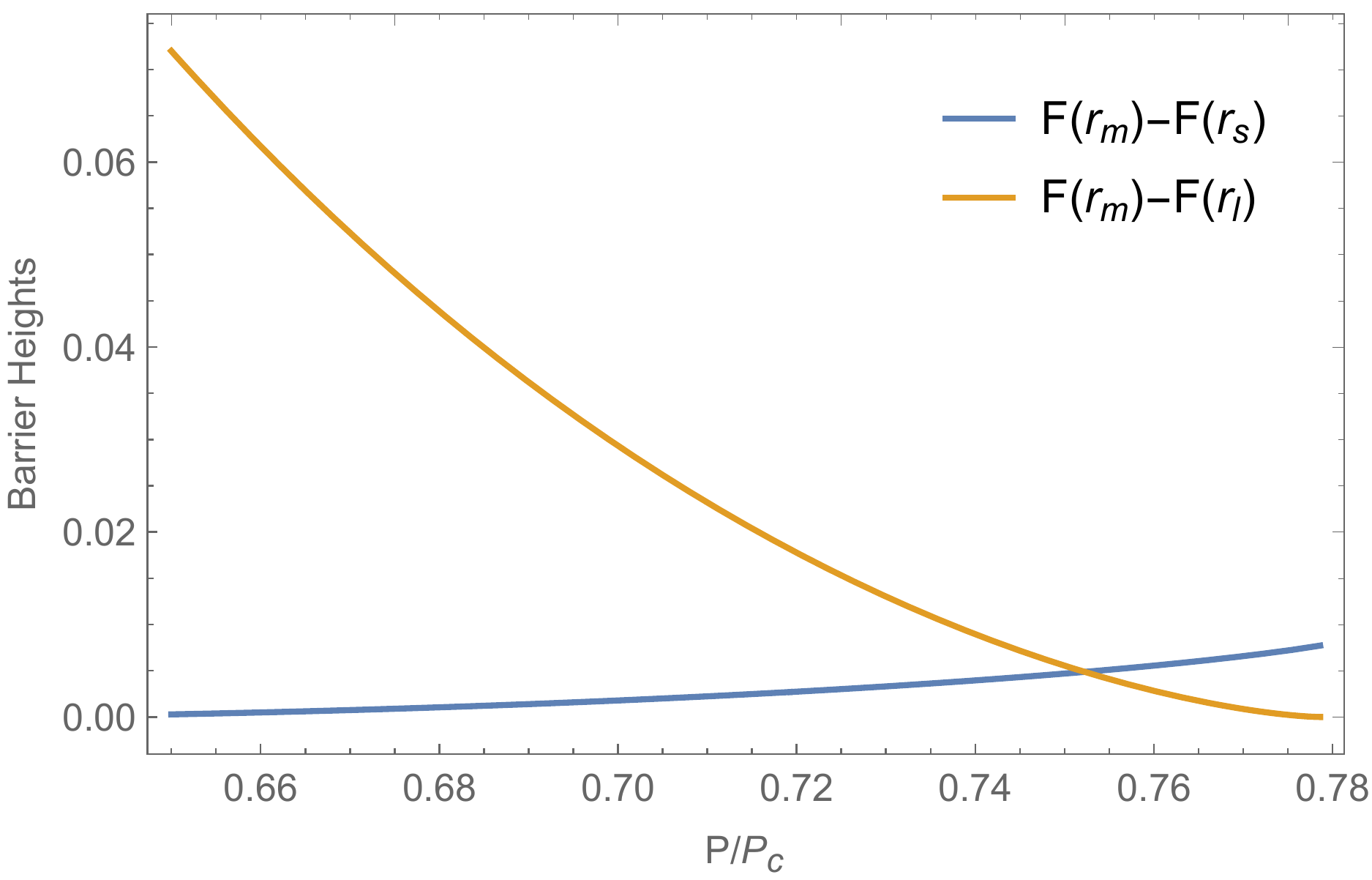}
  \caption{The barrier heights of free energy landscape as the function of the pressure. The barrier height between the small (large) black hole and the intermediate black hole monotonically increases (decreases) with $P$. The electric potential $\Phi=0.5$, the coupling constant $\alpha=0.1$, and the ensemble temperature $T=0.052$.}
  \label{BHPGCE}
\end{figure}

\begin{figure}
\centering
\subfigure[]{\label{MFPTstlPGCE}
\includegraphics[width=4cm]{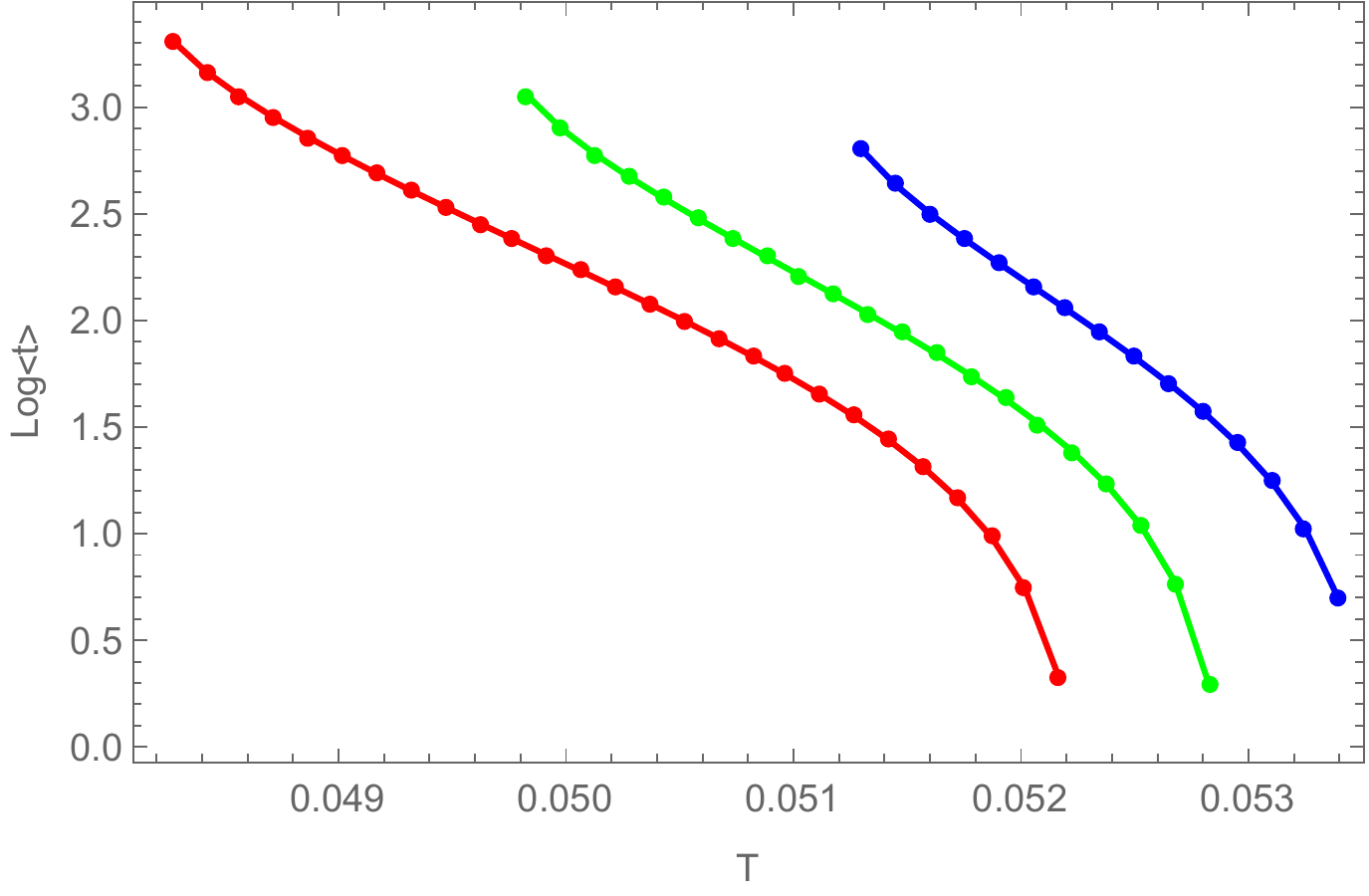}}
\subfigure[]{\label{RelFlucstlPGCE}
\includegraphics[width=4cm]{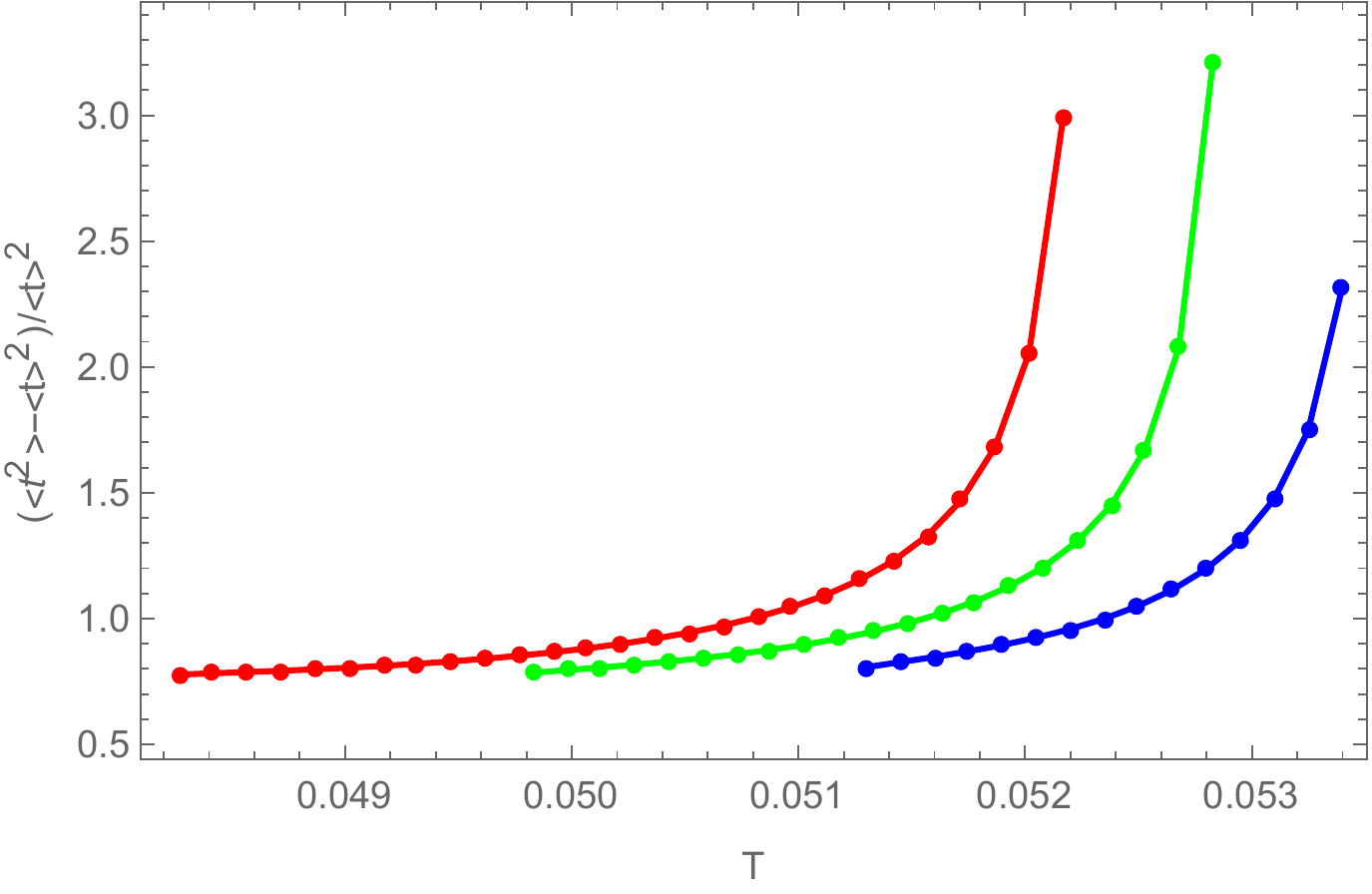}}
\subfigure[]{\label{MFPTltsPGCE}
\includegraphics[width=4cm]{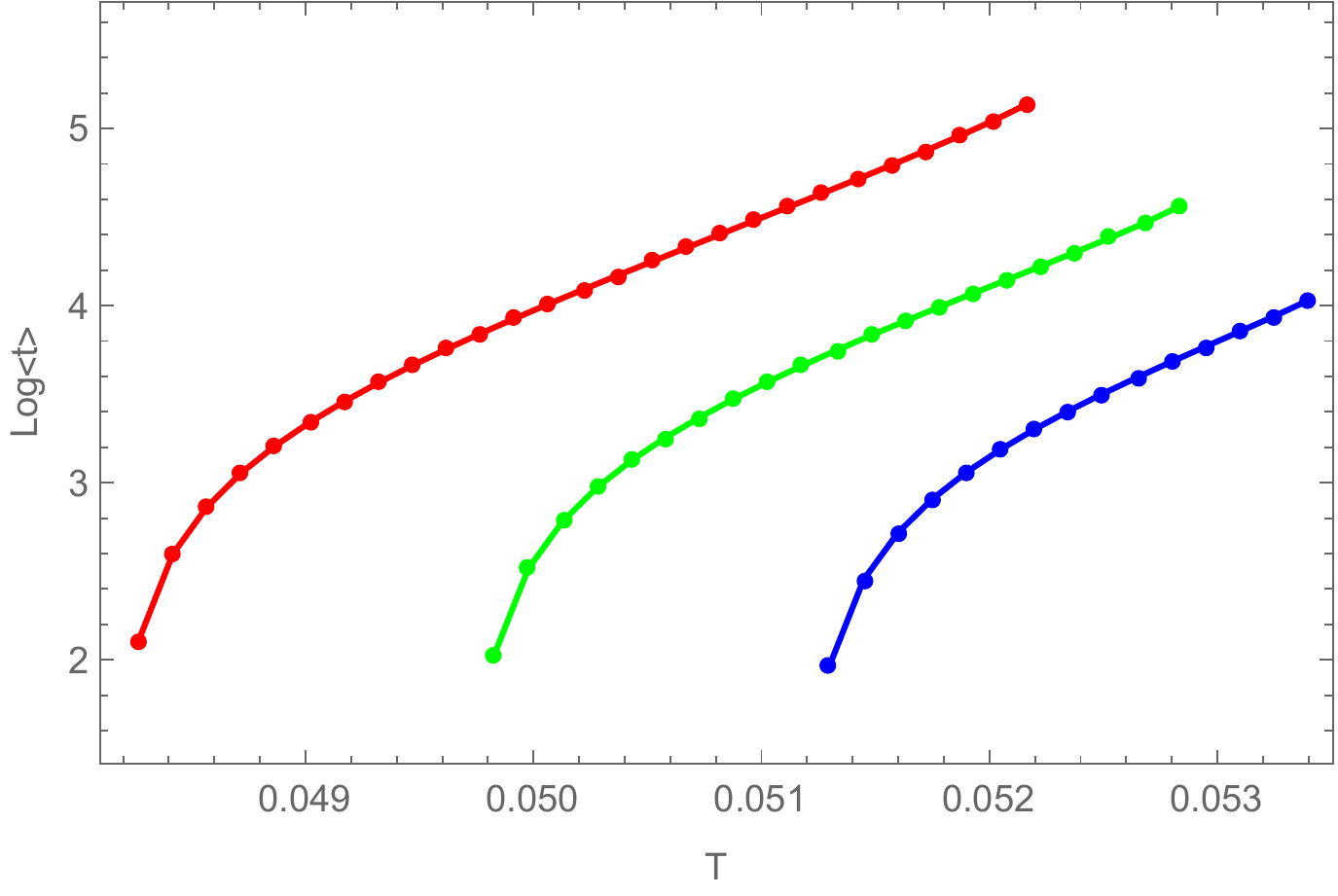}}
\subfigure[]{\label{RelFlucltsPGCE}
\includegraphics[width=4cm]{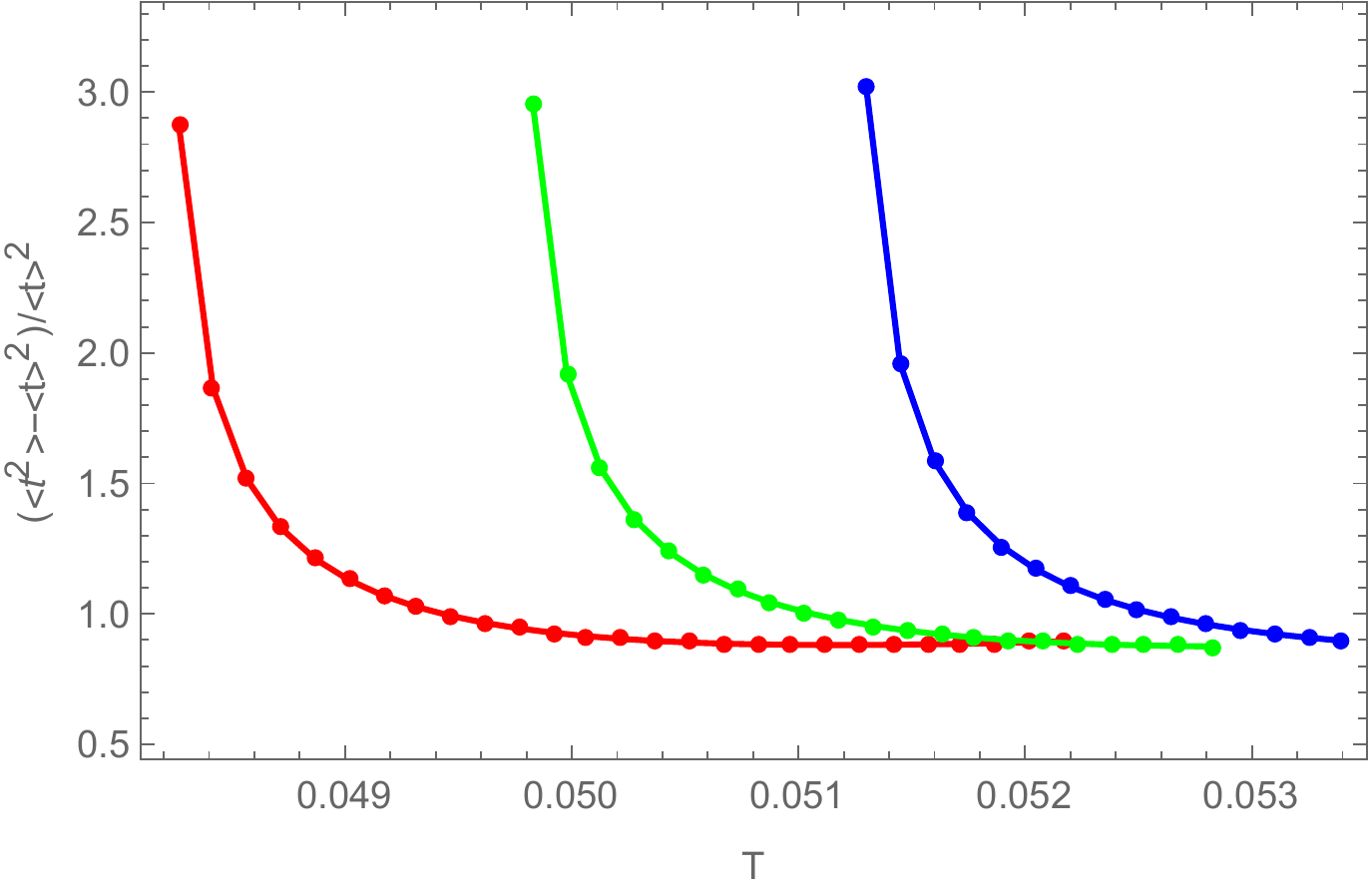}}
\caption{MFPT and the relative fluctuation of the state transition in the grand canonical ensemble as a function of temperature $T$ for the different pressure $P$.
The panels (a)-(d) shows the same quantities as in fig.\ref{MFPTGCEalpha}.
In each panel, the red, green, and blue curves correspond to $P=0.65P_c$, $0.7P_c$, and $0.75P_c$.
The electric potential $\Phi=0.5$, and the GB coupling constant $\alpha=0.1$, which are kept fixed.}
\label{MFPTGCEP}
\end{figure}

In Fig.\ref{FPlotP}, the free energy $F$, black hole mass $M$, the product $TS$ of entropy and temperature and the electric potential related term $\Phi^2 r_h$ as the functions of $r_h$ with different electric potential are plotted.
We can see that the black hole mass or energy is nonmonotonic with respect to $r_h$, In fact, certain small size black hole is preferred (minimum of M vs $r_h$). We can also see that the entropy multiplied by the temperature or electric potential term ($Q \Phi$) monotonically increases as $r_h$ increases. In fact, low energy or mass or high entropy or low electric potential related contribution ($-Q\Phi$) are always preferred when acting alone. Since both mass and entropy are nonlinear functions of the black hole size and the electric potential related contribution is linear in black hole size, the competition among the three  can generate interesting behaviors and possible new phases unexpected from the action of the mass or the entropy or the electric potential related contribution alone. As shown in Fig.\ref{FPlotP}, when the effective pressure $P$ or the equivalently absolute value of the cosmological constant $\Lambda$ is small, the small sized black hole is less preferred from the mass versus $r_h$ under shallower slope while when the effective pressure $P$ or the equivalently absolute value of the cosmological constant $\Lambda$ becomes bigger, the small sized black hole is more stable because the mass versus $r_h$ has steeper slope. Considering the entropy or electric potential related contribution alone, we see that it always prefers the large size black hole.The entropy and electric potential related contribution are not explicitly dependent on the effective pressure $P$ or the equivalently absolute value of the cosmological constant $\Lambda$   from the analytical expression and as shown in Fig.\ref{FPlotP}. The grand thermodynamic potential as the result of the competition among the mass (energy), the entropy and the electric potential related contribution leads to two possible stable black hole phases in certain parameter regimes, a small size black hole favored by energy or mass and a large size black hole favored by the entropy and electric potential related contribution. In other words, the grand thermodynamic potential has two minimum representing the two phases of the black hole, a low energy, low entropy and low electric potential related contribution phase of the small black hole  where the energy element of the grand thermodynamic potential in the competition with the entropy and electric potential related element wins and a high energy,  high entropy and high electric potential related contribution phase of the large black hole where the entropy and the electric potential related contribution element of the grand thermodynamic potential energy in the competition with the energy element wins. From the changes of the slope of the mass versus $r_h$ with respect to the effective pressure $P$ or the equivalently absolute value of the cosmological constant $\Lambda$, we can see from the free energy landscapes in Fig.\ref{FPlotP} that the black hole phases change from the large black hole preference (changing to shallower basin) to the small black hole preference  (changing to deeper basin) as the the effective pressure $P$ or the equivalently absolute value of the cosmological constant $\Lambda$ increases.

Notice that the the black hole mass and entropy multiplied by T as well as the electric potential related contribution are all much larger than that of the resulting grand thermodynamic potential as shown in Fig.\ref{FPlotP}. Therefore, the grand thermodynamic potential landscape and the black hole phases are the results of the delicate balance among the three large numbers or scales, the mass, the entropy and the electric potential related contribution,  giving rise to a much smaller number or scale of the grand thermodynamic potential. Due to the scale difference of the grand thermodynamic potential (small) in comparison to the energy, entropy and electric potential related contribution (large), accurate quantifications of the energy and entropy as well as the electric potential related contribution are essential for obtaining the accurate information of the grand thermodynamic potential and the thermodynamics of the black hole. Small errors in estimating the energy, entropy or the electric potential related contribution can generate relatively larger errors in the grand thermodynamic potential estimation. In other words, it is difficult to switch from one state to another when considering the energy/mass or entropy or electric potential barrier alone. For example, it is very difficult to switch from a small size black hole state to the large size black hole state from the energy/mass perspective since the barrier ($\Delta E$, where $\Delta E$ is the energy barrier from initial to the final state) for switching is high compared to the thermal temperature and it is difficult to realize the transition through the themral motion. On the other hand, it is very difficult to switch from a large size black hole state to the small size black hole state from the entropy or electric potential related contribution perspective since the barrier ($T\Delta S$ or $Q \Delta \Phi$,  where $T \Delta S$ is the entropy barrier from initial to the final state and $Q \Delta \Phi$ is the electric potential barrier from the initial to the final state) for switching is high compared to the thermal temperature and it is difficult to realize the transition through thermal motion. However the competition among the mass/energy, the entropy and the electric potential related contribution due to their intricate balance can lead to the significantly reduced grand thermodynamic potential and the associated grand thermodynamic potential barrier between the small and large size black holes. The grand thermodynamic potential barrier height is in the order of $T$ which makes the thermal transitions possible between the the small and large size black holes.

\subsubsection{Summary}

The free energy landscapes at different physical parameters are presented in Fig.\ref{FPlotalpha}, Fig.\ref{FPlotPhi}, and Fig.\ref{FPlotP}. It is shown that varying the parameter $\alpha$, $\Phi$, or $P$ leads to more significant changes in the large black hole than the small black hole. The barrier heights in the free energy landscapes when varying different physical parameters are plotted in Fig.\ref{BHalphaGCE}, Fig.\ref{BHPhiGCE}, and Fig.\ref{BHPGCE}. The numerical results of the kinetics through the MFPT and its relative fluctuation for different $\alpha$, $\Phi$, and $P$ are presented in Fig.\ref{MFPTGCEalpha}, Fig.\ref{MFPTGCEPhi}, and Fig.\ref{MFPTGCEP}, respectively. When $\alpha$ increases, the range of the ensemble temperature, where the double well shaped free energy landscape topography appears, moves to the left. When $\Phi$ increases, the temperature range also moves to the left. When $P$ increases, the range moves to the right. These behaviors are similar to that in the canonical ensemble. The thermodynamic quantities of different types of black hole are also listed in Table \ref{TablealphaGCE}, Table \ref{TablePhiGCE}, and Table \ref{TablePGCE}, respectively.

The plots of the free energy, black hole mass, the product of entropy and temperature, as well as the electric potential related term in Fig.\ref{FPlotalpha}, Fig.\ref{FPlotPhi}, and Fig.\ref{FPlotP} show that the free energy is the result of the delicate balance and competition between the three relatively large numbers, the energy, entropy multiplied by temperature, and the electric potential related term compared to $kT$. Low energy (mass) and low entropy can give rise to a stable thermodynamic state in terms of free energy minimum (energy/mass preferred) while the high energy (mass) and high entropy (entropy preferred) can also give rise to a stable state in terms of free energy minimum. The comparable free energy barrier with respect to $kT$ makes it possible for the switching from small size black hole state to the large size black hole state under thermal fluctuations and vice versa.

By means of the similar analysis, one can observe that the dependence of the MFPTs on the different physical parameters $\alpha$, $\Phi$, and $P$ are also similar to that in the canonical ensemble. In other words, in the grand canonical ensemble, when the GB coupling coupling constant increases, or the electric potential increases, or the pressure decreases, the barrier height between the small (large) black hole state and the intermediate black hole state becomes lower (higher). This leads to the behavior that the small black hole state is more easier to escape to the large black hole state while the inverse process becomes more harder when varying the parameters, i.e. the small (large) black hole state will becomes less (more) stable.

For different physical parameters, the MFPT of the state transition process from the small black hole to the large black hole is a monotonous decreasing function of the ensemble temperature $T$ as shown in Fig.\ref{MFPTstlalphaGCE}, Fig.\ref{MFPTstlPhiGCE}, and Fig.\ref{MFPTstlPGCE}. The reason is that the barrier height between the small black hole state and the intermediate black hole state is also a decreasing function of the ensemble temperature. Moreover, as shown in Fig.\ref{MFPTltsalphaGCE}, Fig.\ref{MFPTltsPhiGCE}, and Fig.\ref{MFPTltsPGCE}, for the process from the large black hole and the small black hole, the MFPT is a monotonous increasing function of the temperature $T$, which is the consequence of the increasing barrier height between the intermediate black hole and the large black hole when increasing the ensemble temperature. It is concluded that the MFPT with different parameters $\alpha$, $\Phi$, $P$ has the universal tendency when increasing the ensemble temperature.

The numerical results shown in Fig.\ref{RelFlucstlalphaGCE}, Fig.\ref{RelFlucstlPhiGCE}, and Fig.\ref{RelFlucstlPGCE} indicate that the relative fluctuation of first passage time for the process from the small black hole to the large black hole is relatively large when the temperature approaches $T_{max}$. When the temperature approaches $T_{max}$, the barrier height between the small black hole and the intermediate black hole approaches to zero. In this case, the transition process from the small black hole to the large black hole is dominated by the thermal fluctuation. This is the reason that the relative fluctuation is large at the higher temperature. As shown in Fig.\ref{RelFlucltsalphaGCE}, Fig.\ref{RelFlucltsPhiGCE}, and Fig.\ref{RelFlucltsPGCE}, the relative fluctuation for the process from the large black hole to the small black hole is relatively large at $T=T_{min}$. The reason behind is that the barrier height between the large black hole and the intermediate black hole approaches to zero at $T=T_{min}$ and the state switching process is dominated by the thermal fluctuation at the lower temperature. Therefore, we can conclude that the MFPT and its fluctuations are closely related to free energy landscape topography in the grand canonical ensemble through the barrier heights and the ensemble temperature.

\section{Conclusion}

In summary, based on the free energy landscape, we study the thermodynamic phase transition and the underlying kinetics of the four dimensional charged GB AdS black holes both in canonical ensemble and in grand canonical ensemble. It is shown that the free energy landscape topographies in the canonical ensemble and in the grand canonical ensemble have the same shape of double basins with each representing one stable/unstable black hole phase (the small or the large black hole). From the free energy landscape topography, we show that there is a small/large black hole phase transition which is determined by the equal depths of the basins.

Furthermore, we demonstrate the underlying kinetics of the phase transition under the thermal fluctuations by studying the time evolution of the probability distribution of the state in the ensemble as well as the mean first passage time and its fluctuation of the state switching and phase transitions. The results show that the final distribution is given by the Boltzmann law while the mean first passage time and its fluctuations are closely related to the free energy landscape topography through the barrier heights and the ensemble temperature. The temperature dependence of the mean first passage time and the kinetic fluctuations with different physical parameters is also studied in detail.

In the canonical ensemble, when the GB coupling coupling constant increases, or the electric charge increases, or the pressure decreases, the small black hole state is more easier to escape to the large black hole state meanwhile the inverse process becomes more harder.
In the grand canonical ensemble, when the GB coupling coupling constant increases, or the electric potential increases, or the pressure decreases, the small black hole state is more easier to escape to the large black hole state meanwhile the inverse process becomes more harder.

This provides a complete description of the kinetics of the phase transition of the four dimensional GB black hole both in the canonical ensemble and in the grand canonical ensemble. Our studies shown that the free energy is the result of the delicate balance and competition between the two relatively large numbers, the energy and entropy multiplied by temperature compared to $kT$. Low energy (mass) and low entropy can give rise to a stable thermodynamic state in terms of free energy minimum (energy/mass preferred) while the high energy (mass) and high entropy (entropy preferred) can also give rise to a stable state in terms of free energy minimum. The comparable free energy barrier with respect to $kT$ makes it possible for the switching from small size black hole state to the large size black hole state under thermal fluctuations and vice versa.

At last, we should point that the present work concentrates on the four dimensional GB black holes. It has been shown that the high dimensional GB black holes have much richer features  \cite{Cai:2001dz,Cai:2013qga,Zou:2013owa,Wei:2014hba,Ghosh:2019pwy,Zhou:2020vzf}, such as more than one intermediate states. Recently, the neutral small/large GB AdS black hole phase transition in the five dimensional GB gravity was studied in \cite{Wei:2020rcd} by using the formulism of the free energy landscape. However, the phase transition and the kinetics of the high dimensional charged GB AdS black holes based on the free energy landscape deserve further study in the future.

 \end{document}